# Transporte Aéreo:

# Economia e Políticas Públicas

Alessandro Oliveira

# Transporte Aéreo:

# Economia e Políticas Públicas

1ª Edição

**Índices  para  catálogo  sistemático:**

1.   Transporte aéreo: Aspectos econômicos
        387.71

# Sumário

## A Regulação Econômica do Transporte Aéreo ..........79

## Princípios Balizadores da Regulação e das Políticas Públicas do Transporte Aéreo

## Notas

## Bibliografia

# Apresentação

A aviação civil é um importante instrumento para a alavancagem do crescimento econômico, dado o seu papel no escoamento produtivo e na locomoção das pessoas, e instrumento fundamental para a integração, tanto interna, em países com dimensões continentais, como o Brasil, como externa, interconectando diferentes países.

No mundo todo se presencia um momento de grande expansão do setor, com transformações significativas na estrutura e no acesso a esses serviços. No Brasil, essas transformações não foram diferentes. O número de passageiros cresceu no mercado doméstico a taxas mais que duas vezes superiores a da renda nacional. Os benefícios dessa expansão só foram possíveis com a reestruturação da indústria num regime mais propício a concorrência, em que menores preços atraíram mais usuários.

Até 1980, em todo o mundo, o setor era planificado, fosse através de uma única empresa nacional, situação tipicamente encontrada em países de área territorial menor, ou por meio de malha aérea dividida em sub-monopólios concedidos a algumas poucas empresas, este caso, mais evidente em países territorialmente maiores. A construção da infra-estrutura aeroportuária, com altos investimentos afundados, cabia ao Estado, que criava as condições para a viabilização da operação dos serviços aéreos.

Nesse regime de quase monopólio, a oferta ditava a demanda. As empresas monopolistas das suas rotas equalizavam a sua questão econômico-financeira através de preços uniformes, por meio de subsídios entre rotas e, por conseguinte, entre aeroportos. Observava-se que, nas rotas mais densas e nos horários mais convenientes, eram cobrados preços semelhantes a outras opções menos atraentes, em que as empresas voavam com maior ociosidade. Ineficiência que criava custos para a expansão da economia ao gravar seus setores mais dinâmicos dependentes dos serviços aéreos.

Impulsionados pela necessidade de oferecer incentivos de dinamismo ao setor, os Estados Unidos e a Europa tomaram medidas de desregulamentação econômica, flexibilizando as alocações de rotas e de preços. Essas medidas permitiram a entrada de novos atores econômicos que de posse de tecnologias e capital mais baratos, ofereciam serviços a preços mais módicos.

Em termos de consumo do serviço, os resultados foram muito promissores, pois a indústria poderia melhor aproveitar as suas economias de rede. Todavia, essas economias de rede dependem da utilização conjunta e articulada de várias plataformas de infra-estrutura, tais como, aeroportos e espaço aéreo.

Nos anos 1990, com o crescimento econômico contínuo, jamais visto nas economias de mercado, conjugado com o êxito na queda das tarifas (efeitos renda e preço), todos os países lograram dinamizar seu setor de aviação civil com recordes de volume de passageiros.

Essas mudanças ensejaram diversas transformações no papel do agente regulador da aviação civil. Uma agência de aviação civil tem como principal função corrigir uma falha de mercado advinda da assimetria de informação entre usuários e provedores do serviço quanto à segurança operacional. O usuário do transporte aéreo, muito mais do que em outros modais, incorreria em altos custos informacionais para poder observar corretamente a segurança operacional de cada serviço que adquire. Mesmo que os provedores incorressem nesse custo, a possibilidade de risco sistêmico, quando um provedor age incorretamente, torna ineficiente as ações individuais não cooperativas dos provedores. Dessa forma, uma agência que credencie esses parâmetros geraria ao setor uma significativa externalidade positiva.

Contudo, além da segurança operacional, já discutida, atualmente o regulador é chamado a zelar pela maximização do bem-estar do consumidor em termos de quantidade e qualidade dos serviços ofertados pelos operadores de transporte. De modo a cumprir com esta finalidade, é necessário lidar com outra falha de mercado, a saber: as economias de rede. Na concorrência, os agentes não conseguem informações suficientes para internalizar, nas suas decisões, as decisões dos outros agentes. Cabe ao regulador criar mecanismos de incentivo para essa internalização, seja por meio da redução das barreiras regulatórias à entrada e da garantia de acesso aos aeroportos seja na precificação diferida das plataformas. Sem isso teremos níveis ineficientes de congestionamento com excesso de atrasos e cancelamentos de vôos.

Cumpre destacar que, diante do cenário de rápida expansão do setor, nestas últimas décadas, também se observou a redução do número de acidentes aéreos. Os mercados de tecnologia de aviação e de controle de espaço aéreo tiverem incentivos para prover tecnologias que facilitavam a segurança para essa nova realidade. Emerge, então, a importância do regulador aéreo de se adiantar nas necessidades de uso e de implementação dessas tecnologias e de preparar o capital humano que possa se utilizar dessas.

Diante desta nova realidade, o planejamento da infra-estrutura aeroportuária se transforma radicalmente de indutor de oferta para antecipador da demanda. Neste cenário, cabe, mais uma vez, ao regulador liderar essa transformação, normatizando preços eficientes e promovendo a concorrência entre sítios aeroportuários.

A expansão desse novo paradigma para o setor de transporte aéreo internacional encontra entraves, oriundos de uma filosofia de proteção dos mercados domésticos e por questões de soberania. O serviço do transporte internacional é regido por Acordos de Serviços Aéreos (ASA), que possuem a natureza jurídica de um Tratado, mas na prática, funcionam como instrumentos regulatórios, vez que controlam o acesso ao mercado (capacidade, pontos, rotas e liberdades do ar), a designação das empresas que utilizaram essa capacidade e tarifas (dupla aprovação, país de origem, dupla desaprovação ou liberdade tarifária). Dependendo da forma como está desenhado o ASA entre dois ou mais países, o fluxo do transporte aéreo internacional pode ser restringido, por meio de quotas.

Êxitos de liberalização na área internacional aconteceram onde a integração econômica de países motivou a concorrência, como, por exemplo, a política de Céus Abertos na Europa e na região da Australásia, em que os critérios de reciprocidade foram flexibilizados em troca de uma maior pujança para o setor. Governos desses países observaram que tal flexibilização traria ganhos para todos. Mais uma vez cabe ao regulador, agora, implementar um modelo de liberalização que faça uma transição gradual do modelo protecionista. Dado o afastamento geográfico do Brasil, essas oportunidades mostram-se vitais para o desenvolvimento do setor de turismo e para o escoamento das nossas importações e exportações.

Por fim, a criação da Agência Nacional de Aviação Civil (ANAC), em 2005, inaugurou uma nova fase no marco regulatório do setor aéreo brasileiro. Como ente autônomo, cabe a Agência implementar essa nova realidade, assim como estabelecer proposições de mudança de política.

Portanto, é mais do que oportuna a publicação deste livro, do professor Alessandro Oliveira, reconhecido economista e pioneiro nas pesquisas quantitativas da Organização Industrial da aviação civil no Brasil. Seus estudos têm sido uma fonte inspiradora para todos que atuam no desenvolvimento da regulação econômica desse setor. Logo, uma publicação como esta, que reúne grande parte da sua bibliografia, é muito bem-vinda e, certamente, permitirá que reguladores e pesquisadores possam contar com uma obra de referência para aprimorar seus conhecimentos do setor.

Ronaldo Seroa da Motta

Diretor da Agência Nacional de Aviação Civil, ANAC

*Rio de Janeiro, dezembro 2008*

# Prefácio

É muito fácil encontrar pessoas apaixonadas por aviação. Toda a história de como o homem conseguiu projetar aparelhos mais pesados que o ar e com capacidade de voar é, por si só, um apaixonante enredo. Sem dúvida, o transporte aéreo atrai paixões de muitas pessoas, a começar por aqueles que viveram ou ainda vivem o setor aeronáutico diariamente. Mas também atrai pelo fascínio natural que irradia, inclusive nas novas gerações – no encanto de crianças que ainda enxergam no avião uma das grandes maravilhas do mundo, mesmo nestes tempos dos *video games* de última geração. Ainda mais em se tratando de um País com o legado de um Santos Dumont, e que é berço de um ITA e de uma Embraer, além de possuir uma ampla gama de engenheiros especializados do mais alto gabarito. É, portanto, mais do que natural que o tema seja ponto focal em vários círculos de conhecedores e interessados em geral.

A aviação comercial é, igualmente, uma área na qual os aficionados costumam enveredar com bastante maestria. Toda a questão do nascimento da aviação civil, do surgimento das primeiras companhias aéreas e do estabelecimento de um regulador que logo foi incumbido de tentar arrefecer a competição entre elas – além de cuidar da sua atividade-fim, de fiscalização da segurança de vôo e de regulação técnica –, das famílias e modelos de aeronaves utilizados ao longo da história, etc. São todos temas que suscitam acalorados debates dentre aqueles que se interessam pelo assunto. Ainda mais porque o transporte aéreo da atualidade é também repleto de temas igualmente instigantes – ao mesmo tempo que polêmicos –, como a questão da abertura de capital de companhias aéreas para estrangeiros, a alocação de *slots* em aeroportos saturados, o fomento à aviação regional, os direitos de cabotagem, as alianças estratégicas e o papel de companhias aéreas do tipo "*Low Cost, Low Fare*", dentre outros.

A presente obra foi feita por um desses apaixonados. Mas um tipo diferente desses, talvez até mesmo um estranho no ninho. Para começar, não sou engenheiro, mas economista. Não à toa dedico minha vida ao estudo dos aspectos econômicos do transporte aéreo. Interessante notar que, salvo raras exceções, os economistas no Brasil sempre relegaram o tema "transporte aéreo" para segundo plano. Historicamente, assuntos como os rumos da economia, as crises e os planos de estabilização sempre ocuparam a atenção da maior parte dos cérebros formados em Ciências Econômicas.

Este livro é voltado a questões como demanda e oferta de transporte aéreo. São temas da rotina do setor e que fazem parte da temática principal aqui apresentada, mas em uma abordagem introdutória, voltada ao leitor de primeira viagem – ou melhor, de primeiro vôo. Longe de ser um livro de cunho acadêmico, um manual cuja completude deixaria o iniciante entediado e o especialista frustrado, ou pautado por equações e fórmulas em um formalismo que me custaria o entendimento de uma grande maioria interessada, a presente obra tem a única pretensão de ser uma leitura agradável. Nas palavras do Nobel de Economia Paul Krugman, vencedor de 2008, este livro seria como uma "literatura de aeroporto", isto é aquelas leituras que você pode muito bem fazer enquanto aguarda a chamada para o embarque do seu vôo ou conexão.

Espero que "Transporte Aéreo: Economia e Políticas Públicas", ou TAEPP, para os mais íntimos, possa ser também uma leitura útil para a compreensão do funcionamento do setor aéreo em regime de livre mercado. Isso porque, ainda hoje, há muita incompreensão sobre como os mercados aéreos funcionam e como as companhias aéreas se comportam nesses ambientes. Essa falta de maior entendimento pode ser percebida no âmbito da tomada de decisão governamental, mas também em muitas instâncias empresariais. Em sendo bem-sucedido nessa meta, TAEPP já seria uma verdadeira realização.

Em termos acadêmicos, é também possível usar TAEPP como fonte de informação e conhecimento para alunos e interessados em geral. Pode ser utilizado como material suplementar em treinamentos visando à qualificação de técnicos e gestores do setor, ou em cursos acadêmicos de transportes, turismo e mesmo negócios. Por ser repleto de referências à legislação do transporte aéreo, pode igualmente ser uma referência para interessados da área de direito. Adicionalmente, como sua abordagem dos problemas do transporte aéreo em geral centra-se na perspectiva da autoridade regulatória, pode ser útil como um material introdutório aos aspectos de regulação e desregulação econômica, sem a formalística acadêmica típica da literatura da área.

Minha intenção, ao concretizar esse livro, é traduzir na linguagem o mais simples e objetiva possível as questões que vêm sendo intensamente discutidas não apenas no âmbito decisório de empresas aéreas e instituições que cuidam do setor aéreo, mas também no âmbito da academia. Para isso, conto com o auxílio de uma gama de estudos e artigos científicos publicados no âmbito do NECTAR, Núcleo de Economia dos Transportes, Antitruste e Regulação. O NECTAR é o centro de pesquisas do ITA que fundei em 2004. Não me canso de dizer com orgulho que o NECTAR é a única instituição de pesquisas da América Latina exclusivamente voltada para a economia dos transportes, com ênfase no modal aéreo.

O NECTAR veio resgatar a longa tradição de atuar em pesquisas na área de economia da aviação no âmbito do ITA. Essa linha de pesquisa foi extremamente pujante no final da década de 1970 e início dos anos 1980, sob coordenação e influência dos professores Michal Gartenkraut e Luís Paulo Rosenberg. Dissertações clássicas foram produzidas na época, como as dos professores Arnoldo Cabral, "*Um modelo oligopolístico do mercado de transporte aéreo brasileiro*" (1979), Protógenes Porto, "*Distorções na Proporção de Insumos em um Oligopólio: Aplicação ao Caso de Transporte Aéreo*" (1981) e Dario Rais Lopes, "*Avaliação dos Custos Operacionais no Desempenho das Empresas Brasileiras de Transporte Aéreo Domestico Regular*" (1982).

Com o NECTAR, desde 2004, foram produzidas dissertações de mestrado que reativaram a linha de pesquisa, como a de Débora Lovadine "*Estudo empírico da conduta competitiva das Companhias Aéreas Brasileiras*" (2006), Renée Ferraz, "*Um estudo da chegada de passageiros em sistemas de reservas no contexto do Overbooking*" (2006), Humberto Bettini, "*Macroeconomic, regulatory and market determinants of airline capacity setting*" (2006), Moisés Vassallo, "*Simulação de Fusões: Aplicação ao Transporte Aéreo*" (2007) e Natália Ferreira, "*Guerras de Preço no Transporte Aéreo: Competição Saudável ou Predação? Aplicação de Modelo de Parâmetro de Conduta Competitiva*" (2008). Aliás, sobre as atividades do NECTAR, o Professor Michal recentemente se manifestou em uma mensagem enviada para um conjunto de professores do ITA: "*Sempre estive convencido que o ITA é o lugar certo, sem prejuízo de atividades nas demais instituições, para constituir o centro articulador de ensino, estudos e pesquisas em Transporte Aéreo. Nada além do que já estava previsto no Plano Smith-Montenegro. Alessandro: com o seu esforço pessoal, você tornou este sonho uma realidade. Por isso, a você os maiores agradecimentos*". Uma mensagem como essa deve ser guardada como um daqueles diplomas que se obtém apenas uma vez na vida. É uma felicidade poder compartilhar isso com todos.

A realização deste livro consta do planejamento estratégico do NECTAR, elaborado em 2004. Tratava-se de um dos objetivos mais importantes a serem realizados dentro dos quatro anos iniciais de constituição do núcleo. Olhando para trás, penso que muito foi conquistado. Temos hoje uma marca (NECTAR) forte e bem focada, um bom reconhecimento dos pares no meio acadêmico e profissional. Enfim, um nome e uma reputação a zelar. Tivemos, conforme planejado com a Fundação de Amparo à Pesquisa do Estado de São Paulo, FAPESP, em 2004, o estabelecimento do convênio com a autoridade antitruste (SEAE), o desenvolvimento de um portfólio de projetos de iniciação científica e mestrado, um portfólio de publicações, quer seja de *papers* já publicados, *papers* em construção, *papers* em fase de submissão, artigos em fase de *revise-and-resubmit*. Nossa página na *internet* em constante melhoria e atualização, a formação de uma rede de pesquisadores supra-ITA, um conjunto expressivo de premiações, dentre muitas outras atividades. Penso que o NECTAR hoje, seja por meio de seus *alumni*, pesquisadores, ou por meio de seus artigos disponíveis na *internet*, tem um papel positivo no planejamento e execução da regulação do transporte aéreo no Brasil.

Fico feliz por ter tido esses quatro anos de trabalho no projeto de concepção e constituição do NECTAR, e penso que a FAPESP – via seu assessor *ad hoc*, que recentemente avaliou o projeto e suas realizações – também se manifestou positivamente. Ponto não para mim, que fui mero articulador de todo o processo. Ponto sim, para todos aqueles que contribuíram para que isso se concretizasse, a começar pelos reitores Michal Gratenkraut e Reginaldo Santos, os chefes de Divisão Cláudio Jorge Alves e Carlos Müller, os colegas de trabalho no dia-a-dia, e, é claro, os alunos. É uma conquista coletiva, parabéns a todos nós.

Trocando em miúdos, é, assim, uma extrema felicidade poder concretizar essa etapa da constituição do núcleo. Para se chegar à versão final do livro, foram necessárias muitas horas de trabalho pela madrugada afora. Sendo sincero, muitas vezes cheguei a literalmente acordar sobre os teclados. Mas sem a ajuda e o apoio de inúmeras pessoas e instituições, não teríamos concretizado esse sonho. Para começar, agradeço ao Instituto Tecnológico de Aeronáutica (ITA), à Fundação de Amparo à Pesquisa no Estado de São Paulo (FAPESP), pelo apoio. Não posso deixar de agradecer aos amigos Frederico Turolla, Lucia Helena Salgado e Cristian Huse, pela força e parceria. Agradeço também a Humberto Bettini, Moisés Vassallo, Natália Ferreira, Débora Lovadine, Yuri Kretzmann, Rodrigo Oliveira, Carlos Müller, Cláudio Jorge Alves, Fridhilde Manolescu, Dante Aldrighi, Márcio Nakane, Erivelton Guedes, José Maria Silveira, Apóstole Lack Chryssafidis, Adalberto Febeliano, Clarice Rodrigues, Juliano Norman, Ronaldo Seroa, Marcelo Guaranys, Maria Nazareth da Silva, Erika Monteiro, Thelma Harumi Ohira, dentre inúmeros outros colegas. Agradeço a Deus, a Ângela Marques, Aldo Barcia, Aparecida Oliveira, José Ribeiro Soares, a minha mãe, Elena, e meu pai, Pedro Coimbra (in memorian), aos quais devo agradecimentos eternos.

Alessandro Oliveira

*São José dos Campos, janeiro de 2009.*

*A minha esposa Rachel e minha filha Maria Clara.*

# Introdução

O transporte aéreo é uma indústria que faz parte do cotidiano de milhões de pessoas no País. Por um lado, é um bem de consumo para centenas de milhares de famílias brasileiras, que compram passagens aéreas para desfrutarem suas férias, para cuidarem de seus interesses particulares, sejam eles referentes a saúde, educação e treinamento, cultura, lazer, congressos, etc. Por outro lado, é um bem industrial, um verdadeiro insumo produtivo de pequenas, médias e grandes corporações responsáveis pelo desenvolvimento econômico nacional.

Em uma pesquisa realizada em conjunto por Peter D. Hart Research e o Winston Group, em maio de 2008, junto a pouco mais de mil passageiros nos Estados Unidos, revelou-se que 66% deles declaravam-se "satisfeitos" com o transporte aéreo, apesar de uma parcela significativa também ter demonstrado alguma frustração com o modal. Entretanto, dentre os insatisfeitos apontados pela pesquisa, uma proporção relevante era de passageiros mais freqüentes, ou seja, aqueles que mais realizam viagens por ano e que, portanto, mais demandam o sistema aéreo e aeroportuário como um todo. De fato, dentre os passageiros que viajam cinco vezes ou mais por ano, 48% se mostravam insatisfeitos com o setor. Isso não representa a maioria dos viajantes freqüentes, mas constata uma realidade que pode ser facilmente extrapolada para o Brasil da atualidade: há ainda muito que se aperfeiçoar a gestão e as políticas públicas do setor aéreo até que se alcance a finalidade última que é termos o consumidor bem atendido em conjunto com a geração de riqueza e progresso técnico.

O transporte aéreo no Brasil nas últimas décadas passou por uma reviravolta regulatória que permitiu que as empresas passassem a concorrer entre si, tendo por objetivo o aumento da competitividade do setor e o bem-estar do passageiro. De fato, a chamada "Política de Flexibilização da Aviação Comercial", dos anos 1990 e 2000, expôs à competição nossas *legacy carriers* – companhias que conviveram com o regime regulatório estrito – tornando o mercado das companhias aéreas visivelmente mais livre em termos de precificação, acesso e mobilidade. Entretanto, a concentração dessa indústria e das rotas e aeroportos mais rentáveis nas mãos de poucas empresas tem se mostrado um elemento dissipador desses potenciais ganhos, e enseja o acompanhamento de perto por parte das autoridades. A alocação de direitos de pouso e decolagem – os chamados *slots* – tornou-se questão regulatória fundamental na tentativa de se garantir a manutenção dos ganhos advindos com a Política de Flexibilização. Conceder *slots* passou a ser o mesmo que conceder direitos de exercício de poder de mercado, com sérias implicações sobre as relações de oferta e de demanda e sobre o desempenho da indústria.

O conceito chave para se compreender a dinâmica competitiva do setor de transporte aéreo, as questões de poder de mercado das firmas estabelecidas e das potenciais intervenções regulatórias em ambientes livres mas com alta concentração é o conceito de ***contestabilidade de mercado***. Informalmente, contestabilidade é a extensão na qual a provisão de um bem ou serviço está aberta para fornecedores alternativos. Assim, tem-se que um mercado seria contestável se novos fornecedores podem entrar (e sair) facilmente, constituindo-se como alternativas reais de consumo. A simples ameaça de tal entrada já serviria para gerar uma disciplina nas atuais firmas instaladas, que acabariam apresentando comportamento – ou conduta – distinta daquela que teriam se não houvesse a possibildade de entrada. Assim, a característica de "contestável" de um mercado serviria como um afrouxamento das condições de monopólio ou oligopólio, inibindo aumentos abusivos de preços muito acima dos custos. A livre entrada e saída derrubaria qualquer tentativa de monopolização nesse sentido.

Em termos formais, a Teoria dos Mercados Contestáveis foi desenvolvida no início dos anos 1980 pelos clássicos estudos do economista norte-americano William Baumol. Ele definiu "contestabilidade" não como um sinônimo de concorrência, mas como se referindo a uma situação na qual um provedor encara uma ameaça que considera crível de competição. A credibilidade da entrada torna-se, portanto, um conceito importante no arcabouço desenvolvido. Esses estudos permitiram chegar à conclusão de que a regulação econômica que gera barreiras à entrada de novos concorrentes e o controle de preços seria desnecessária no sentido de evitar comportamentos abusivos das empresas, dado que se houver crença na livre entrada e saída, isso por si só já restringiria as práticas das firmas estabelecidas (incumbentes) e protegeria o consumidor. Foi intensamente utilizada como argumentação para a desregulação do mercado de transporte aéreo nos Estados Unidos que culminou no *Airline Deregulation Act* de 1978.

> *De vinte em vinte anos, a história da regulação do transporte aéreo nos Estados Unidos passava por importantes reformas regulatórias. Curiosamente, as alterações sempre aconteciam em anos terminados por oito. Em 1938, houve o Civil Aeronautics Board Act, que criou o então regulador, o CAB. Em 1958, aconteceu o Federal Aviation Agency Act, que constituiu a FAA, órgão responsável pela regulação e fiscalização da segurança. Por fim, em 1978, houve o Airline Deregulation Act, que promoveu a completa desregulação econômica do setor.*

Conta Martin (2000) que a Teoria dos Mercados contestáveis foi apresentada ao mundo acadêmico como um direcionamento para a conduta dos reguladores, sempre que a regulação se fizesse necessária. Os conselhos da Teoria da Contestabilidade seriam os de permitir liberdade de entrada e saída, a flexibilidade de preços e de assegurar acesso equânime aos competidores. No decorrer dos trinta anos que se seguiram ao *Airline Deregulation Act* (ADA) do Presidente Jimmy Carter, muito se investigou, estudou e analisou acerca do comportamento de firmas em mercados aéreos liberalizados. Muitas lições foram aprendidas. Sabe-se hoje em dia que os mercados aéreos não são perfeitamente contestáveis. Os próprios Baumol e o ex-presidente do *Civil Aeronautics Board* (CAB), Alfred Kahn, reconheceram essa evidência *ex-post*. Muito pelo contrário, mesmo em regime de plena liberdade estratégica, com desregulação econômica em estágio avançado, as companhias aéreas podem sempre encontrar formas de erigir barreiras à entrada a novos competidores e de exercer poder de mercado às custas do consumidor. O bloqueio à entrada em aeroportos congestionados, por exemplo, é um dos mais importantes tipos de barreiras criadas pelas companhias aéreas. E a captura das atenções do regulador no exercício da regulação de rotina é outra forma de barreira.

O presente livro visa gerar conhecimento sobre a evolução do teor de contestabilidade dos mercados no transporte aéreo regular doméstico de passageiros no Brasil. A aplicação a esse segmento da indústria será utilizada na maioria dos casos aqui estudados. Buscará entender como o regulador evoluiu no sentido de flexibilizar o mercado e conceder maior liberdade de atuação das operadoras na expectativa de beneficiar o passageiro. Buscará, sempre que possível, embasar e pautar as discussões em estudos e pesquisas sobre o transporte aéreo realizadas no Brasil.

O livro está assim organizado: no primeiro capítulo, há uma apresentação das características econômicas do setor aéreo, dando ao leitor um apanhado de informações e estatísticas da indústria e sua relação com a economia como um todo. No segundo capítulo, entra na questão da regulação e das políticas públicas para o setor, com ênfase nos aspectos evolutivos ao longo das últimas décadas. No capítulo terceiro, enfatiza o estudo do desempenho de mercado das companhias aéreas, com o foco em questões da concentração e dos aspectos de contestabilidade e de competitividade do setor. A seguir, apresenta um capítulo com três estudos de caso de demanda e oferta de transporte aéreo, na tentativa de ilustrar ao leitor, de forma informal, mas devidamente embasada, uma visão geral de como esses aspectos vêm sendo tratados no âmbito da academia. Por fim, o livro é encerrado com um conjunto de "princípios balizadores da regulação e das políticas públicas do transporte aéreo", onde utiliza uma abordagem de Economia Normativa para tecer considerações sobre os possíveis caminhos para o setor.

Será extremamente estimulante tê-los a bordo para esta jornada. Apertem os cintos e boa viagem!

# Uma Indústria Vital
# para a Economia Brasileira

**Decolagem: Características Econômicas do Setor**

O transporte aéreo é um dos setores que freqüentemente são apontados como "estratégicos" tanto por governos quanto por analistas setoriais. Esta qualificação é, em geral, devido a algumas de suas principais características econômicas. Por exemplo, o transporte aéreo é um verdadeiro "insumo produtivo" para centenas de milhares de empresas pelo Brasil afora, dado que as maiores corporações o utilizam intensamente para deslocamento rápido de empresários, executivos, técnicos, carga, correspondência. Deslocamento nesse caso significa mobilidade, agilidade, eficiência e, por decorrência, a indução de negócios, o fechamento de contratos, enfim, o crescimento econômico. Outro importante diz respeito à integração da Amazônia e o desenvolvimento sustentável. Sabe-se que o modal terrestre nunca irá atender satisfatoriamente as necessidades de locomoção intra-região amazônica e, dado o caráter estratégico que a região possui para o desenvolvimento do País, seja no âmbito econômico, seja na questão da segurança das fronteiras, seja no próprio atendimento das necessidades locais, tem-se no transporte aéreo uma importante fonte de potencialidades de alavancagem do progresso com respeito ao meio ambiente.

Em qualquer nação, a credibilidade do funcionamento do sistema aéreo é fator imprescindível para os custos e riscos associados aos investimentos no País. Problemas com o transporte aéreo geram efeitos em cascata, *spillovers,* negativos importantes por toda a economia e por toda a sociedade. Os "apagões aéreos" de 2006 e 2007 são ilustrativos disso. Durante esses episódios, ficou marcante que a credibilidade do transporte aéreo foi bastante afetada por conta do expressivo volume de vôos atrasados e cancelados e pela sensação de maior insegurança devido aos trágicos acidentes aéreos da ocasião. Por decorrência, efeitos lesivos relevantes foram induzidos sobre a economia, o turismo, o ambiente de negócios, a qualidade percebida e vida cotidiana dos passageiros.

Dentre as características do transporte aéreo, pode-se destacar as seguintes: importância na economia, alavancagens da cadeia produtiva, inserção internacional do País e vulnerabilidade e choques externos, impacto nas contas externas, posição efeito de integração e desenvolvimento ao longo do território nacional.

**Transporte Aéreo e a Economia**

O transporte aéreo tem participação relevante na economia do País, respondendo, em termos de receita de vôo das companhias aéreas regulares brasileiras, nos mercados doméstico e internacional, por algo em torno de 14,3 bilhões de reais[1]. Isso representa cerca de 0,55% do Produto Interno Bruto (PIB) daquele ano, que ficou na casa dos 2,6 trilhões de reais[2].

Estimativas do SNEA, Sindicato Nacional das Empresas Aeroviárias, e citadas por Marchetti et al (2001), calculavam em 18 bilhões de dólares anuais o total dos efeitos econômicos do transporte aéreo para o País, no final da década de 1990. Essa estimativa é mais abrangente do que a métrica acima, dado que computa impactos *diretos, indiretos* e *induzidos* da indústria. Considerando que, do montante de 18 bilhões de dólares anuais, aproximadamente seis bilhões seriam de impactos diretos do setor aéreo - isto é, salários, combustível, etc. – tem-se um fator de impacto (FI) igual a 3.

Isso significa que, para cada mil reais (ou mil dólares) gerados em atividades diretamente vinculadas à aviação, são gerados outros dois mil por toda a economia como efeito de transbordamento (*spillovers*). O turismo, por exemplo, é um dos setores mais impactados por efeitos indiretos do transporte aéreo.

> *Os Impactos Diretos são aqueles que representam as atividades econômicas que não ocorreriam com a ausência do aeroporto, isto é, o emprego, a renda e outros benefícios gerados por aqueles que trabalham no aeroporto ou contribuem para as suas atividades, como a administração aeroportuárias, as empresas aéreas, o controle de tráfego aéreo, etc. Os impactos Indiretos seriam aqueles gerados pelas empresas prestadoras de serviços para as instituições e firmas diretamente envolvidas com o aeroporto, como por exemplo, os restaurantes, as lojas de conveniência, os serviços de limpeza. Por fim, os Impactos Induzidos contabilizam o efeito multiplicador do setor da economia nos outros setores. As principais diretrizes para o cálculo de impactos na área aeroportuária estão colocadas em um documento da Federal Aviation Administration (FAA), de 1992, denominado "Estimating the Regional Economic Significance of Airports".*

O transporte aéreo de passageiros no Brasil apresentou crescimento fantástico ao longo das últimas décadas. Duas métricas de tamanho de mercado corroboram essa afirmação: o ASK (em inglês, *available seats-kilometers*, ou assentos-quilômetros disponíveis) e o RPK (em inglês, *revenue-passenger kilometres*, ou passageiros pagantes-quilômetros transportados). O ASK é uma medida de capacidade produtiva de transporte das companhias aéreas, enquanto o RPK é um indicador de tráfego, de fluxo de passageiros. Desde 1970, tanto o ASK quanto o RPK passaram de aproximadamente 4 e 2 bilhões, respectivamente, para valores em torno de 60 e 40 bilhões em 2007. O gráfico abaixo mostra essa evolução.

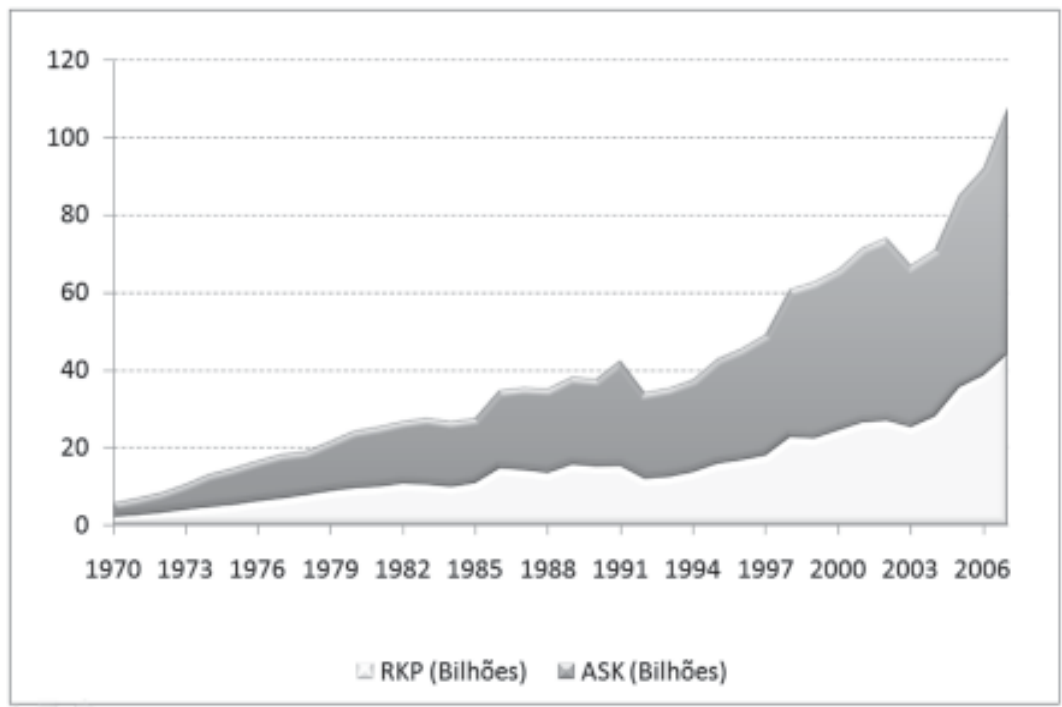

Figura 1 – Evolução do Tamanho da Indústria (Fonte: ANAC)

Os dados da Figura 1 mostram um crescimento considerável tanto do ASK quanto do RPK, evolução esta que foi advinda de fatores como o crescimento e desenvolvimento do próprio país, os investimentos históricos em infra-estrutura aeroportuária e infra-estrutura de controle de espaço aéreo realizados ao longo do tempo, o investimento das próprias companhias aéreas, além da desregulação do setor dos anos 1990 – que proporcionou uma maior popularização do modal aéreo.

Adicionalmente, é fato que o setor aéreo apresenta considerável elasticidade à renda da demanda. Isso significa que, em períodos de crescimento econômico, o tráfego aéreo cresce mais do que proporcionalmente à renda (ou PIB), e assim, reforçando o efeito do próprio crescimento econômico. Historicamente, temos a "regra de bolso" de 2:1, ou seja, cada 2% de crescimento do tráfego aéreo, são advindos de 1% de crescimento do PIB no mesmo período. Esse fenômeno tende a ser ainda mais acentuado em atividades correlatas ao transporte aéreo. Por outro lado, em períodos recessivos, ele tende também a contribuir mais do que proporcionalmente com a queda nos indicadores de atividade econômica.

A Figura 2 exibe as taxa de crescimento percentual anual do PIB e de setores selecionados – dentre eles, o transporte aéreo de passageiros domésticos, medido pelo RPK. Com ela, é possível observar que, entre 1970 e 2007, o transporte aéreo cresceu em média 8,4% ao ano, superando o crescimento do consumo de energia elétrica (6,2% ao ano), das vendas de automóveis (4,8% ao ano), o PIB real da construção civil (3% ao ano) e o próprio PIB real brasileiro (3,9% ao ano).

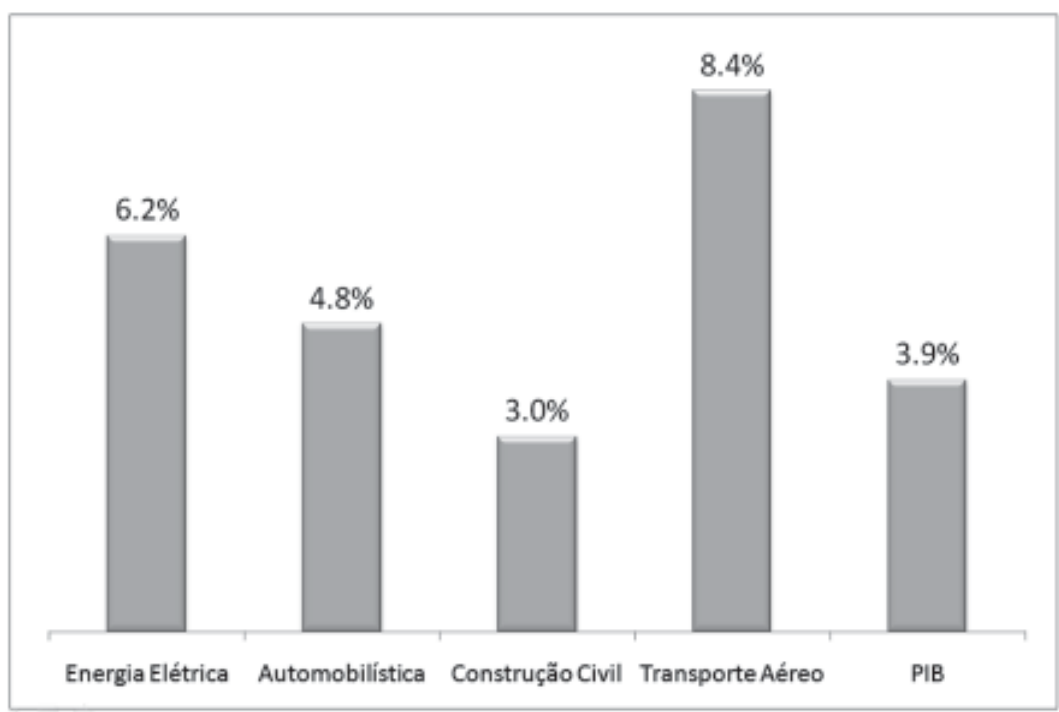

Figura 2 – Taxa de Crescimento Anual Média (1970-2007). Fonte: Ipeadata.

É possível entender os padrões de crescimento desses setores e da economia como um todo por meio de uma desagregação dos mesmos dados dos setores acima listados, ao longo das décadas. Isso é feito por meio da Tabela 1:

Tabela 1 – Taxa de Crescimento Anual Média por Década.

| Período | Energia Elétrica | Automobilística | Construção Civil | Transporte Aéreo | PIB |
|---|---|---|---|---|---|
| 1970-1979 | 10.2% | 10.4% | 9.2% | 15.6% | 7.7% |
| 1980-1989 | 5.8% | -3.3% | -0.3% | 5.1% | 2.0% |
| 1990-1999 | 3.8% | 5.4% | 1.8% | 4.0% | 2.1% |
| 2000-2007 | 2.4% | 6.8% | 1.2% | 7.7% | 2.9% |

Fonte: Ipeadata

A Tabela 1 permite visualizar que o transporte aéreo era um dos setores da indústria brasileira com crescimento mais pujante no período do Milagre Econômico, no final da década de 1960 e primeira metade dos anos 1970. Esse crescimento acelerado do transporte aéreo voltou a ser destaque em meados dos anos 2000, quando o setor voltou a crescer a taxas anuais bastante superiores às taxas de crescimento do PIB, atingindo uma média anual de 7.7% entre 2000 e 2007, mesmo com os problemas de gargalos de infra-estrutura e "apagões" de 2006 e 2007. Em 2005, o transporte aéreo atingiu a incrível marca dos 26,7% de crescimento anual (RPK), um salto que foi mais de oito vezes o crescimento do PIB real daquele ano. Esta foi a maior relação crescimento do RPK por crescimento do PIB fora de períodos de planos de estabilização de todo o período – ou seja, ao longo de quase 40 anos. Anteriormente, o recorde em períodos desse tipo era a relação 6:1 (isto é, 6% de RPK para cada 1% de PIB) em 2001, ano da entrada da Gol no mercado brasileiro.

**O Transporte Aéreo, Seus Recortes e a Fatia do Bolo**

Não existe apenas uma única indústria do transporte aéreo. Existem várias. O setor é, na verdade, composto por uma multiplicidade de produtos e fornecedores basicamente constituídos pelas questões "*quem transporta?*" e "*o que transporta?*" e "*de onde para onde transporta?*". Algumas das opções de configuração dessas perguntas são apresentadas a seguir:

✈ Quem transporta: empresas aéreas regulares, empresas aéreas de fretamento, empresas de táxi aéreo, etc. (operadoras de aviação comercial), pessoas físicas (aviação geral, com aeronaves de menor porte), instituições militares (aviação militar).

✈ O que transporta: passageiros, carga, correio.

✈ De onde para onde: doméstico (origem e destino no País), regional (origem e destino no País, mas envolve não ligações principais ou "tronco"), internacional (origem ou destino fora do País).

Desta forma, tem-se que, em primeiro lugar, há que se distinguir a aviação civil da aviação militar. Dentro da aviação civil, pode-se subdividir em aviação comercial (com vôos regulares ou não-regulares, também conhecidos como fretamento ou "*charter*") e aviação geral, que envolve aeronaves de menor porte (alguns deles chamados de "jatinhos"). Outras classificações são denominadas de "sub-setores do transporte aéreo", como o transporte de passageiros e transporte de cargas, doméstico e internacional. Mas há outros recortes possíveis, como de linhas aéreas "nacionais" (isto é, aquelas referentes aos mercados "tronco" ou as mais densas, as principais ligações) e linhas aéreas "regionais". Recentemente, vem sendo também utilizado o conceito de "sub-regional" para expressar algumas alternativas de vôos regionais internacionais dentro do continente sul-americano. Há também as divisões feitas a partir de critérios de tamanho da etapa de vôo – vôos de "curto", "médio" e "longo" percurso". Por fim, em termos de segmentação de mercado, pode-se pensar o transporte aéreo como sendo constituído por segmentos de consumidores de acordo com a motivação da viagem, que pode ser negócios, lazer, motivos pessoais os mais diversos, etc., ou de acordo com o tipo de modelo de negócio utilizado pelas companhias aéreas no mercado – do tipo "Preço Baixo, Custo Baixo" ou "*Low Cost, Low Fare*", do tipo "*Hub-and-Spoke*" ou "*Network*", várias formas híbridas possíveis, etc.

Enfim, pode-se afirmar que há um conjunto muito amplo de possibilidades de recortes do transporte aéreo para se efetuar análises de mercado. Os analistas devem, portanto, ser cuidadosos no sentido de explicitar o tipo de recorte que está sendo efetuado, sob o risco de gerar confusão e interpretações errôneas de suas análises dos mercados aéreos.

Dado que o setor pode ser visto sob várias óticas, vamos aqui adotar a prática de enfocar o chamado sub-setor de transporte aéreo doméstico regular de passageiros na maioria dos casos e exemplos. Para se compreender a representatividade deste sub-setor dentro da indústria de aviação comercial como um todo, pode-se visualizar a Tabela 2, que contém a importância relativa de cada sub-setor, considerando os segmentos doméstico e internacional, os passageiros em vôos regulares, vôos fretados, o transporte de correio e frete, etc. Esta tabela foi construída por meio da desagregação das receitas de cada sub-setor apresentada pelo Anuário Estatístico de 2007, volume II, da Agência Nacional de Aviação Civil:

Tabela 2 – Receitas do Transporte Aéreo por Sub-Setor (em Bilhões de R$, 2007)[3]

| Item | Doméstico | Internacional | Total |
| --- | --- | --- | --- |
| Passagens | 9.28 | 2.62 | 11.90 |
| Fretamentos | 0.28 | 0.03 | 0.32 |
| Correios | 0.21 | 0.00 | 0.21 |
| Carga | 0.77 | 0.99 | 1.75 |
| Outros | 0.14 | 0.01 | 0.15 |
| Total | 10.68 | 3.65 | 14.33 |

Como pode ser visto na Tabela 2, tem-se que o transporte aéreo regular de passageiros domésticos e internacionais pelas companhias aéreas gerou uma receita de 11.90 bilhões de reais em 2007. O total de receitas geradas pelas companhias aéreas brasileiras naquele ano foi de 14,33 bilhões de reais. Isso equivale a 11.8% do PIB de serviços de transporte, armazenagem e correio, calculado pelo IBGE, que ficou em torno de 121 bilhões de reais. Ou seja, um a cada 8 reais produzidos em serviços referentes a transportes no país são relacionados diretamente com as companhias aéreas regulares. A Figura 3 exibe a composição do faturamento de acordo com as porcentagens de participação:

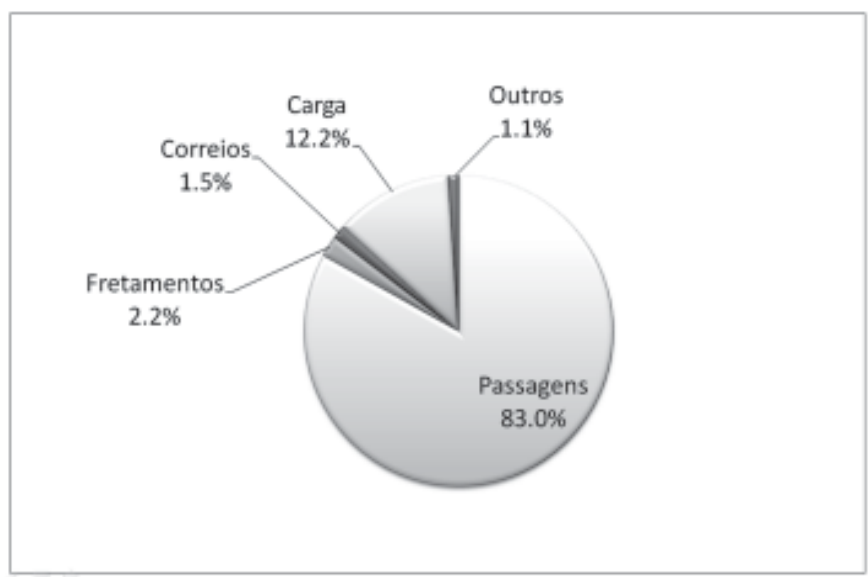

Figura 3 – Composição das Receitas do Transporte Aéreo por Sub-Setor (2007)[4]

Pode-se identificar que as receitas da indústria do transporte aéreo são geradas, basicamente, pelo transporte de passageiros (83,0% do total de receitas), seguido pelo transporte de cargas (12,2%). É importante também destacar a relevância da segmentação entre transporte doméstico e internacional, dado que há uma divisão de aproximadamente um quarto (¼) para três quartos (¾) do total de receitas entre os mercados internacional e doméstico, respectivamente – 10,68 bilhões de reais no primeiro e 3,65 bilhões no segundo. As receitas com fretamentos representam 2,2% do total, sendo que a maior parte desse faturamento diz respeito aos vôos não-regulares domésticos (280 milhões de reais, ou 2,0% do total). A Figura 4 Figura 3permite visualizar a desagregação doméstico-internacional entre os principais sub-setores do transporte aéreo brasileiro:

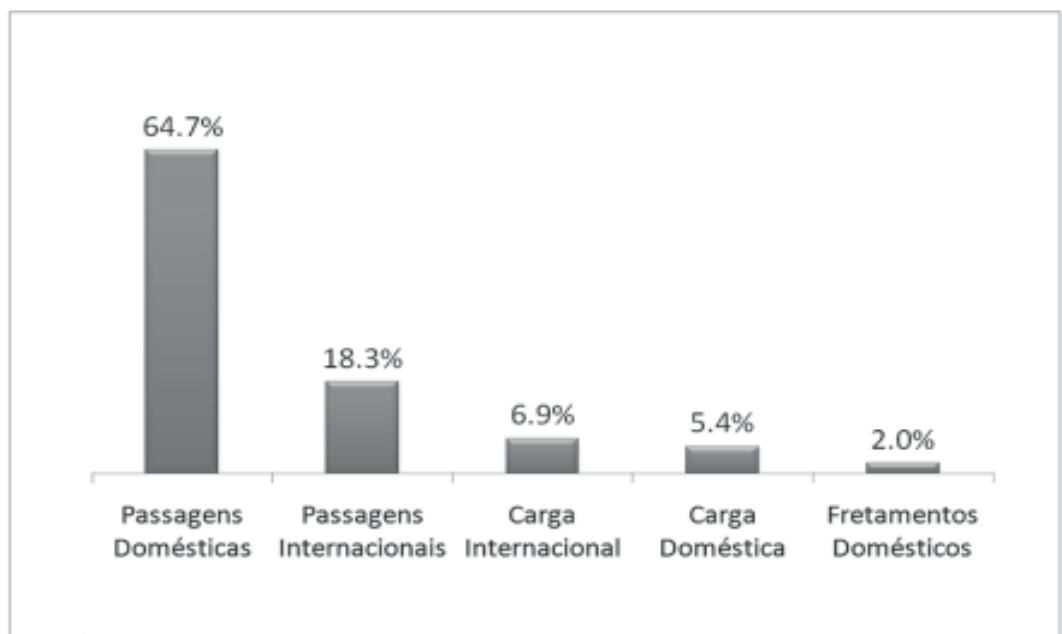

Figura 4 - Composição das Receitas do Transporte Aéreo por Sub-Setor e Segmento (2007)[5]

## A Cadeia Produtiva do Transporte Aéreo e suas Alavancagens

O transporte aéreo é um setor claramente inserido e vinculado a uma cadeia produtiva. Essa constatação pode não parecer tão trivial em primeira instância, dado ser essa cadeia produtiva bastante ampla, visto que o transporte aéreo é insumo de produção para qualquer empresa ou corporação no País, seja para transporte de passageiros ou de carga. Assim como outros setores de serviços que compõem a infra-estrutura do País, como energia elétrica e telecomunicações, o transporte aéreo possui a característica de possuir "demanda derivada". Isto significa que tem a sua demanda atrelada aos fatores condicionantes dos produtos à jusante da cadeia. De fato, a grande maioria dos passageiros ou empresas utiliza-se do transporte aéreo com um meio de atingir um fim último, e não como um bem final. Em outras palavras, ninguém consome transporte aéreo por si só, com base apenas em seus atributos. Em geral, há que se ter uma utilidade relacionada com o destino da viagem (realização de um negócio, visita a parentes, turismo, etc.) para que o consumo do bem "transporte aéreo" seja adquirido. Por modal aéreo, a "viagem dos sonhos" das pessoas está em geral relacionada com o que será feito no destino e não no decorrer do percurso. Tem-se caracterizada, portanto, a característica de demanda derivada inerente ao setor.

A Figura 5 apresenta o conjunto de *interfaces* com as quais o transporte aéreo mantém estreita interação. Qualquer problema com alguma dessas *interfaces*, que compõem o conjunto de *stakeholders* do setor, a indústria provavelmente terá sérios problemas para se manter competitiva e mesmo viável. A montante, constituindo os "insumos produtivos" do transporte aéreo, temos os setores de aeroportos e de controle de tráfego aéreo, os fabricantes de aeronaves, produtores e distribuidores de combustível de aviação, escolas de formação de pilotos, etc. Já a jusante, temos os "clientes" do transporte aéreo, que podem ser o setor turístico, como a rede hoteleira, os *resorts*, o setor postal e as corporações em geral. Neste caso, temos o transporte aéreo como um bem intermediário na cadeia produtiva. Se pensarmos o cliente do transporte aéreo como sendo as famílias ou o indivíduo com sua cesta de consumo, aí teremos o setor como produzindo um "bem final" na economia.

Adicionalmente, há o conjunto de bens substitutos (outros modais de transporte) e bens complementares (produtos que o cliente adquire para realizar suas viagens, como, por exemplo, o serviço de táxi até o aeroporto) e que fazem parte da cesta de consumo da maioria dos passageiros.

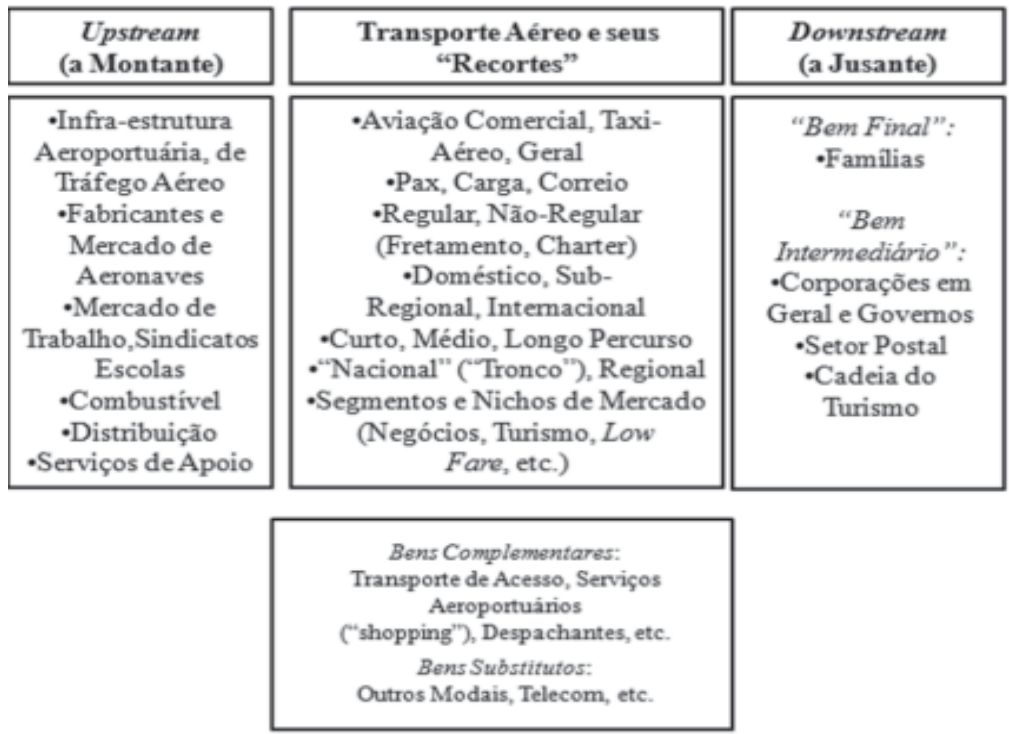

Figura 5 – A Cadeia Produtiva do Transporte Aéreo

O serviço de telecomunicações pode ser considerado tanto um bem substituto quanto complementar ao transporte aéreo: por um lado, no curto prazo, na medida em que as telecomunicações avançam, menor será a necessidade das pessoas se deslocarem. No médio e longo prazos, entretanto, as telecomunicações criam ambientes propícios a novos negócios e maior integração entre regiões, o que gera a necessidade de maior número de viagens. Como bem complementar, as telecomunicações facilitam a vida do cliente do transporte aéreo, com a disponibilização da oferta de vôos *on-line*, a emissão do *e-ticket*, etc. Na medida em que esses serviços são facilitados, a tendência é de geração de maior número de viagens e a inserção de um maior número de consumidores. Resumo da ópera: no longo prazo, as telecomunicações são um importante fator indutor de transporte aéreo, permitindo ampliações do tamanho e abrangência do mercado.

**Interação com a Indústria Aeronáutica Nacional**

O transporte aéreo nacional sempre foi bastante cotado como potencial mercado para a inserção de aeronaves e tecnologia produzidas pela Indústria Aeronáutica Nacional, mais precisamente a Embraer. Com quatro décadas de experiência em projeto, manufatura, comercialização e pós-venda de aeronaves, a empresa é atualmente o terceiro maior fabricante do mundo, atrás apenas da americana Boeing e da européia Airbus. Até o momento, sua produção monta a cerca de cinco mil aviões para companhias aéreas de quase oitenta países, nos cinco continentes. Historicamente a inter-relação da Embraer com o transporte aéreo doméstico no Brasil começou com o SITAR, Sistema Integrado de Transporte Aéreo Regional onde, a partir de meados dos anos 1970, as autoridades governamentais promoveram um redesenho da aviação regional brasileira, tendo por base o fomento às operações em cidades de pequeno e médio porte, com utilização de um dos maiores sucessos comerciais da Embraer, a aeronave EMB-110, o Bandeirante.

> *O Bandeirante (EMB-110) foi o primeiro avião comercializado pela então estatal Embraer, a aeronave que abriu as portas da empresa ao sucesso no mercado de fabricantes. O Bandeirante é um avião bimotor com capacidade de 15 a 21 passageiros, para uso civil ou militar. O projeto do Bandeirante possibilitou à Embraer acumular conhecimento imprescindível para os projetos seguintes.*

Com o passar dos anos, a inserção de novas aeronaves da Embraer na aviação comercial se esgotou. Até meados dos anos 2000, a última aeronave utilizada por empresa aérea regular havia sido o ERJ-145, utilizado pela extinta Rio Sul em ligações mais densas como a ponte aérea Rio de Janeiro – São Paulo. Essa falta de coordenação entre os dois setores da economia brasileira – o de aviação comercial e o de manufatura de aeronaves – sempre foi apontada como um problema grave a ser equacionado por meio de políticas públicas. A Tabela 3 mostra a composição da frota de aeronaves de passageiros das companhias aéreas regulares no Brasil, onde até 2007, o fabricante nacional ocupava apenas o quarto lugar, e tão-somente com modelos antigos e com menor capacidade de assentos, como o Bandeirante e o Brasília (EMB-120).

Tabela 3 – Composição da Frota das Companhias Aéreas Regulares (Passageiros)

| Fabricante | País de Origem do Fabricante | Frota em 31/12/2007 | Tamanho Médio das Aeronaves |
|---|---|---|---|
| Boeing | Estados Unidos | 119 | 157 |
| Airbus | Europa | 102 | 169 |
| Fokker | Holanda | 27 | 97 |
| Embraer | Brasil | 21 | 28 |
| Aerospatiale | Europa | 20 | 50 |
| Cessna | Estados Unidos | 7 | 9 |
| Let | República Tcheca | 7 | 19 |
| Douglas | Estados Unidos | 5 | 285 |
| **Indústria** | | **308** | **135** |

Levando-se em consideração o expressivo potencial de crescimento do transporte aéreo no País, sobretudo no que diz respeito à aviação regional, tem-se que a falta de interação entre os setores de transporte aéreo e de manufatura de aeronaves faz com que o País deixe de explorar efeitos de encadeamentos dinâmicos sobre o crescimento econômico e sobre a competitividade do sistema produtivo. Isso porque, na linguagem da literatura de desenvolvimento econômico, se trata de ***setores-chave*** da economia, capazes de produzir todo um movimento de indução ao longo das cadeias produtivas. Um "setor-chave" seria aquele que apresenta maior poder de encadeamento para frente (horizontal) e/ou de encadeamento para trás (vertical), de forma que uma variação dos níveis nesse setor terá efeitos multiplicadores sobre a renda acima da média dos setores. O "encadeamento para trás" seria o efeito de um aumento do investimento no setor sobre os setores fornecedores dos insumos, localizados a montante na cadeia produtiva.

Já o "encadeamento para frente" seria o efeito indutor e/ou multiplicador gerado sobre os compradores do produto do setor, localizados a jusante. Um setor-chave teria a propriedade indutora de competitividade e externalidades positivas de forma a propiciar uma aceleração do processo de crescimento. São setores que, portanto, mereceriam a atenção do planejamento público no fomento, na concessão de linhas de créditos, etc.

Como vimos, a manufatura de aeronaves está a montante na cadeia produtiva do transporte aéreo. Temos então que a alavancagem dos dois setores pode ter efeitos de encadeamentos importantes sobre os mesmos e sobre a economia como um todo – a começar pelos os setores a jusante como o turismo. A boa notícia da segunda metade da década de 2000 é que os novos jatos da Embraer – a família dos *E-Jets*, aeronaves com capacidade entre 70 e 110 assentos – começaram a apresentar demanda por parte das operadoras nacionais, com pedidos efetuados pelas empresas BRA, Azul e Trip.

*Com os E-Jets, a Embraer buscou expandir seus mercados com aeronaves para a aviação comercial de maior capacidade de assentos. Os E-Jets são a família de aeronaves com capacidade entre 70 e 122 passageiros. É constituída pelos modelos E-170, E-175, E-190 e E-195. O jato E-170, de 70 a 80 assentos, foi o primeiro da família de aeronaves. O E-175 tem capacidade entre 78 a 88 assentos e fez seu vôo inaugural em junho de 2003. O E-190 possibilita configurações entre 98 a 114 assentos, voou pela primeira vez em 2004, e foi homologado em agosto de 2005. Suas primeiras entregas foram para a companhia Low Cost, Low Fare norte-americana jetBlue, seguida pela Air Canada, principal companhia aérea do país da maior concorrente da Embraer, a Bombardier. O E-195 pode transportar entre 108 e 122 lugares, foi certificado pela ANAC em 30 de junho de 2006. Os E-Jets estão em operação em mais de 45 companhias áreas e em mais de 30 países. São aeronaves com alto índice de "comunalidade" dentro da família. Essa característica significa que as aeronaves podem quase sempre ser operadas com mesmas tripulações, mesmos equipamentos de terra e mesmo almoxarifado de peças de reposição para os quatro modelos. O alto grau de comunalidade entre as quatro aeronaves resulta em uma excepcional redução nos custos de manutenção e de peças de reposição, bem como menos tempo necessário no treinamento de pilotos para operar diferentes modelos.*

Com os pedidos de *E-Jets*, as companhias aéreas nacionais passam a figurar na carteira de entregas futuras de aeronaves, e muito provavelmente a Embraer irá ocupar papel de destaque no cenário do transporte aéreo doméstico do País. Sua política de "*right-sizing*" – isto é, o pleito pela adoção, por parte das companhias aéreas, de aeronaves com capacidade que seja "exata" para as condições de mercado, evitando capacidade ociosa fora dos padrões esperados – aplica-se muito apropriadamente à conjuntura do setor no Brasil, onde, ao contrário da maioria dos mercados pelo mundo afora, o tamanho médio das aeronaves vem apresentando trajetória crescente. Em 2007, a aeronave média tinha capacidade em torno de 135 assentos, segundo o Anuário Estatístico da ANAC, volume I. Para se ter uma idéia dessa tendência de aumento do tamanho médio das aeronaves no mercado doméstico, a Figura 6 apresenta a evolução recente do *mix* de aeronaves de passageiros da aviação regular no Aeroporto Leite Lopes/Ribeirão Preto. O *mix* é representado pelas faixas definidas no estudo "Demanda Detalhada dos Aeroportos Brasileiros" realizado pelo Instituto de Aviação Civil[6] (2005).

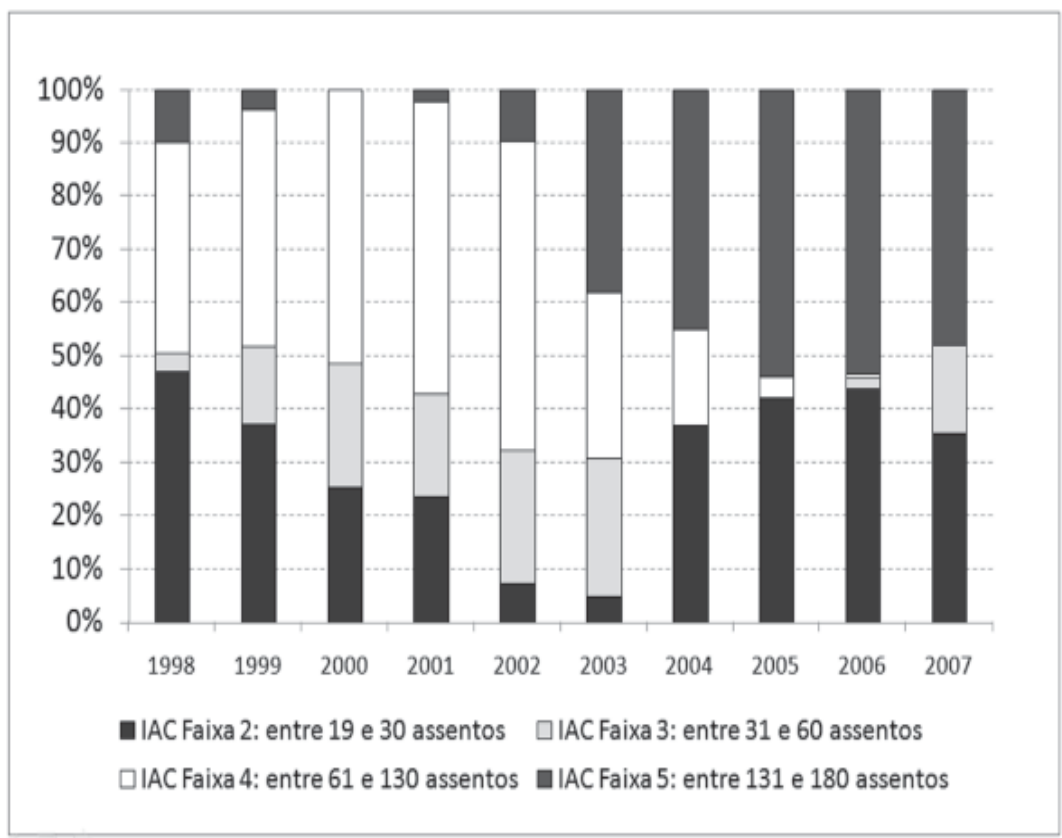

Figura 6 – Evolução do *Mix* de Aeronaves no Aeroporto Leite Lopes[7]

O IAC, em seu estudo de 2005, previa que havia uma tendência de queda da movimentação de aeronaves da Faixa 4 (aeronaves com capacidade entre 61 e 130 assentos) no Aeroporto Leite Lopes, o que levaria a sua participação para 5% em 2025. Como a Figura 6 deixa claro, temos que, consistente com a tendência recente do transporte aéreo no País, esse fenômeno de queda na participação das aeronaves de Faixa 4 já havia sido observado desde 2005. Em outras palavras: houve uma antecipação das previsões do IAC em exatos 20 anos. Isso se deu por conta do fim das operações da aeronave Fokker 100 (108 lugares) e da relativa demora na inserção de aeronaves a jato Embraer

de capacidade similar, como o Embraer E-190. Entretanto, a tendência a partir de 2009, é de retomada das operações dessa Faixa, dada a recente inserção da Embraer no mercado brasileiro com as aquisições por parte das companhias aéreas Azul e Trip.

**Especialização e Qualificação da Mão-de-Obra**

Uma indústria ou setor é chamado "capital-intensivo" se o mesmo utiliza, em seu processo produtivo, mais recursos do insumo capital do que outros fatores de produção, relativamente a outros setores da economia. O transporte aéreo é claramente uma indústria com tecnologia capital-intensiva. Dessa forma, trata-se de um setor que demanda mão-de-obra altamente qualificada, como pilotos, engenheiros, pessoal de manutenção, etc. e, mais importante, necessita de uma estrutura permanente de qualificação, treinamento e reposição de pessoal.

A Figura 7 apresenta a evolução do número de funcionários das companhias aéreas regulares, por categoria de trabalhadores empregados.

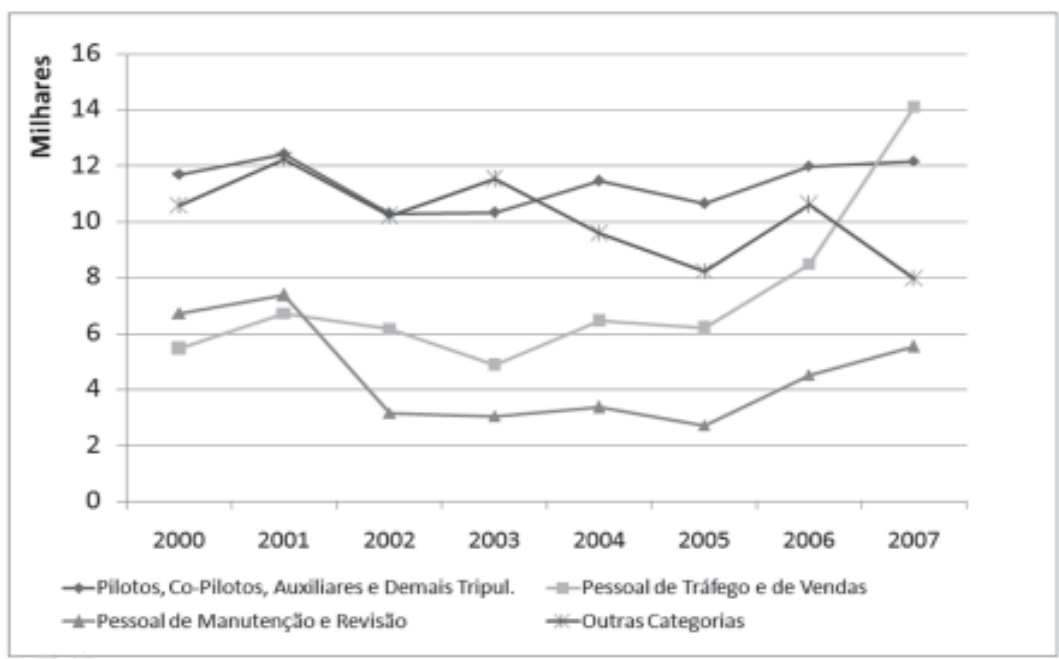

Figura 7 – Evolução do Número de Funcionários por Categoria

A evolução do número de empregos de pilotos, co-pilotos e auxiliares de vôo está diretamente relacionada com o crescimento da demanda por transporte aéreo. Segundo estimativas do Núcleo de Economia dos Transportes, Antitruste e Regulação (NECTAR/ITA), para cada 10% de crescimento do RPK, o setor aéreo terá um aumento de 2,2 % nas ofertas de vagas para essas categorias.

No caso da categoria de pessoal de tráfego e de vendas, a relação acima reportada é bem mais acentuada: para cada 10 % de crescimento do RPK, teremos 13 % de crescimento no número dessas ofertas de emprego. Isso era de se esperar, dado o "*boom*" de contratações nesse segmento de trabalhadores, que, de 2006 para 2007 se tornou maior mesmo que a categoria de funcionários diretamente relacionados com o vôo (pilotos, co-pilotos e auxiliares de vôo). Esse "*boom*" é fruto da maior competitividade alcançada recentemente pelo setor, onde as companhias aéreas estão cada vez mais centrando esforços na conquista e fidelização do cliente – o que desloca a arena da competição com foco nos aspectos operacionais para o foco na área comercial. A tendência é que as companhias aéreas se tornem cada vez mais orientadas ao *marketing* na busca da geração de valor para seus produtos e marcas. O fortalecimento do *staff* nessas áreas é uma decorrência natural disso. Além disso, há uma política de redução do número de funcionários terceirizados das empresas aéreas, muitos dos quais trabalham na área de vendas, como *telemarketing*, por exemplo. Na TAM, por exemplo, o número de terceirizados caiu de 5161 em 2006 para 455 em 2007, aumentando-se o *headcount* nessa área[8].

### Demanda e Potencial de Mercado

Mais de 45 milhões de passageiros foram transportados em 2007 por companhias aéreas regulares brasileiras para destinos nacionais e internacionais (fonte: Anuário Estatístico da ANAC, 2007, volume II) – o que representa 0,25 viagens anuais para cada brasileiro. Isso em tese significa que o brasileiro demora, em média, quatro anos para realizar uma viagem. O problema com esse indicador está no fato de que os números de passageiros transportados reportados pela Agência Nacional de Aviação Civil não representa o efetivo número de pessoas distintas que entraram no sistema – ou, em linguagem informal, o número de carteiras de identidade ou CPFs –, mas sim o número de viagens realizadas – o chamado PAX. É natural de se esperar que apenas uma parcela reduzida das mais de 45 milhões de viagens tenha sido realizadas por diferentes viajantes. É também consenso o fato de que um mesmo CPF – isto é, o mesmo indivíduo – pode entrar no sistema de uma ou mais companhias aéreas várias vezes ao longo de um ano.

A demanda do transporte aéreo pode ser melhor entendida quando utilizamos a seguinte expressão do tráfego de passageiros (PAX):

[1]    PAX = Número de Clientes (C) x Número de Viagens por Cliente (V)

A expressão acima diz que o tráfego (PAX) é igual ao número de viagens realizadas pelos clientes das companhias aéreas em um dado período do tempo. Para se promover essa decomposição, há que se conhecer o número de clientes em potencial e multiplicar pelo número médio de vezes que um cliente em potencial vai viajar. Enquanto PAX é uma variável de fluxo, que mede o tráfego de passageiros, temos que C (número de clientes) é uma variável de estoque, é a carteira de clientes das companhias aéreas, um verdadeiro inventário do número de pessoas abrangidas pela zona de influência dos aeroportos e de fato influenciadas pelo modal aéreo e pelas estratégias de atração das companhias aéreas. Já V (número de viagens por cliente) é um fator que se ajusta de forma a igualar as duas outras dimensões, dado que mede a taxa de geração de viagens de cada membro em C.

Note que C não pode abranger a população como um todo, mas apenas um subconjunto desta, dado que o transporte aéreo é um setor bem distante do conceito de universalização, onde todos teriam a possibilidade de demandar o bem. Entretanto, a própria condição de abrangência do modal aéreo por toda a sociedade pode ser internalizada na equação de PAX, que assim ficaria:

[2]    PAX = População (POP) x Percentual de Clientes na População (C%)
x Número de Viagens por Cliente (V)

Desta forma, temos que o tráfego é função do tamanho da população (POP) e também da participação dos clientes do modal aéreo nessa população (C%).

A participação dos clientes mede a propensão a fazer parte da carteira de clientes das companhias aéreas, ou seja, a propensão a viajar pelo modal aéreo. Trocando em miúdos, C% responderia à seguinte pergunta: "*Dos atuais 190 milhões de habitantes, quantos considerariam de fato viajar de transporte aéreo como uma alternativa viável?*". Como dito, sabe-se que, no Brasil, uma grande parcela da população está afastada do mercado aéreo, sobretudo por questões de renda. Assim, localidades ou extratos sociais com renda baixa teriam menor propensão a viajar pelo modal aéreo (baixo C%). Para que haja um incremento do tráfego de passageiros – isto é, um aumento do PAX – em um dado aeroporto, cidade, região metropolitana, estado ou País, etc, é preciso que pelo menos uma das três condições abaixo se materializem:

✈ **Aumento em POP:** se duas cidades possuem a mesma propensão a viajar pelo modal aéreo, aquela que tiver a maior POP terá maior PAX. Se o conceito de POP utilizado for o de PEA – População Economicamente Ativa, isto é, apta e disposta a trabalhar -, então essa correlação deve ser mais intensa. Obviamente, se as propensões a viajar (C%) são distintas, daí terá maior PAX a que tiver o maior efeito combinado das duas variáveis. O crescimento populacional é um fator natural de expansão do transporte aéreo e da economia em geral, e não é mera coincidência o fato das localidades com maior contingente populacional possuírem também o maior montante de riqueza gerada – apesar de que nada se pode falar, *ex-ante*, da riqueza ou renda *per capta*.

✈ **Aumento em C%:** a propensão a viajar – medida pelo tamanho relativo da carteira de clientes – é alavancada com fatores como: o aumento da renda *per capta*, a queda nas taxas de juros que facilitem o acesso ao crédito e a financiamentos de compras de passagens, o aumento da escolaridade média da população, o aumento da informação em relação ao modal aéreo (promoções, propagandas publicitárias e esforços de vendas, por exemplo), o incremento da qualidade do acesso ao aeroporto (ex. uma nova via de acesso ou um trem de alta velocidade com estação no aeroporto), etc.

✈ **Aumento em V:** este caso significa que os atuais passageiros (carteira atual de clientes) aumentam sua freqüência de vôo. Pode-se afirmar que praticamente todos os fatores listados para C% podem ser listados também como determinantes de V, tais como o aumento na renda, na escolaridade, etc. Adiciona-se talvez os fatores que geram um maior hábito de consumo, como a promoção e publicidade direcionadas ao público-alvo existente. Exemplo: uma concessão de milhagem bônus afeta apenas os atuais clientes de uma companhia aérea. Mas pode-se esperar que os dois fatores, V e C% cresçam concomitantemente em diversas situações. Na medida em que a população economicamente ativa cresce, acaba por acontecer também o fenômeno da ascensão de novas gerações, mais habituadas às novas tecnologias de *e-commerce* – o chamado "Novo Turista", por exemplo –, renovam-se as chances de aumento da propensão a viajar (V) por transporte aéreo. De fato, o transporte aéreo é um setor historicamente proativo nesse sentido, sobretudo por conta do pioneirismo das companhias aéreas *Low Cost, Low Fare*, e, por decorrência, com um excelente posicionamento na *internet*.

No Brasil desde 2005, mais exatamente quatro anos após a entrada da Gol no mercado doméstico e no mesmo ano em que TAM e Varig desistiram de seu acordo de compartilhamento de aeronaves iniciado em 2003, tem-se observado um retorno do crescimento acelerado do mercado aéreo. Fatores que contribuíram para isso são de cunho microeconômico, como a desregulação econômica e o "Efeito Gol" – efeito de entrada de uma empresa de baixo custo que induziu maior competitividade no setor – e de cunho macroeconômico, como a queda na taxa de juros e as melhorias na economia e na distribuição de renda. Há um debate sobre se os fatores macroeconômicos foram mais importantes do que os microeconômicos. Seguindo a linha pró-fatores macroeconômicos, o Ministro da Fazenda Guido Mantega, em plena vigência de episódios de apagão aéreo de 2007, atribuiu os constantes problemas com atrasos dos vôos na aviação regular brasileira à "prosperidade" da economia brasileira na ocasião: "*Mais gente viajando, mais aviões, mais rotas*". Nessa corrente está claramente embutida a noção de transporte aéreo como "termômetro" da economia.

Independentemente das causas da aceleração do transporte aéreo de passageiros (do RPK), tem-se que, muito provavelmente o que aconteceu desde 2005, foi um aumento em paralelo dos *drivers* tanto de POP, como de V e de C% da expressão [2] acima. Isso pode ser avaliado por meio de uma excelente pesquisa com questionários (*survey*) junto a passageiros recentemente realizada pela Secretaria de Transportes do Estado de São Paulo[9]. Na pesquisa, foram entrevistados mais de vinte mil passageiros no período de 23/11/2005 e 09/12/2005 em um conjunto de aeroportos administrados por INFRAERO e DAESP (Departamento Aeroviário do Estado de São Paulo). A seguir são apresentados um conjunto de gráficos ilustrativos dos resultados da pesquisa no que tange o perfil e os hábitos de consumo dos clientes do modal aéreo nos aeroportos Internacional de Congonhas/SP e Guarulhos/SP.

Em primeiro lugar, quanto ao motivo da viagem. A Figura 8 permite destacar a pujância do tráfego a partir de viagens a negócios no Aeroporto de Congonhas. Mesmo em se tratando de um período de início de férias escolares e da alta estação turística (a coleta foi realizada entre novembro de dezembro de 2005), tem-se que esse aeroporto apresentou uma amostra com 72% dos passageiros viajando a negócios. Já em Guarulhos, podem-se observar os importantes efeitos das viagens turísticas no período: o motivo "Lazer" foi selecionado por 52% dos respondentes. Isso demonstra claramente as distinções entre os perfis dos dois aeroportos paulistanos: Congonhas mais relacionado ao segmento *business* com elevada disposição a pagar pelo serviço aéreo e baixa sensibilidade ao preço, e Guarulhos acomodando um *mix* mais variado de viagens a negócios e turísticas, tanto domésticas quanto internacionais.

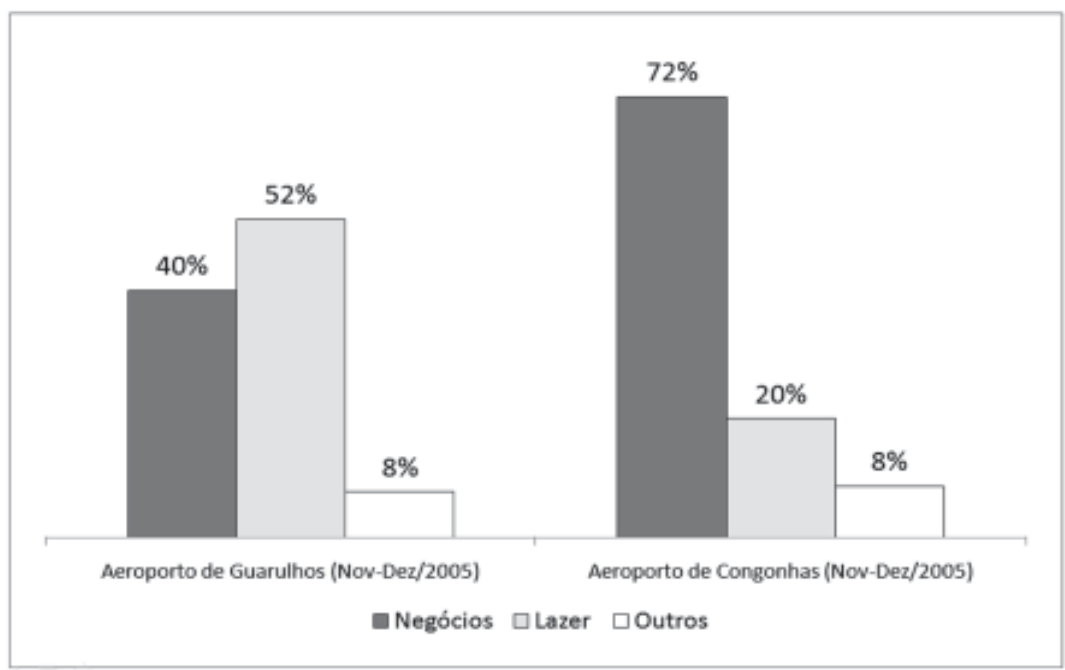

Figura 8 – Motivo das Viagens de Passageiros na Região Metropolitana de São Paulo[10]

A Figura 9 apresenta a distribuição de freqüências de viagens aéreas dos passageiros paulistanos, segundo a amostra obtida pela *survey*. Pode-se perceber o elevado percentual de passageiros que voa com nenhuma ou pouca freqüência ao longo de um ano. As respostas "primeira vez" e "de 1 a 6 vezes por ano" foram classificadas como advindas de "viajantes novatos" e "viajantes eventuais", respectivamente. Elas compõem mais do que 50% da amostra. Isso é consistente tanto com o fato de que tanto a métrica C% (tamanho relativo do mercado para transporte aéreo) como V (propensão a viajar dos atuais passageiros) ainda são reduzidas para um transporte aéreo que ainda encontra-se em estágio inicial de crescimento, ainda longe da maturidade. As elevadas taxas de crescimento que o setor vem apresentando, combinadas com o fato de que uma razoável parcela da população ainda não tem o transporte aéreo em seu portfólio de consumo, demonstram que se trata de um mercado com excelentes oportunidades de expansão e que podem ser bastante atrativos para investidores nacionais e internacionais.

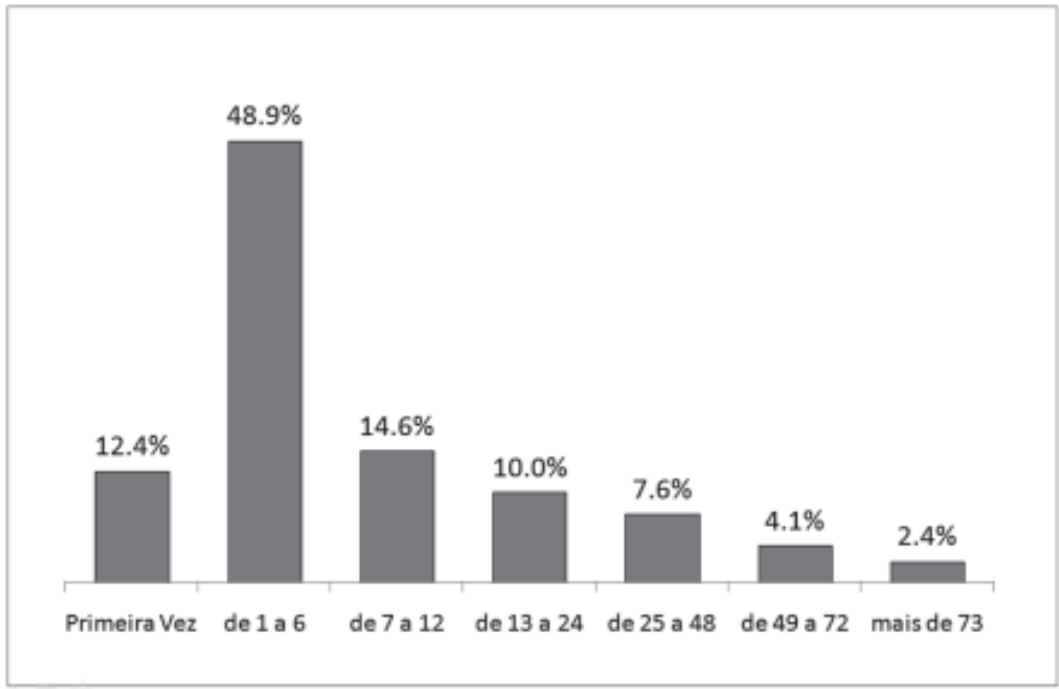

Figura 9 – Freqüência de Viagens Aéreas, Passageiros na Região Metropolitana de São Paulo[11]

*Uma evidência importante de que o mercado aéreo doméstico de passageiros no Brasil é bastante atrativo está no interesse explícito, manifestado ou levado às conseqüências, de empresários famosos do setor aéreo internacional, como David Neeleman (fundador da Jetblue e da Azul Linhas Aéreas) e Richard Branson (fundador da Virgin Atlantic, Virgin Blue e Virgin America), em atuar no mercado brasileiro. No caso do David Neeleman, sua companhia aérea, a Azul Linhas Aéreas, anunciou a quatro cantos que faria o maior investimento para o lançamento de uma companhia aérea na história da aviação mundial, em uma capitalização para US$ 200 milhões. "Nenhuma empresa começou com tanto dinheiro como nós temos" - dizia o empresário, em 2008. A alavancagem dessa magnitude de capitais internacionais revela que o mercado brasileiro é interessante aos investidores.*

Para completar, tem-se que o número de viagens realizadas por ano pelo passageiro (V), na mediana da distribuição, é igual a seis, ou seja, uma a cada dois meses, para os passageiros da amostra, isto é, os já inseridos no sistema aéreo.

A Figura 10 apresenta os rendimentos por tipo de viajante de acordo com a freqüência anual de vôo. Pode-se perceber que a participação dos "viajantes freqüentes" (isto é, aqueles com mais de sete viagens por ano) é inicialmente baixa em comparação ao número de viajantes "novatos" (aqueles que estão viajando pela primeira vez). Na medida em que os salários aumentam, a proporção dos "viajantes freqüentes" aumenta consideravelmente a ponto de se tornar a maior fatia. Como vimos, foram classificados como "viajantes eventuais" aqueles que declararam efetuar entre uma a seis viagens por ano.

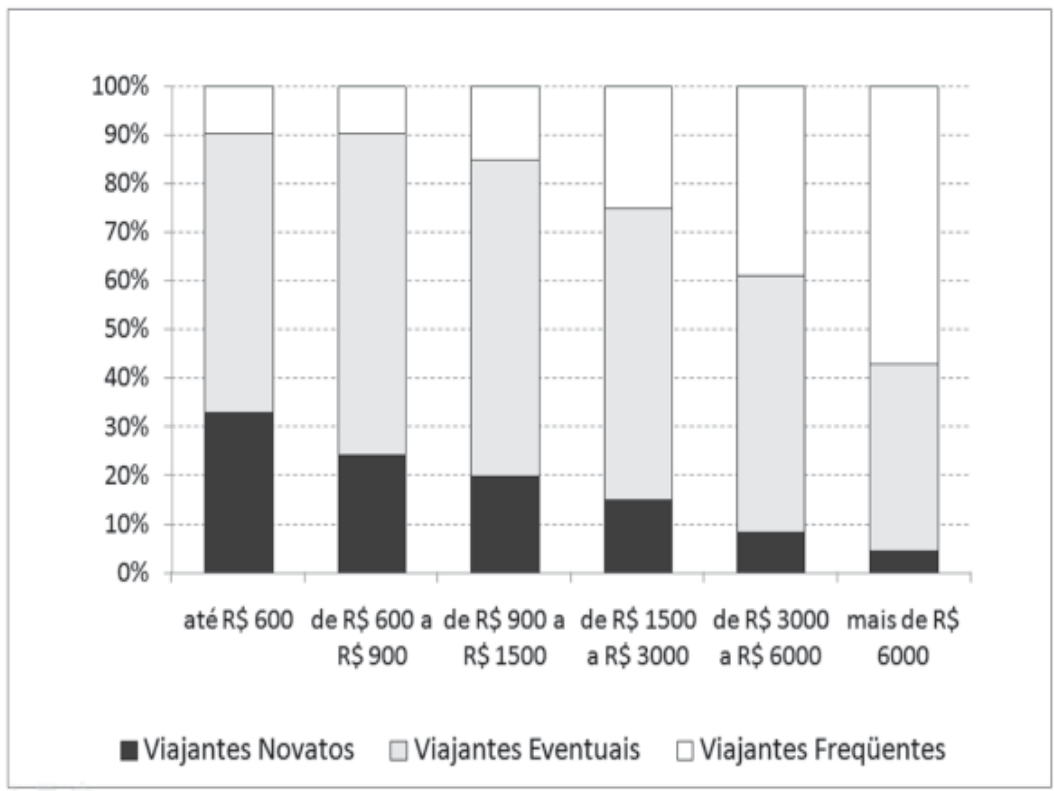

Figura 10 – Rendimentos por Tipo de Viajante na Região Metropolitana de São Paulo[12]

## Comportamento Cíclico, Sazonalidade e Vulnerabilidade a Choques

O setor de transporte aéreo é conhecido por seu comportamento cíclico e por seus aspectos de sazonalidade ao longo do ano. A Figura 11 apresenta o ciclo de longo prazo do setor, enquanto que a Figura 12 o aspecto de sazonalidade dentro do ano "médio". Enquanto o segundo retrata o ciclo "ao longo de um ano", o primeiro retrata os ciclos "com o passar dos anos".

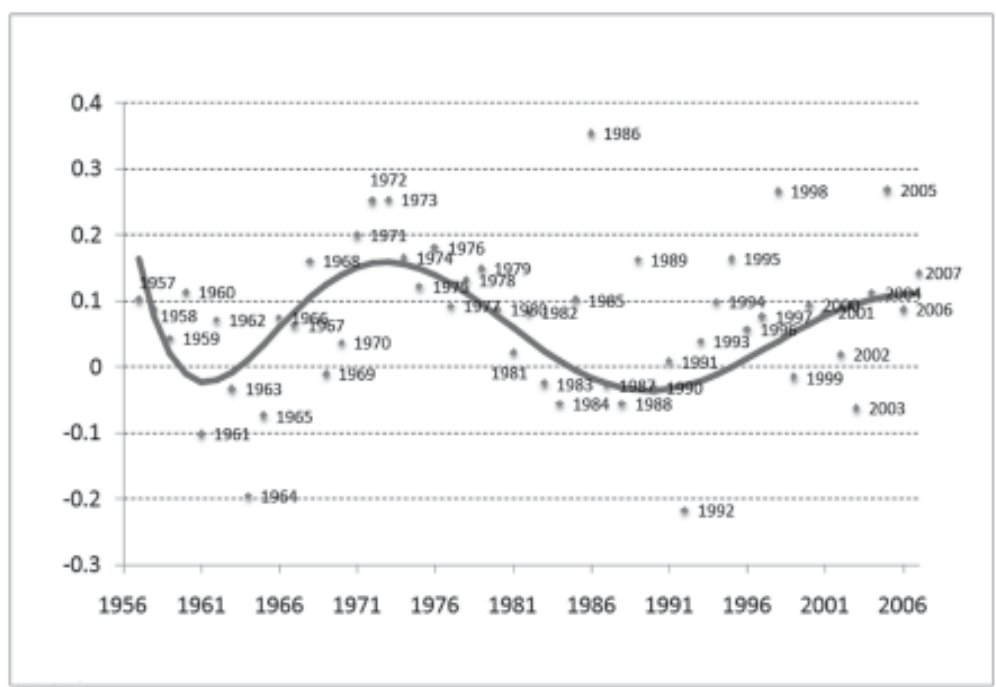

Figura 11 – Taxas de Variação do RPK Doméstico das Companhias Aéreas Regulares[13]

Um exame detalhado da Figura 11 permite chegar à conclusão de que o transporte aéreo no Brasil atual atravessa um momento de alta do ciclo, tendo alcançado as taxas médias de crescimento mais elevadas dos últimos 15 anos. É muito difícil, entretanto, prever qual vai ser a duração dessa fase de alta do ciclo, dado que a mesma depende do comportamento da economia como um todo. O último "vale" observado foi observado ao longo da década de 1980 e início da década de 1990, marcada pelos sucessivos choques macroeconômicos que impactaram profundamente o desempenho do setor. Já o último episódio

de "alta" ou "cume" do ciclo foi observado durante o "Milagre Econômico" do final dos anos 1960 e início dos anos 1970.

Já a Figura 12 mostra claramente o papel que a alta estação turística tem tanto sobre o mercado de passageiros doméstico quanto o internacional. O índice apresentado pelo gráfico contém a média anual igual a 100, referente ao período entre 2000 e 2007. Temos, portanto, que os meses de janeiro, julho e dezembro – meses de férias escolares e de viagens turísticas – são os que em geral apresentam os picos de movimentação de passageiros (RPK) ao longo de todo o ano.

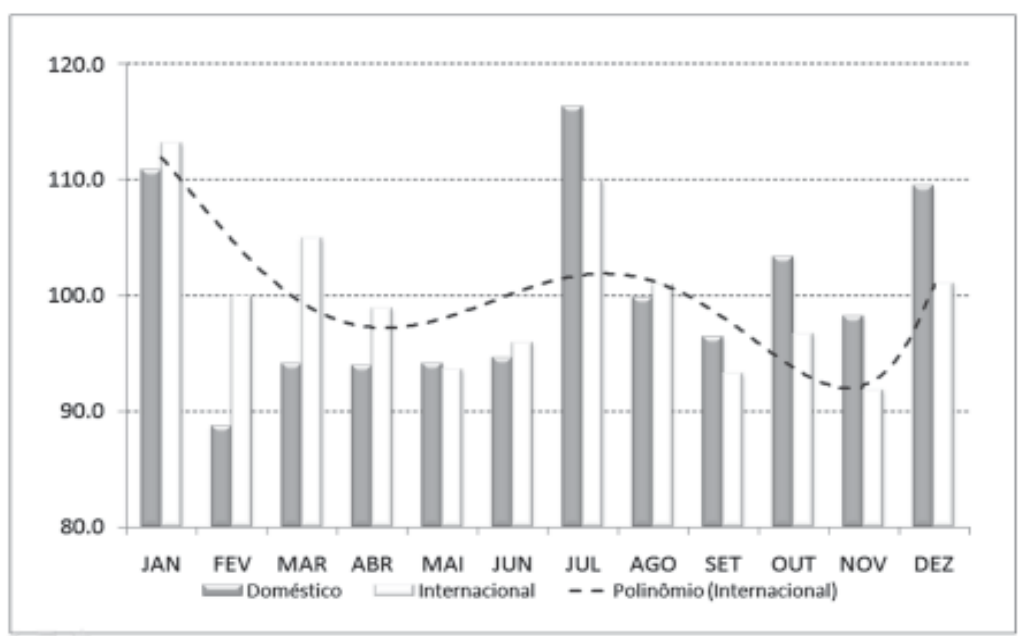

Figura 12 – Índice de Sazonalidade Média – Passageiros Domésticos e Internacionais[14]

Combinada com a situação de movimentação cíclica e sazonal da demanda – que, por si só, pode ser considerado um fator de vulnerabilidade do setor, dado que a oferta total de assentos – o tamanho e composição da frota de aeronaves – costuma ser fixa no curto prazo, temos a instabilidade pelo lados dos custos das companhias aéreas. Se, por um lado, os choques macroeconômicos e de cunho internacional – como crises, movimentos no PIB, taxa de juros, câmbio, etc. – costumam afetar a demanda por transporte, o mesmo se dá com os custos operacionais. A Figura 13 exibe a correlação entre os custos médios das companhias aéreas e a taxa de câmbio efetiva real. A alta sensibilidade tanto às condições de oferta (no caso, os choques em custos incorridos pelas desvalorizações cambiais dos anos 1990 e 2000), quanto às condições de demanda, é um traço marcante do setor – fator que comprime a rentabilidade esperada de companhias aéreas por todo o mundo. O coeficiente de correlação entre as duas séries é da ordem de 0,75, podendo ser considerado elevado e indicativo da forte associação positiva entre elas.

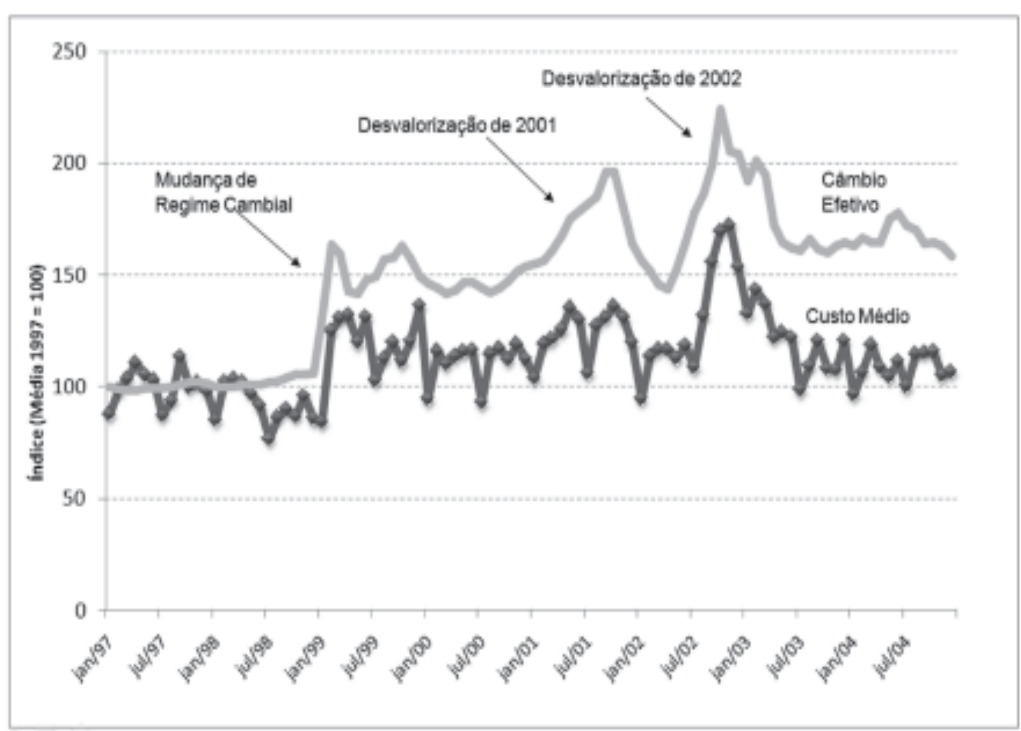

Figura 13 – Evolução de Câmbio Efetivo e Custos Médios das Companhias Aéreas

**Preços como Fator de Competitividade**

Outro fator a se considerar com relação à importância econômica do transporte aéreo, é que o item "passagens aéreas", em geral, tem importante participação nas POF (Pesquisa de Orçamentos Familiares) dos índices de inflação, e, portanto, tem influência direta na evolução dos preços da economia.

Por exemplo, no passado, houve períodos de "guerras de preços" entre operadoras (episódios em 1998, 2001, 2002, 2005), onde o setor foi um dos que mais contribuiu com a queda da inflação apontada pelos institutos de pesquisa – o que certamente influencia a eficiência dos setores que têm no transporte aéreo um de seus insumos produtivos. A grande evidência desse fator de competitividade ao longo da cadeia está no fato de que inúmeras instituições públicas e privadas possuem políticas de controle de orçamentos de gastos com passagens aéreas, sendo que algumas adotam critérios de "sempre viajar com a mais barata", dentre outros mecanismos de controle.

Para efeitos de análises de preços do transporte aéreo, uma excelente fonte é a base de dados do Instituto Brasileiro de Geografia e Estatística denominada de SIDRA (Sistema IBGE de Recuperação Automática). Nela, é possível obter desagregações dos índices de preço setoriais que compõem o índice de inflação ao consumidor, como, por exemplo, o IPCA. As figuras a seguir apresentam esses índices para o transporte aéreo, em comparação com outros setores da economia. Os índices de inflação foram transformados em índices de preço com valor igual a 100 estabelecido para junho de 2006.

A Figura 14 apresenta uma comparação dos índices de preços do transporte aéreo em contraposição a setores regulados da economia, basicamente energia elétrica, planos de saúde e telefonias fixa e móvel (celular). Pode-ser perceber, que, comparando-se o índice do transporte aéreo em contraposição a setores regulados da economia, basicamente energia elétrica, planos de saúde e telefonias fixa e móvel (celular). Pode-ser perceber, que, comparando-se o índice do transporte aéreo com os demais, tem-se que o mesmo chegou a ficar em último lugar em termos de contribuição com a inflação – entre o final de 2006 e início de 2007 –, mas que passou a apresentar altas consideráveis em meados de 2008, atingindo o valor de 110, aproximadamente, em setembro de 2008. Com esse desempenho, a evolução dos preços do transporte aéreo ficaria, nessa janela de análise, abaixo apenas do setor de planos de saúde.

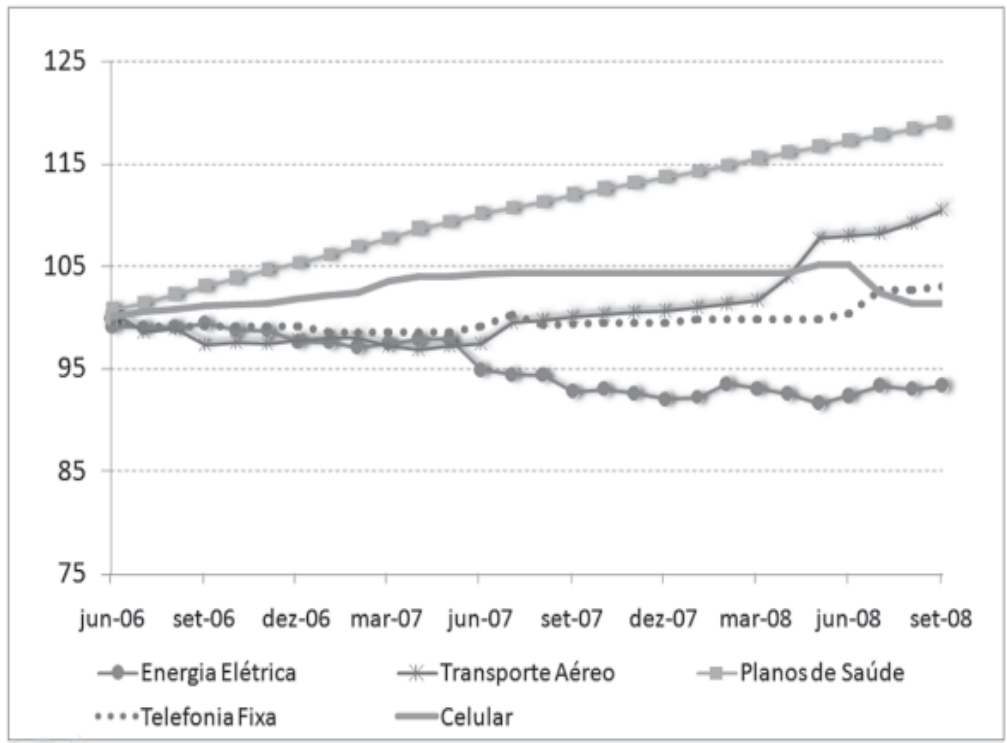

Figura 14 – Índices de Preços: Transporte Aéreo Versus Setores Regulados

A Figura 15 mostra a contraposição da evolução dos preços do transporte aéreo em relação a outros itens do setor de transportes. O comportamento relativo dos preços das passagens aéreas foi, em geral, parecido com o apresentado na Figura 14, apresentando-se relativamente baixo no início do período e crescendo acima da média em 2008. Importante notar que o transporte aéreo apresentou variação bem abaixo dos sucessivos aumentos observados para o transporte rodoviário inter-estadual, que é o concorrente mais direto das companhias aéreas brasileiras que operam no segmento doméstico.

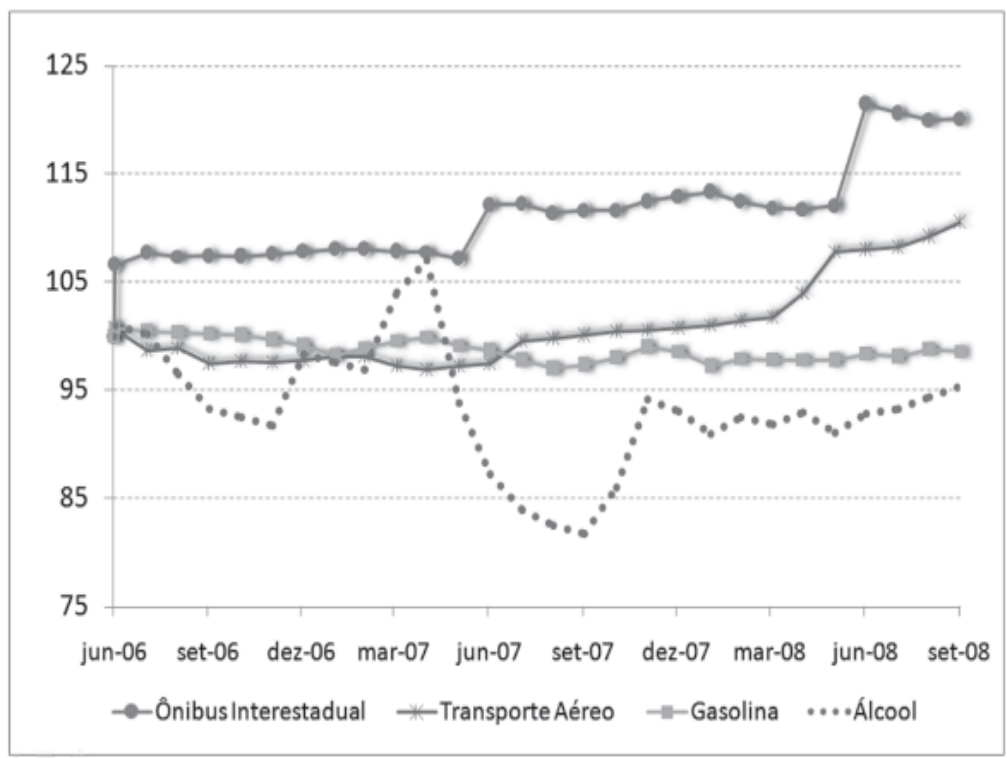

Figura 15 – Índices de Preços: Transporte Aéreo Versus Outros Itens de Transportes

### Escala de Vôo: Guerras de Preços e Elasticidade da Demanda

O fenômeno conhecido como "guerras de preços" é muito comum em transporte aéreo. Apesar de ser muito difícil qualificar, na prática, se um determinado mercado encontra-se, ou não, em guerra de preços, por todo o mundo, de tempos em tempos, a mídia anuncia a ocorrência desses episódios competitivos entre companhias aéreas. Autores como Morrison e Winston (1996) apontam que uma guerra de preços poderia ser contabilizada sempre que os preços apresentassem queda superior a 20%. Na Ponte Aérea Rio de Janeiro-São Paulo, as guerras de preço se tornaram mais comuns com a desregulação de fins dos anos 1990. Alguns exemplos:

→ 1998: em março, TAM e as empresas do chamado *Pool* da Ponte Aérea

reduzem em mais de 24% suas tarifas cheias; em maio, TAM concede descontos de até 34% da tarifa cheia; e em setembro, Vasp e Transbrasil concedem descontos de até 42%;

✈ 2001: em março, TAM e Transbrasil oferecem descontos de até 53% (com restrições de horário), e Vasp equipara suas tarifas cheias ao desconto máximo da concorrência;

✈ 2002: em janeiro, Gol entra na ligação com preços 54% menores que as tarifas cheias da maioria das incumbentes; em abril, TAM concede descontos de até 57%, Gol reage oferecendo descontos de até 23% sobre os seus já reduzidos preços e Vasp acompanha.

Para se avaliar se vale a pena ou não entrar em uma guerra de preços, as companhias aéreas devem analisar se haverá uma resposta satisfatória da demanda à queda abrupta das tarifas, ou do forte aumento da disponibilidade das classes tarifárias de valor mais baixo. Uma métrica bastante conhecida para isso é o que os economistas chamam de "elasticidade-preço da demanda". A elasticidade-preço da demanda avalia o percentual de incremento (ou de decréscimo) do tráfego de passageiros acarretado por dado um decréscimo (ou um incremento) percentual nos preços. Trata-se de um valor negativo e sem unidade de medida – o que facilita a comparabilidade entre mercados. Por exemplo, se a elasticidade-preço da demanda é igual a -2, tem-se que, para cada queda de preços de 10%, haverá a um aumento do tráfego existente em 20%. Mercados onde a elasticidade-preço da demanda está entre -1 e 0 são chamados *inelásticos* a preço. Nesses mercados, como a ponte aérea, as guerras de preço provavelmente não proporcionarão grandes volumes de tráfego novo, apenas permitirão que as empresas capturem fatias de mercado umas das outras (*market stealing* versus *market creation*). A Figura 16 apresenta estimativas de elasticidade-preço da demanda por transporte aéreo no Brasil:

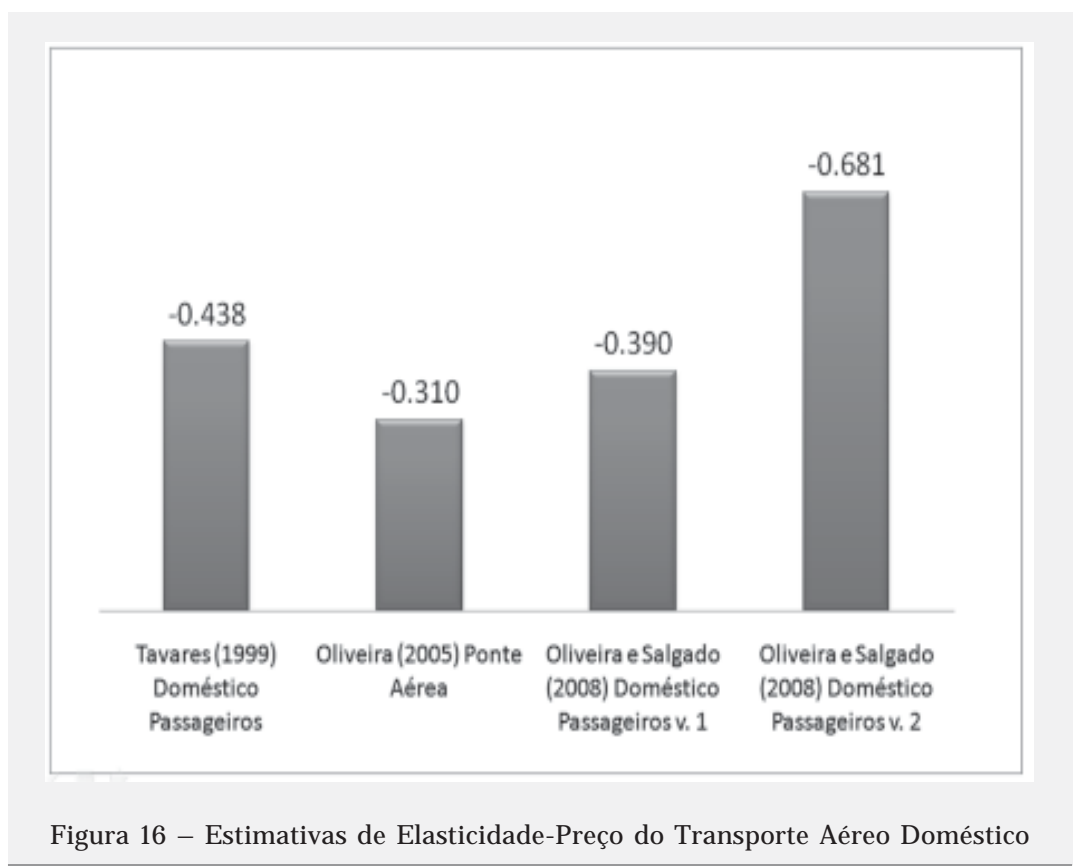

Figura 16 – Estimativas de Elasticidade-Preço do Transporte Aéreo Doméstico

**Diferenciação de Produto e Modelos de Negócio**

Foi-se o tempo em que o produto comercializado pelas companhias aéreas era visto como uma verdadeira *commodity*. O termo "*commodity*" retrata os bens cuja qualidade é bastante homogênea, onde não há espaço para diferenciação de produto que gere preços diferenciados. Em geral são transacionados em bolsas de mercadorias, como o petróleo, o ouro, a prata, o café, o algodão ou a soja. Em um passado não tão distante, pensava-se que o serviço de transporte de um passageiro de um ponto A para um ponto B em um dado espaço geográfico era basicamente o mesmo, qualquer que fosse a companhia aérea e seu posicionamento no mercado. Podia ser, então, analisado como se fosse uma *commodity*.

Não há como negar que, dependendo da situação competitiva e do quão elevado espera-se que seja o grau de substitutibilidade entre as ofertantes, pode vir a ser razoável assumir que a inexistência de atributos de diferenciação, ou, o que é equivalente, a homogeneidade do produto. Tome o caso de competição entre American Airlines e United Airlines, nos Estados Unidos, em mercados onde detêm aproximadamente o mesmo número de freqüências de vôo ou *market share*. Numa situação como essa, de concorrência entre duas grandes empresas com serviços equivalentes em termos de conveniência, qualidade e disponibilidade, pode-se argumentar que o serviço é "comoditizado". Mesmo assim, há questões relativas ao programa de milhagem de cada uma, onde os passageiros freqüentes acabam por se fidelizar e formar fortes preferências com relação à empresa previamente escolhida.

Em contraste, a competição no setor aéreo hoje em dia é analisada sob a ótica da rivalidade entre firmas que são distintas entre si, que possuem atributos diferenciados e que são efetivamente percebidos pelo consumidor. Estes atributos distintos adviriam de:

↬ vantagens competitivas ao nível da rota, como, por exemplo, diferentes escalas de operação, com diferentes números de freqüências diárias de vôo, o que gera assimetrias com relação às disponibilidades das ofertas ao consumidor, gerando diferentes distâncias entre horários de vôo ofertados e horários de partida desejados – conhecidas como "*schedule delay*" –; diferentes apelos aos passageiros com viagens por motivos de negócios; diferentes padrões de serviços de atendimento ao consumidor, serviço de bordo e atendimento durante o vôo, o tipo de aeronave, etc.

↬ vantagens competitivas ao nível do aeroporto e da cidade, como por exemplo, tamanhos diferentes de rede doméstica e internacional; participação em alianças globais de companhias aéreas; número de cidades atendidas pela companhia aérea; níveis de propaganda na zona de influência do aeroporto; características do programa de milhagem; restrições verticais com relação aos agentes de viagem, atendimento antes e depois da viagem, etc.

Ademais, tem-se que, com o advento das chamadas "Companhias Aéreas de Custo Baixo, Preço Baixo" ("*Low Cost, Low Fare*"), e sua penetração cada vez maior nos mercados aéreos em todo o mundo, a tendência da literatura tem sido a de considerar este setor como típico de produto diferenciado, dada a nítida distinção entre os padrões de serviço e a forma de atuação das novas entrantes e as incumbentes, companhias aéreas baseadas em redes ("*network carriers*"). A coexistência de distintos modelos de negócios, com diferentes estratégias de posicionamento e de *marketing* levando a abordagens alternativas dos vários segmentos de consumidor, trouxe à tona as questões de heterogeneidade do produto no setor aéreo.

### Escala de Vôo: A Entrada da TAM na Star Alliance

Uma aliança é um acordo entre duas ou mais companhias aéreas para promover integração, apoio mútuo, cooperação e coordenação de ações. Por meio das alianças, viabiliza-se uma estrutura de rede de altíssima conectividade e conveniência para passageiros internacionais e para a constituição de pacotes de viagens internacionais. As mais importantes alianças entre companhias aéreas em todo mundo são aquelas constituídas em nível internacional, denominadas de "alianças globais". As três maiores são a Star Alliance, a SkyTeam e a Oneworld. Existem também alianças entre operadoras aéreas do segmento de carga. As alianças vêm ampliando substancialmente a sua extensão global, incrementando o número de associadas. Segundo o Tourism Features International, as três principais alianças combinavam um total de 29 membros em 2003. Atualmente, esse número remonta a 42 membros (2008), sem incluir as companhias aéreas consideradas "afiliadas" ou "associadas". As três alianças, respondiam, juntas, por quase 80% da capacidade de vôo das companhias aéreas (medida em ASKs), 78% do total de RPKs mundiais e 73% do total de passageiros transportados por todo o mundo (Tourism Features International). Seguem as características dessas alianças (2008):

✈ Oneworld: Fundada em 1999. 10 membros, 20 afiliadas. Fundadoras: American Airlines, British Airways, Canadian Airlines, Cathay Pacific Airways e Qantas Airways.

✈ SkyTeam: Fundada em 2000. 11 membros e 3 associadas. Fundadoras: Air France, Delta, AeroMexico e Korean Airlines.

✈ Star Alliance: Fundada em 1997. 21 membros e 3 regionais. Fundadoras: United Airlines, Air Canada, Lufthansa, Thai Airways e SAS-Scandinavian Airlines.

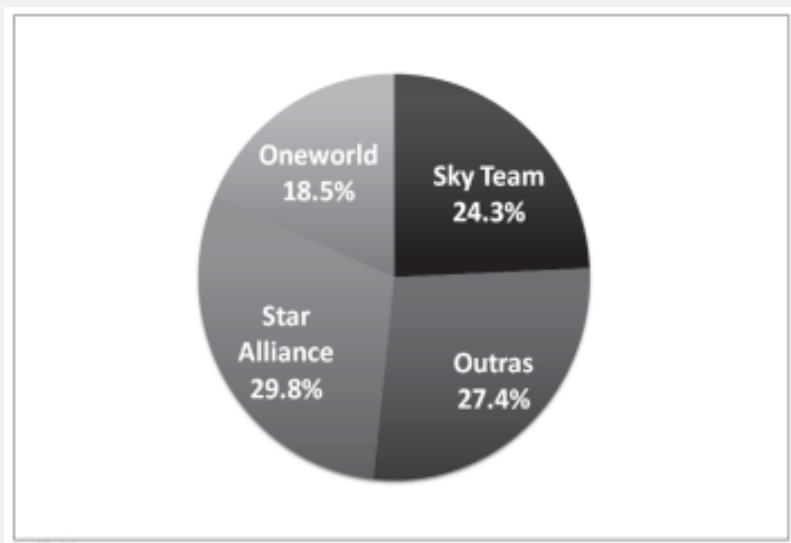

Figura 17 - Market Share de Passageiros Transportados pelas Alianças
(Fonte: Tourism Features International, 2008)

Há uma corrida para incluir empresas de mercados emergentes, como a China, a Índia e o Brasil. Enquanto a China Southern Airlines se juntou ao SkyTeam, a Star Alliance incluiu a Air China e a Shanghai Airlines, além de incluir a Air India como um membro futuro. No Brasil, a Varig foi praticamente uma das fundadoras da Star Alliance, quando se juntou à aliança cinco meses após a sua constituição, em 1997. Atualmente a empresa, sob administração da Gol, não faz mais parte da aliança. A TAM associou-se à Star em 2008. Em notícia do Estado de São Paulo, de 8/10/2008, intitulada "TAM entra na Star Alliance", comentava-se que "*De acordo com o presidente da TAM (...) a entrada na Star Alliance vai implicar em um acréscimo de receita de US$ 60 milhões ao ano*". Adicionalmente, "*o aumento de tráfego proporcionado pela aliança é estimado em 2 a 3 pontos percentuais no mercado doméstico*". Isso acontece porque os benefícios inerentes à aliança são percebidos pelo passageiro como um atributo a mais de qualidade, aumentando a sua disposição a pagar e propensão a viajar. Obtém-se integração dos programas de milhagem, compartilhamento de salas VIP em aeroportos de todo o mundo, conectividade mais fácil, despacho de bagagem único, dentre vários benefícios para passageiros.

## Modelo de Negócios "Low Cost, Low Fare"

O fenômeno da expansão das empresas chamadas de "Custo Baixo, Preço Baixo" ("*Low Cost, Low Fare*", ou simplesmente, "*Low Cost*") foi um dos acontecimentos mais marcantes do transporte aéreo mundial nas últimas décadas. Com expressivas taxas de crescimento e preços consideravelmente mais baixos que os então vigentes, as companhias aéreas baseadas nesse modelo de negócios acabaram por popularizar consideravelmente o transporte aéreo como meio de transporte em diversos países ou bloco de países. A Figura 18 a seguir apresenta a evolução da participação de mercado dessas companhias no mercado norte-americano.

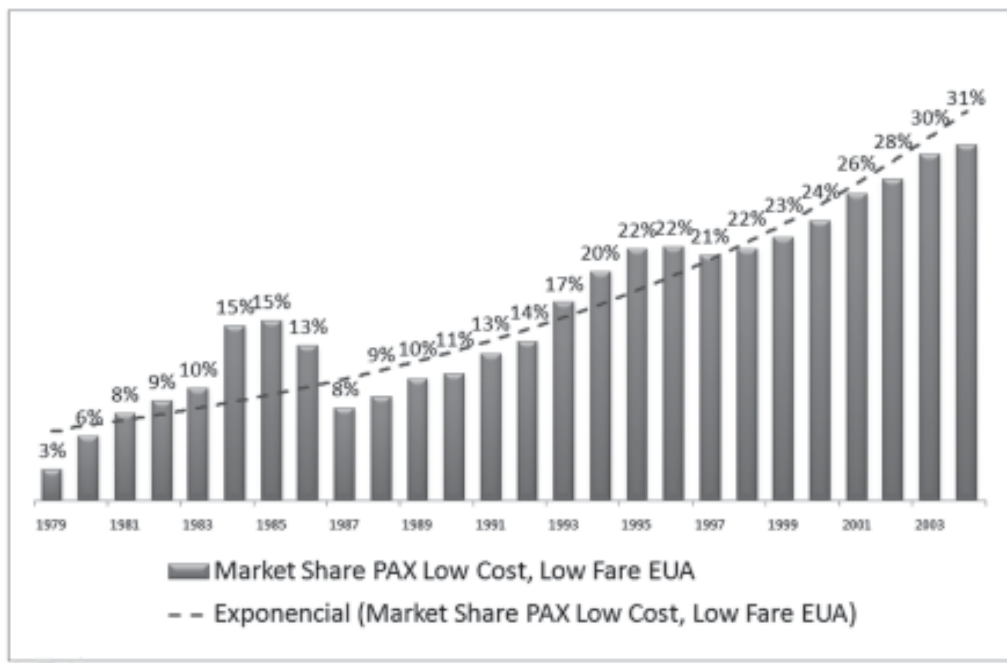

Figura 18 – *Market Share* das Companhias *Low Cost* dos Estados Unidos[15]

Como se pode perceber, mesmo em se tratando de fatias de mercado, ou seja, de percentuais, tem-se um crescimento exponencial das companhias "*Low Cost, Low Fare*", como mostra a linha de tendência apresentada no gráfico. Isso significa que não apenas houve uma captura considerável do mercado antes sob domínio das chamadas "*legacy carriers*" – as grandes empresas que já operavam em âmbito nacional desde o período regulatório, pré-*Deregulation Act* de 1978. Mas também significa que essas empresas souberam criar novos mercados e viagens mais do que proporcionalmente, na medida em que a economia crescia impulsionando o transporte aéreo. O crescimento do modelo foi tamanho que, desde 2004, três entre quatro passageiros domésticos nos Estados Unidos possuem alguma alternativa real de vôo do tipo *Low Cost*.

Devido ao seu pioneirismo e sucesso, quando se fala em "*Low Cost, Low Fare*", o nome Southwest Airlines inevitavelmente vem logo à cabeça. A Southwest foi a empresa pioneira desse tipo de operação em nível mundial: "*We took a great idea and made it fly*", é o slogan apresentado na autobiografia da empresa, em seu *website*. Sua importância é tamanha que o pesquisador Steve Morrison, da Northeastern University, chegou a estimar, por meio de modelagem econométrica, que o efeito completo da presença da Southwest Airlines nos mercados aéreos norte-americanos em termos de economias para o consumidor era da ordem de 13 bilhões de dólares em 1998, ou 20% da receita das companhias aéreas regulares de passageiros daquele ano (Morrison, 2001).

As características do paradigma representado pelo modelo de negócios da Southwest Airlines são as seguintes:

✈ vôos para rotas densas, com etapas curtas e configuração de rede ponto-a-ponto, sem transferências ou conexões, com ênfase no uso intensivo das aeronaves;

✈ estratégia agressiva de precificação;

✈ pioneirismo na ênfase no *e-ticketing* (bilhetes emitidos eletronicamente);

✈ ênfase nas vendas pelo *website* próprio, sem recurso a agentes de viagem ou sites de terceiros, como Orbitz e Travelocity;

✈ radical simplificação do serviço de bordo, corte de *frills*, sem perder a    atenciosidade;

✈ padronização da frota de aeronaves, em geral configuradas com única  classe;

✈ uso de aeroportos secundários menos congestionados;

O resultado do modelo criado pela Southwest Airlines foi a geração de operações aéreas com custos unitários que são até 70% mais baixos do que os custos das grandes companhias aéreas americanas (Bennett and Craun, 1993). A Southwest pode ser considerada a empresa aérea com maior taxa de crescimento e a mais lucrativa dos Estados Unidos e uma das mais bem-sucedidas da aviação mundial. A empresa é a única grande companhia aérea da história dos Estados Unidos que nunca apresentou prejuízo, contam

Boguslaski, Ito e Lee (2004). Para completar, é persistentemente a empresa com os mais baixos índices de reclamação por passageiro registrados pelo Departamento de Transportes daquele país. A Figura 19 apresenta a evolução do *market share* da Southwest ao longo da década de 1990.

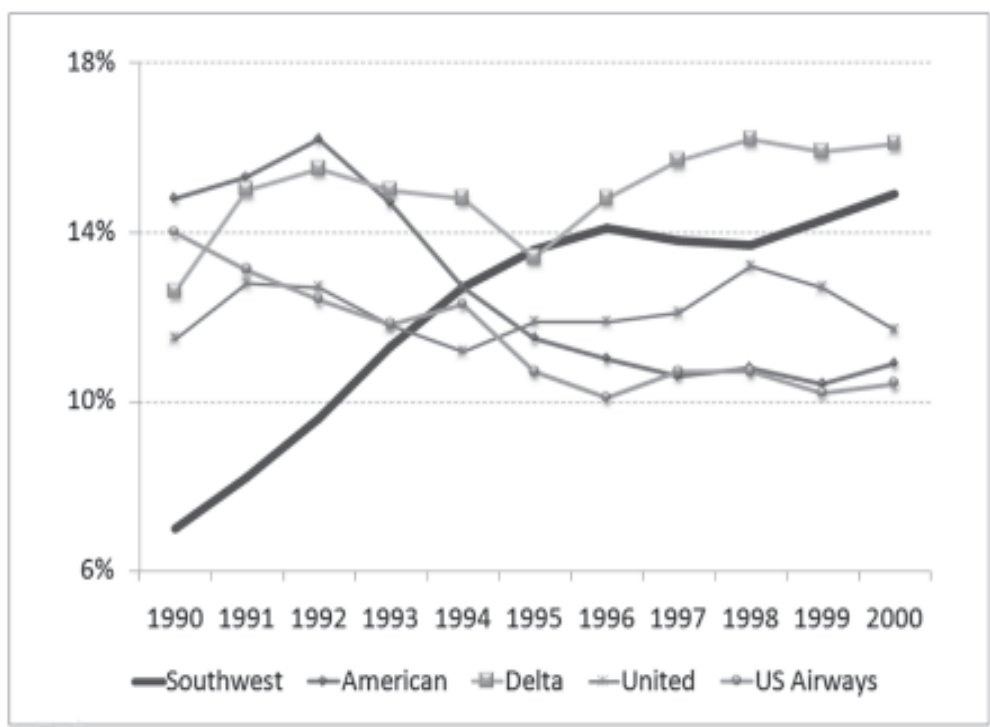

Figura 19 – Crescimento do *Market Share* Da Southwest Airlines ao longo da Década de 1990[16]

Por fim, tem-se o chamado "Efeito Southwest". Este foi um termo criado pelo Departamento de Transportes dos Estados Unidos para descrever o extraordinário fator indutivo de viagens aéreas resultantes da entrada da Southwest Airlines em novos mercados. Com preços consideravelmente inferiores do que as empresas aéreas instaladas nas localidades, a Southwest conseguia efetivamente estimular nova demanda, incrementando o tráfego existente em 2, 3, 4 vezes. Em um dos casos mais incríveis do Efeito Southwest, tem-se a entrada da empresa em Providence, Rhode Island. Naquela cidade, o transporte aéreo cresceu de 100 mil passageiros/ano para 800 mil passageiros ano após a entrada da Southwest (Yeh, 2004). Contam Boguslaski, Ito e Lee (2004), que, depois que a Southwest entrou na rota Oakland–Ontario, em 1989, os preços caíram em 60%, enquanto o tráfego mais do que triplicou. O *Transportation Research Board* analisou as rotas que a Southwest entrou entre 1990 e 1998 e chegou à conclusão de que, na média, as viagens aumentavam em torno de 174% e os preços médios caíam em torno de 54%. A Southwest consegue induzir esse efeito de forte criação de tráfego novo por meio de um forte posicionamento junto aos passageiros que são mais sensíveis a preço, muitos dos quais iriam viajar de carro ou mesmo nem viajariam antes da entrada da Southwest.

Na Europa, a empresa *Low Cost* irlandesa Ryanair, uma das mais perspicazes seguidoras do Paradigma Southwest, também costuma multiplicar por fatores similares o tráfego de um novo aeroporto adicionado em sua malha. A entrada de uma empresa como essas é, portanto, um evento indutor de alavancagens e desenvolvimento por toda a região da nova localidade, dado o incremento na mobilidade das pessoas e indução e facilitação de negócios. Aeroportos em estado de ociosidade concorrem entre si para atrair empresas como a Southwest e a Ryanair em seu *portfolio* de operações de vôo, dados os potenciais *spillovers* gerados com a sua entrada.

No Brasil, a empresa que mais se assemelhou ao modelo ditado pela Southwest Airlines foi a Gol em seus primeiros dois anos de existência (2001-2002). Criada em pleno período de liberalização tarifária do setor doméstico, a Gol Linhas Aéreas entrou no mercado de transporte aéreo com um conjunto de expressivas inovações, como o *e-ticketing*, o serviço de bordo simplificado – por exemplo, as barrinhas de cereal, uma verdadeira revolução, se comparado com sofisticadíssimo serviço de bordo vigente à época –, a frota padronizada com Boeings B737-700 e a ênfase nas tarifas baixas. O seu crescimento, ao longo dos primeiros trimestres, foi avassalador. Provavelmente representa uma das maiores taxas de crescimento da história da aviação mundial, tendo quase que triplicado sua fatia de mercado em pouco mais de quatro anos, passando de 3% em abril de 2001, para 27% em setembro de 2005. Algo que nem a Southwest conseguiu fazer ao longo de três décadas de sucesso no mercado de transporte aéreo norte-americano.

A Figura 20 mostra a evolução do *market share* da Gol no mercado de transporte doméstico de passageiros, destacando os primeiros trimestres de operação comparativamente com as projeções de uma curva exponencial. Pode-se perceber que o ajuste exponencial é praticamente idêntico à evolução observada pela empresa ao longo desses primeiros trimestres. Coincidentemente ou não, esse período ficou conhecido como a fase em que a empresa mais se assemelhava ao padrão "*Low Cost, Low Fare*", com tarifas perceptivelmente mais baixas, lastreadas por custos unitários e *yields* aproximadamente um terço abaixo das empresas instaladas e uma etapa média de vôo que era 20% menor (Oliveira, 2008).

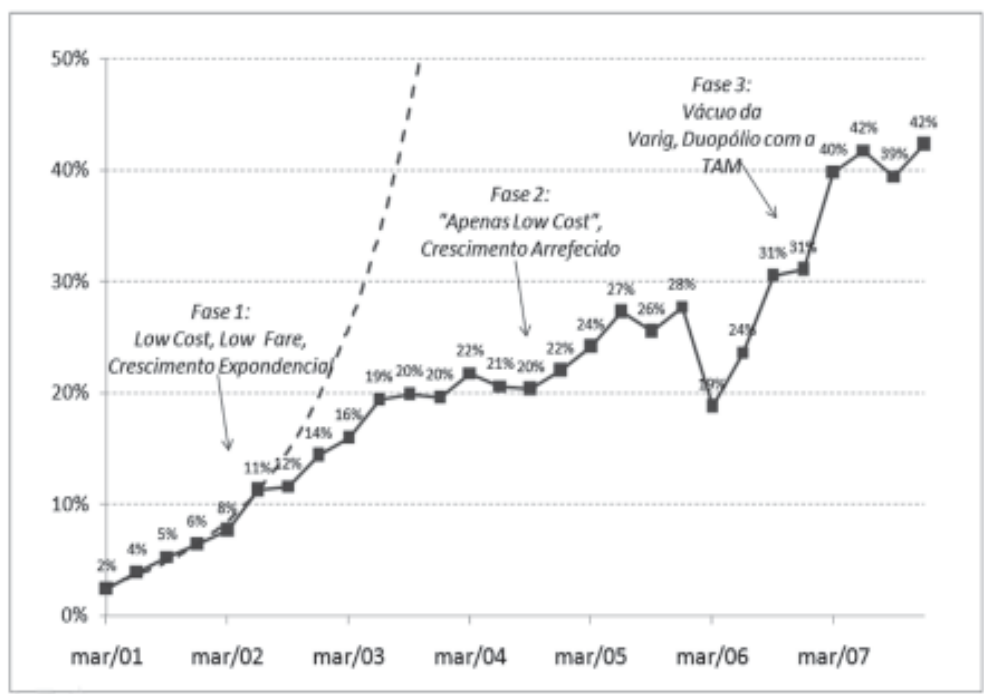

Figura 20 – *Market Share* da Gol no Mercado Doméstico[17]

A Figura 21 apresenta a evolução do numero de passageiros adicionais transportados a cada ano pela Gol. No eixo da esquerda, apresenta-se o número de novas viagens obtidas em um determinado ano (na casa dos milhões). No eixo da direita, apresenta-se a mesma métrica, só que contabilizada por minuto, ou seja, o número de novos passageiros por minuto que a empresa consegue produzir a cada ano que passa[18].

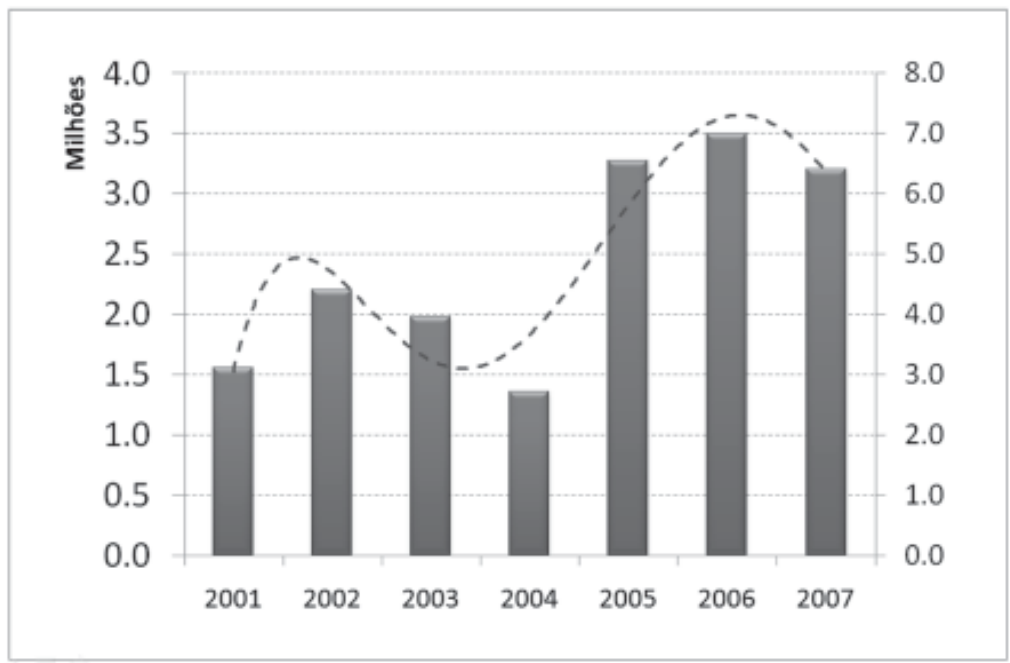

Figura 21 – Passageiros Adicionais por Ano Transportados pela Gol[19]

A partir do último trimestre de 2002, entretanto, a trajetória de crescimento exponencial deu lugar a uma fase de crescimento arrefecido. A forte desvalorização cambial e um conjunto de medidas restritivas impostas em 2003 pelo então regulador, o Departamento de Aviação Civil (DAC) contribuíram para essa queda de performance. Se mantivesse as taxas de crescimento anteriores ao segundo semestre de 2002, a Gol provavelmente atingiria trinta por cento do mercado antes mesmo de completar três anos de operações. De lá para cá, o senso comum diz que a empresa "continua *low cost*, mas deixou de ser *low fare*", em alusão à maior dificuldade de se encontrar tarifas com altos descontos pela empresa. Essa não é uma afirmação testada empiricamente em nenhum estudo até o momento, e, portanto não pode ser considerada como definitiva sobre a questão. É importante frisar que a Gol ainda costuma ser bastante preferida pelo viajante "*budget*", à caça de descontos, em inúmeras situações de compra de passagens aéreas dentro do País e mesmo para o exterior.

**Modelo de Negócios "Network" ou "Hub and Spoke"**

O modelo de negócios das chamadas "Network Carriers" se baseia no conceito de estrutura de rede *hub-and-spoke* (centro-aros). Esse sistema – o mais predominante dos tipos de configuração de rede em transporte aéreo – foi uma das grandes inovações tecnológicas proporcionadas pela desregulação do setor nos Estados Unidos no final da década de 1970, permitindo uma maior racionalização das malhas aéreas então existentes. O sistema pode ser visualizado na Figura 22 a seguir:

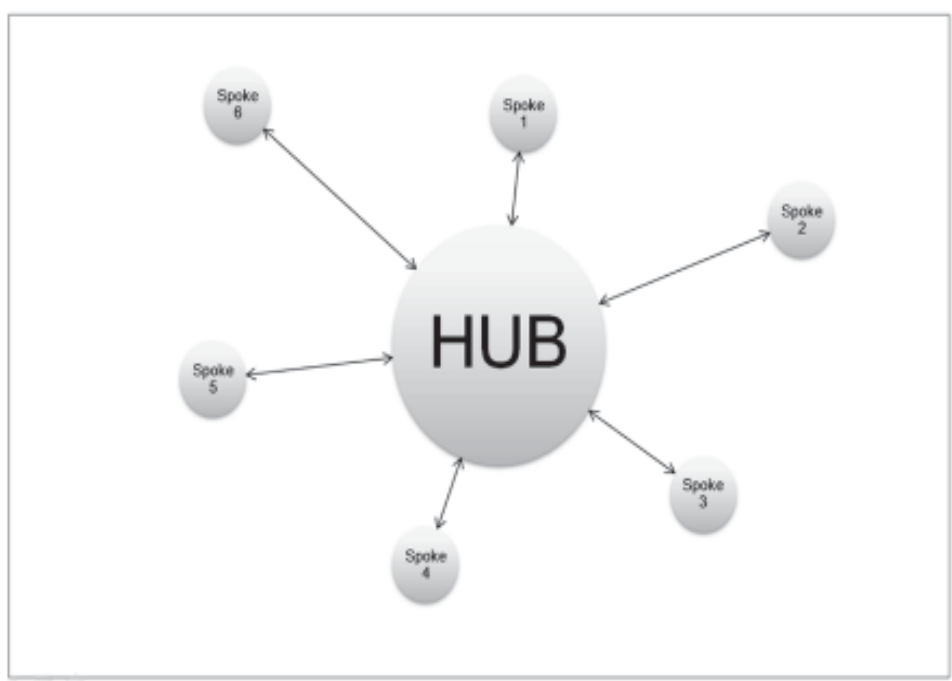

Figura 22 – Configuração de rede "*Hub-and-Spoke*"

Com a estrutura de rede *hub-and-spoke*, as companhias aéreas atribuem a um ou mais aeroportos de sua malha (o "*hub*") como tendo a função de centro de conexões. Os demais aeroportos da malha são os "*spokes*" ("aros"), em geral aeroportos menores ou menos estratégicos para a companhia aérea. Com essa configuração de rede, todos os passageiros passarão pelo aeroporto *hub* da companhia aérea, seja porque ele representa o início ou fim de sua viagem ("tráfego local"), seja porque nele será feita uma conexão rumo ao destino final ("tráfego de passagem pelo *hub*"). Com todos os aeroportos interligados ao *hub*, tem-se, no limite, potencializado o número possível de interconexões na malha, o que incrementa consideravelmente a oferta de vôos das empresas que seguem esse modelo de negócio. Por outro lado, o nível de serviços e a qualidade percebida das empresas perante o consumidor cai, dado que adiciona-se ao tempo de vôo também o tempo de espera pelo vôo de conexão. Há também um uso intensivo do aeroporto *hub*, que acaba por requerer investimentos na melhoria da prestação dos serviços aeroportuários e das áreas comerciais existentes. Os grandes *hubs* nos Estados Unidos, como Hartsfield-Jackson Atlanta, Chicago O'Hare e Dallas/ Ft.Worth, são também os que mais geram volume de negócios.

Para se ter uma idéia da revolução nas malhas das companhias aéreas incorrida pela introdução da configuração em hub-and-spokes, tem-se as figuras a seguir. Elas foram extraídas do clássico livro dos professores Steve Morrison e Clifford Winston, "*The Economic Effects of Airline Deregulation*" de 1986[20]. Retratam a dramática reformulação da estrutura de rotas da companhia aérea Western Airlines antes e depois da desregulamentação norte-americana de 1978. É perceptível, pela análise das malhas com o deregulation, que Salt Lake City e Los Angeles tornaram-se *hubs* da malha da empresa, que antes apresentava uma configuração tipicamente "ponto-a-ponto". Importante perceber como um conjunto de vôos "pinga-pinga" (estrutura do tipo "rede *tour*"), como as que eram observadas entre Salt Lake e Great Falls, foram removidas com a total racionalização da malha.

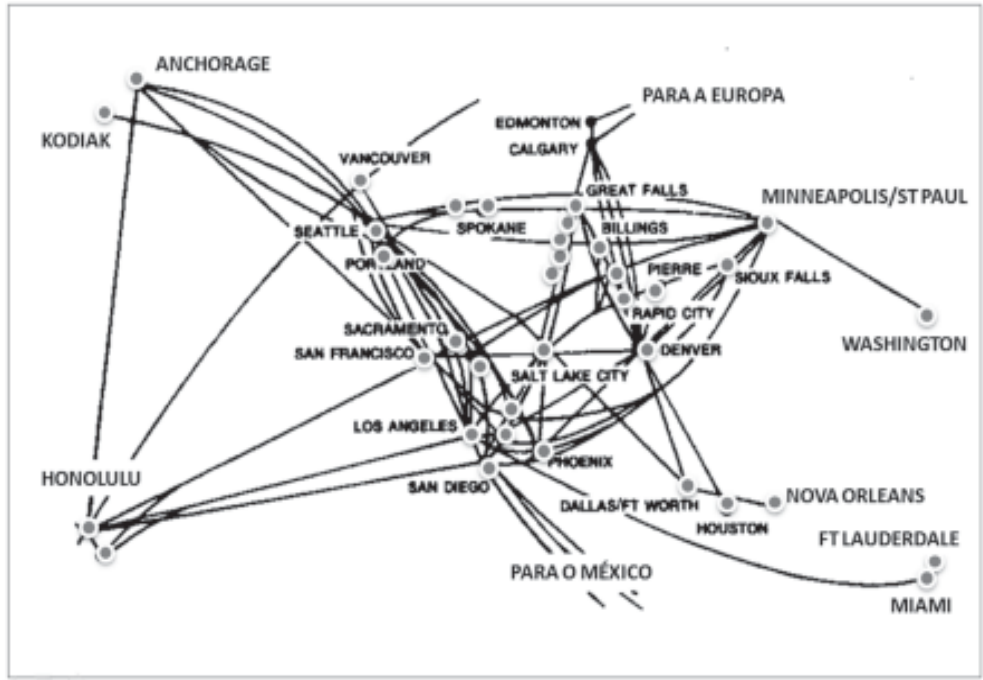

Figura 23 – Malha Aérea da *Northwest Airlines* antes da Desregulação de 1978

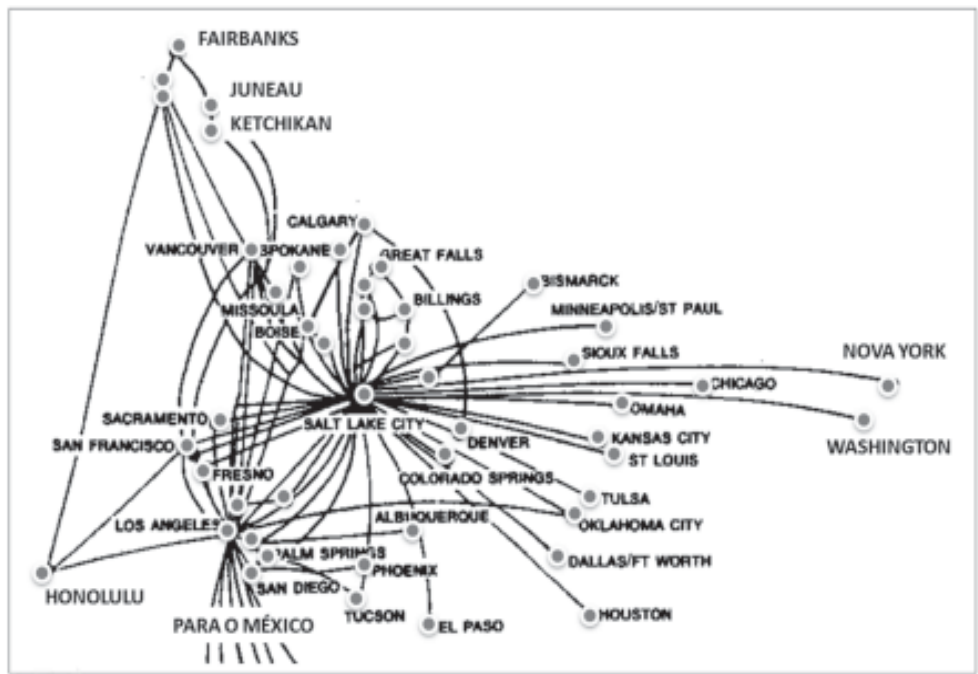

Figura 24 – Malha Aérea da *Northwest Airlines* após a Desregulação de 1978

**Investimentos e Pressão Sobre as Infra-Estruturas**

Como todo setor relacionado com a infra-estrutura – apesar de, a rigor, estar enquadrado dentre os setores de serviços –, o transporte aéreo carece de um constante fluxo de investimentos, de forma a não constituir gargalo ao crescimento econômico. Como se trata de indústria com processo produtivo intensivo em capital, os montantes de investimentos são, em geral, vultosos, apesar de esparsos no tempo, e estas características têm implicações relevantes na economia como um todo. Por exemplo, contam Comegno e Paulino (2003), que, no estado de São Paulo, em 2001, foram anunciados investimentos da ordem de 3.1 bilhões de dólares em transporte aéreo, o que representou 13% do total de investimentos do estado naquele ano.

Como vimos, um transporte aéreo sob o regime de livre mercado pode ser extremamente dinâmico em termos de crescimento econômico. Isso é ainda mais relevante em um caso de mercado aéreo em estágio inicial de amadurecimento, como se trata do Brasil. Nesses mercados, ao contrário de mercados maduros como Estados Unidos e Europa, há um potencial muito maior de crescimento do setor e alavancagem de novos investimentos. Isso acontece porque o norte-americano médio, ou o europeu médio, provavelmente já realizam um excelente número de viagens por ano por modal aéreo. No caso brasileiro, esses números ainda são relativamente baixos. A evidência mais forte disso é o recente crescimento do mercado no País. Esse crescimento acelerado tem ultrapassado todas as expectativas, tanto de analistas setoriais – responsáveis pelas recomendações de investimentos privados – quanto do governo – responsável pelo planejamento, acompanhamento e pelos investimentos públicos.

Para se ter uma idéia desse fenômeno de crescimento acelerado – e que acabou se tornando problemático por conta da forte demanda exercida sobre as infra-estruturas de apoio –, basta contrastar as previsões de transporte de passageiros pelo modal aéreo com o tráfego efetivamente observado. Em 1997, o IV Plano de Desenvolvimento do Sistema de Aviação Civil (PDSAC), desenvolvido pelo Departamento de Aviação Civil, reportou uma previsão efetuada pelo Instituto de Aviação Civil de que o transporte aéreo alcançaria, em 2010, a casa dos 35.3 bilhões de passageiros-quilômetros transportados por ano. No cenário mais otimista de todos, esse número fiaria em torno de 44 bilhões. Pois bem, já em 2007, computando-se apenas o tráfego de passageiros pagantes, essa cifra já foi alcançada. Ou seja, a previsão de tráfego do governo se materializou com 3 anos de antecedência do estipulado. A Figura 25 permite observar o quanto o crescimento do transporte aéreo extrapolou as expectativas das autoridades. Fazendo uma extrapolação da previsão otimista do IAC de 2010 para 2007 (por interpolação geométrica), e comparando com a previsão otimista realizada para o ano de 2000, temos o seguinte gráfico:

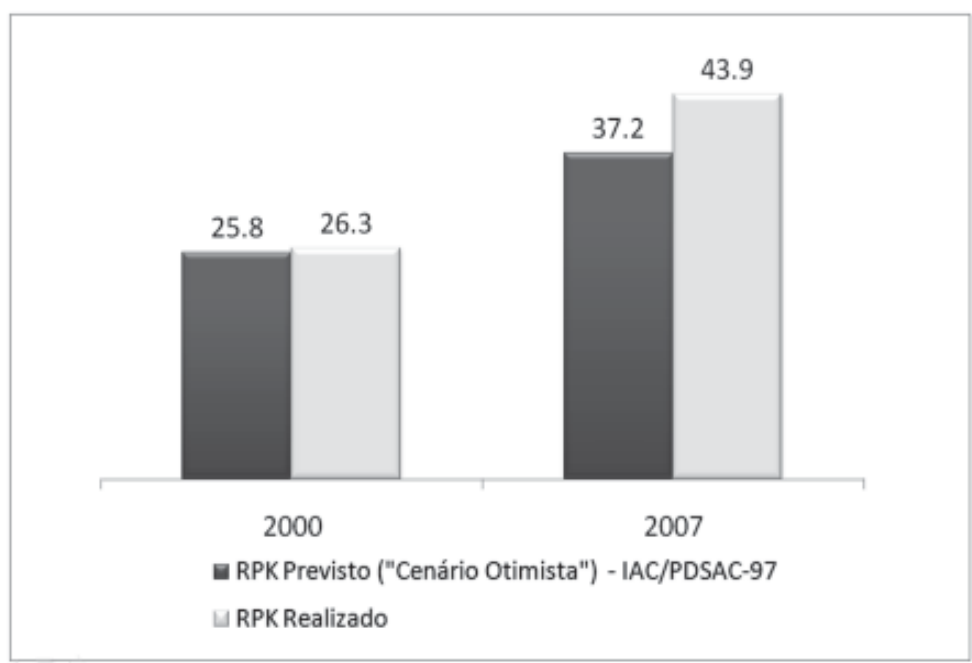

Figura 25 – Comparação entre Tráfego Projetado em 1997 (Cenário "Otimista") e Tráfego Efetivo – anos 2000 e 2007

Pode-se perceber que as projeções do Cenário "Otimista" eram as mais realistas para o início da década de 2000. O rápido crescimento do transporte aéreo fez com que mesmo a previsão otimista ficasse defasada em cerca de 6.7 bilhões de passageiros-quilômetros transportados no ano de 2002. Esses 6.7 bilhões são representativos de aproximadamente um milhão de passageiros adicionais e não previstos no sistema[21]. Com essa magnitude de demanda não esperada e não antecipada nos planos pelas autoridades, tem-se a natural pressão exercida pelo modal aéreo sobre a sua infra-estrutura de apoio – leia-se aeroportos e controle do espaço aéreo. Tem-se que é fundamental garantir um fluxo de investimentos em infra-estrutura que seja consistente com o dinamismo do crescimento dos negócios das companhias aéreas. Para isso é preciso conferir dinamismo equivalente à gestão da operação das infra-estruturas aeroportuária e de controle de espaço aéreo, tornando-as cada vez mais orientadas para o mercado. Sob o risco de termos novos "apagões", como os de 2006 e 2007, em um futuro não tão distante.

**Inserção do País e Contas Externas**

O tráfego internacional de passageiros e de carga é visto, em geral, como importante fator de reforço à inserção do País no contexto internacional e o fortalecimento de suas posições geopolíticas e laços comerciais. Adicionalmente, a recepção de turistas internacionais por modal aéreo é reconhecidamente uma importante fonte de crescimento e desenvolvimento.

O setor tem tecnologia de produção altamente capital-intensiva, dado que as aeronaves adquiridas por uma companhia aérea possuem alto valor agregado. Uma aeronave Embraer E-190, por exemplo, tem um preço de tabela da ordem de 30 milhões de dólares, o equivalente à receita gerada por mais de dois mil carros populares. Isso configura desafios relevantes no que tange à evolução das exportações e importações brasileiras e à sensibilidade às flutuações cambiais. Isto também coloca o transporte aéreo dentre os setores da economia com alto grau de vulnerabilidade a choques externos. Por outro lado, as receitas em moeda estrangeira, auferidas por esse segmento, têm usualmente peso considerável na conta de serviços do Balanço de Pagamentos. No que tange à manufatura de aeronaves, a Embraer foi repetidamente a maior exportadora brasileira entre os anos de 1999 e 2001, e a segunda maior empresa exportadora entre os anos de 2002 e 2004. Sua influência na dinâmica da pauta de exportações brasileiras é, portanto, considerável e marcante.

O Brasil é o maior mercado aéreo da América do Sul, e divide com o México a qualificação de mais importante mercado da América Latina. De acordo com Pasin e Lacerda (2003), o tráfego doméstico e internacional superou a casa dos 75 milhões de passageiros em toda a região em 2000, sendo que a maior participação, 35%, foi transportada por companhias brasileiras (principalmente Varig e TAM). Dentre as principais companhias latino-americanas, ainda considerando valores de 2000, destacavam-se a Varig com maior *market share* global (14,9%), seguida pela Aeromexico (13%), TAM (11,9%) e Mexicana (11,7%). Com a derrocada da Varig, em 2006, a TAM passou a ser a principal operadora do mercado, tendo anunciado sua entrada na aliança global *Star Alliance* em 2008.

A menção de maior e mais importante mercado de toda a região da América Latina é extremamente relevante para a atratividade das companhias aéreas junto aos investidores. Isso é ainda mais potencializado pelo fato das principais companhias aéreas da região possuírem atualmente ações negociadas na Bolsa de Nova York, por exemplo. Naquela instância de acesso aos capitais internacionais, empresas TAM, Gol e Lan Chile, por exemplo, competem pelos mesmos recursos entre si e com as rivais norte-americanas, por exemplo. Dado que o País possui alta potencialidade para vencer a corrida para a designação dos aeroportos como principais hubs internacionais da região, dada a sua maior escala de operações; além disso, o reforço do transporte aéreo no Mercosul é importante na alavancagem do turismo internacional na região; por fim, essa menção facilita o acesso de companhias aéreas ao mercado de capitais norte-americano , bem como, potencialmente, facilita o pleito por uma redução de taxas de arrendamento e seguros de aeronaves, e, portanto nos custos operacionais das companhias aéreas.

*Para abrir o capital, lançando ações nas bolsas de valores, ascompanhias aéreas precisam demonstrar que são transparentes, fortes financeiramente e que operam em mercados sólidos, com boa combinação de crescimento e lucratividade. Ser uma empresa com bom market share no principal mercado da América Latina contribui em muito para a companhia aérea alavancar seu capital, por exemplo, na sua oferta inicial de ações (IPO, Initial Public Offer). Em 2004, por exemplo, a companhia aérea Gol abriu capital simultaneamente em duas bolsas, a Bovespa e a New York Stock Exchange (NYSE), e concorre diretamente com a TAM pela atração dos investimentos para o País. Concorre também com outras companhias aéreas latino-americanas e norte-americanas.*

## Integração Inter-Regional e Desenvolvimento Sustentável

A cobertura do transporte aéreo ao longo do território nacional é fator crucial de políticas públicas de promoção do desenvolvimento sustentável. Por conta de sua extensão, o País não pode prescindir de mecanismos que promovam o crescimento mais equânime e menos economicamente concentrado no Sudeste. O que era no passado alcunhado de "integração nacional", termo exaustivamente usado pelos governos militares, mas sem nenhuma conceituação minimamente embasada em aspectos técnicos de economia regional e urbana, transformou-se nessa necessidade premente de garantir a sustentabilidade do crescimento. O transporte aéreo também pode ser analisado sob essa ótica de políticas públicas. Entretanto, diferentemente de outros setores regulados, como telecomunicações, energia elétrica e saneamento, nunca será viável a uma autoridade governamental ter por meta garantir algo próximo à universalização do transporte aéreo.

Mesmo não sendo possível garantir a universalização completa do setor, pode-se afirmar que há formas de se planejar e alcançar resultados positivos para importantes indicadores de cobertura espacial, tais como o *status* da concentração da malha em poucos aeroportos, os percentuais de conexões de vôo nos principais *hubs* e a participação dos aeroportos regionais no bolo do tráfego aéreo brasileiro.

Ao longo de toda a malha aeroportuária brasileira existem inúmeras realidades a serem melhor entendidas e investigadas, no sentido de propiciar políticas de fomento e melhor alocação dos escassos recursos para investimentos e reformas. Essa miríade de realidades é devido à grande extensão do território nacional e da desigualdade sócio-econômicas entre as regiões, ilustrando bem as distintas situações a que passam todo o sistema aeroportuário brasileiro e no qual as autoridades têm que se debruçar.

***Pouso: As Diferentes Realidades de Congonhas e de Cruzeiro do Sul***

Um dos choques de realidade mais ilustrativos do setor aeroportuário brasileiro pode ser feito pela comparação entre os aeroportos de Congonhas, em São Paulo e Cruzeiro do Sul, no Acre. O primeiro está cravado na região central da maior metrópole brasileira, enquanto o segundo está localizado em plena região Amazônica, no Vale do Juruá. Um, comprimido pela vida urbana e seus problemas – sendo também causador de alguns deles, como o risco de acidentes, o ruído, as emissões de gases poluentes. O outro cercado de natureza, sendo em muitas circunstâncias a única alternativa de aviação de toda a região. A Figura 26 é ilustrativa desses contrastes.

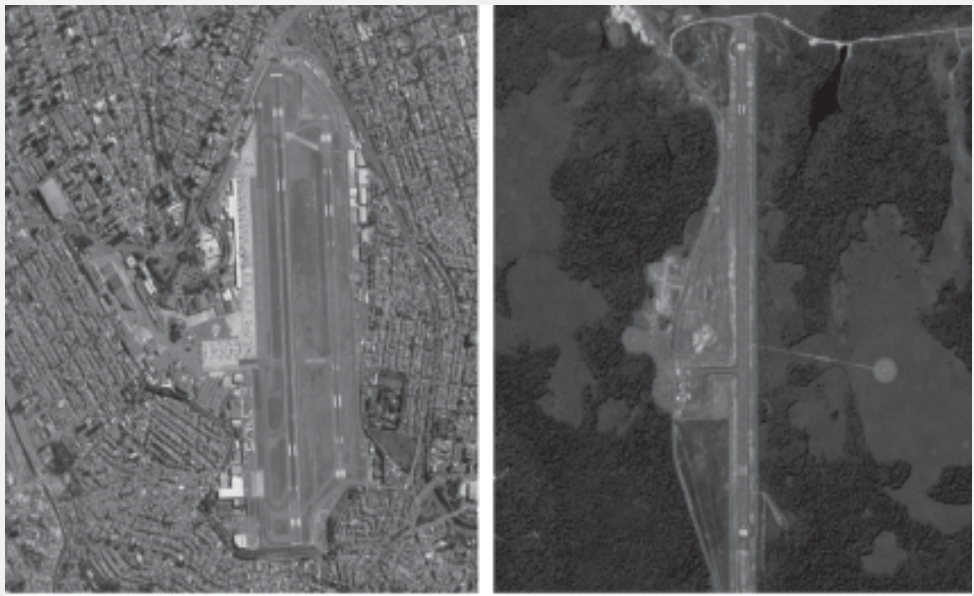

Figura 26 – Dois Aeroportos, Duas Realidades: Congonhas e Cruzeiro do Sul
(Imagem: Google Earth)

Enquanto Congonhas recebe anualmente cerca de quinze milhões de passageiros, Cruzeiro do Sul recebe pouco mais de oitenta mil passageiros (fonte: *website* da INFRAERO, dados de 2006). Cruzeiro do Sul leva três dias para efetuar o mesmo número de movimentos de aeronaves que Congonhas realiza em uma única hora. Mas também há semelhanças entre as duas realidades. Como todo meio de transporte e, por decorrência, de geração de riquezas e crescimento econômico, tanto o aeroporto de Congonhas quanto o de Cruzeiro do Sul buscam se modernizar e fazer reformas. Em um trecho de notícia do Valor Econômico de 06/11/2008, lê-se que "*O governador e o prefeito (...) apresentaram ontem ao Ministério da Defesa um projeto para prolongar em cerca de 1.000 metros as duas pistas - principal e auxiliar - do aeroporto de Congonhas, na cabeceira em direção ao bairro do Jabaquara. (...) os custos de desapropriação em áreas residenciais [ficariam] em R$ 260 milhões, segundo o valor venal dos imóveis, e R$ 400 milhões, pelos valores de mercado*". Em outra notícia, do Panrotas (21/08/2008), temos que "*A previsão é que em setembro o Novo Aeroporto de Cruzeiro do Sul já esteja sendo inaugurado. Depois de algumas adaptações que tornaram o projeto original ainda melhor, a obra está na etapa de conclusão. Na primeira fase foram investidos R$ 24 milhões e nesta segunda, R$ 5,9 milhões. O aeroporto possui design amazônico, único no País, com formato que lembra uma cabana indígena*".

**2**

# A Regulação Econômica
# do Transporte Aéreo

**Decolagem: Da Regulação à Flexibilização**

O setor de transporte aéreo no Brasil passou por duas grandes reformas regulatórias ao longo dos últimos quarenta anos. A primeira foi a introdução da regulação estrita – chamada de "regime de competição controlada" – associada a mecanismos de desenvolvimento regional, entre o final dos anos 1960 e início dos anos 1970. A segunda foi a chamada política de "Flexibilização da Aviação Comercial", introduzida no início da década de 1990 e que teve suas bases implementadas em rodadas, até 2001. Ao longo dessas quatro décadas de políticas governamentais para o setor, esses foram os dois marcos regulatórios mais notáveis, e que influenciam o debate com relação à regulação do setor ainda hoje.

A adequada compreensão das questões concernentes à redefinição do arcabouço regulatório que se segue à constituição da Agência Nacional de Aviação Civil – ANAC, instituída pela Lei n. 11.182, de 27 de setembro de 2005 –, passa por um maior conhecimento das características desses dois marcos regulatórios. É também preciso compreender de forma  adequada a sua evolução ao longo do tempo. Além disso, cumpre entender como se efetivou o papel das demais autoridades governamentais no acompanhamento do setor (autoridades de política macroeconômica, antitruste e de política industrial) e as tendências nesse relacionamento interinstitucional, como forma de dar bases para a propositura de esquemas eficazes de coordenação entre autoridades visando à minimização da sobreposição de papéis e à indução do bem-estar econômico setorial.

Pode-se dividir os últimos quarenta anos de políticas públicas para o transporte aéreo em cinco grandes períodos: Período de Regulação Estrita (1968-1985); Período de Regulação Enfraquecida (1986-1992); Primeira Rodada de Liberalização, (1993-1997); Segunda Rodada de Liberalização (1998-2001); Quase-Desregulação (2001-2002); Re-Regulação (2003-2004); e, por fim, Retomada da Desregulação com Redesenho Institucional (2005 em diante). A Figura 27 explicita os períodos.

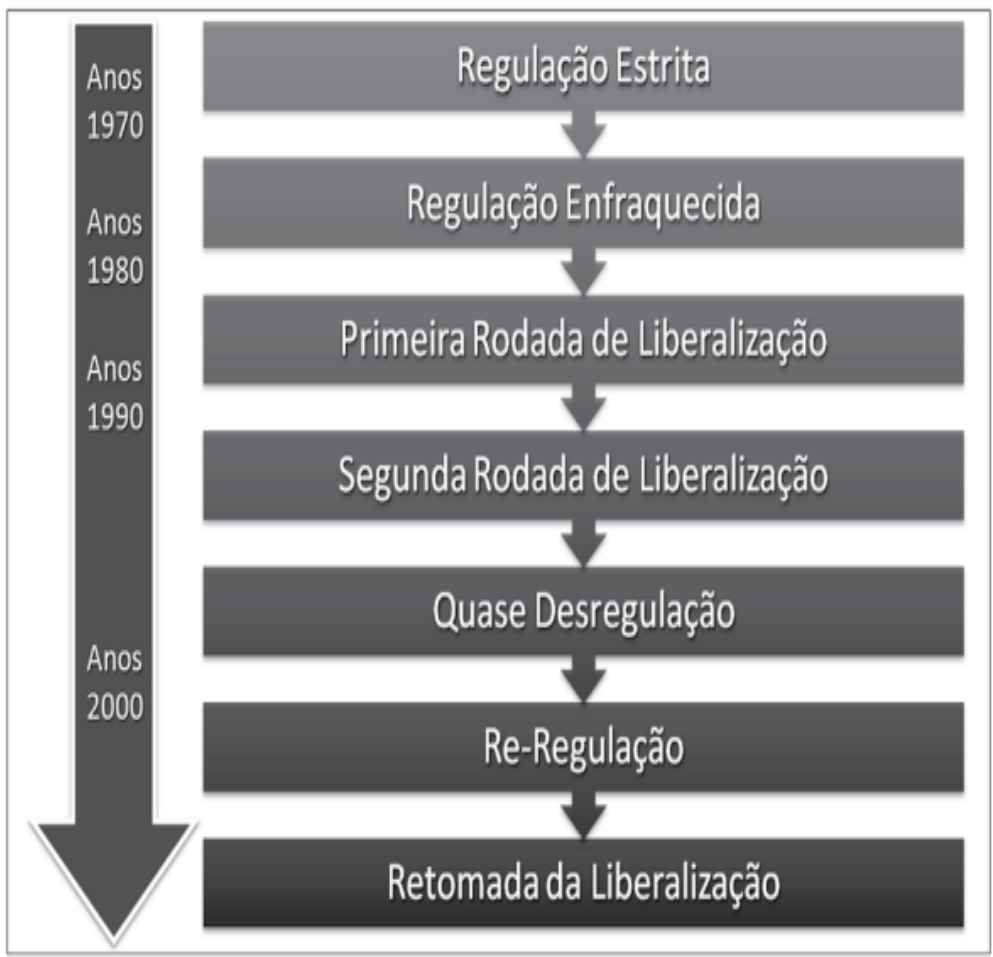

Figura 27 – Evolução da Regulação do Transporte Aéreo Doméstico no Brasil

## Rédeas Curtas e Fomento: Período da Regulação Estrita com Política Industrial (1968-1986)

O período mais representativo da era de regulação estrita do setor foi de 1968 a 1986, no qual as autoridades objetivaram implementar tanto instrumentos de regulação propriamente ditos, como mecanismos de política desenvolvimentista. Denominamos esse período de "Regulação Estrita com Política Industrial". Nele, estabeleceu-se um arcabouço oficial de "4 companhias nacionais e 5 companhias regionais". Essa estrutura foi completamente implementada em 1975-76, com o estabelecimento do Sistema Integrado de Transporte Aéreo Regional (SITAR) — Decreto nº 76.590 de 12 de novembro de 1975, que visava estabelecer uma Política Industrial para o transporte aéreo regional.

> *A Política Industrial é um mecanismo de ação governamental com o objetivo de influenciar quais indústrias irão se expandir na economia e quais terão fomento via subsídios, cortes de tributos, financiamentos privilegiados e outras formas de auxílio. Visa, assim, encorajar o investimento em determinadas indústrias com potencialidades de crescimento - uma tarefa conhecida como "pick winners", ou atribuição dos "campeões nacionais". Pode também objetivar uma proteção a indústria nascente, por exemplo. O objetivo é induzir a vantagem competitiva onde há potenciais alavancagens positivas e/ ou economias de escala. Pode ser levada adiante na forma de lei ou de regulamentação. Há inúmeros debates e críticas sobre a capacidade, e mesmo a necessidade, do governo suplantar a iniciativa privada fazer o processo de escolha de indústrias ou firmas "vencedoras". Críticos argumentam que o governo acabam trocando o "pick winners" por um "pick losers", dado o lobby ativo de grupos de pressão constituídos por firmas com maior poder econômico, mas não necessariamente elegíveis aos incentivos. Age também desalinhando os incentivos à concorrência e à eficiência das firmas no mercado.*

Conta Malagutti (2001) que, na década dos 60, imperava na aviação comercial brasileira uma forte crise econômica, fruto do ambiente político e econômico instável, dos custos de manutenção e da forte concorrência instaurada. A baixa rentabilidade impulsionava o *lobby* das grandes empresas no sentido de chamar à atenção das autoridades quanto ao risco inerente à manutenção da regularidade dos serviços. O *lobby* era uma potencial fonte de captura do regulador, no sentido de usar de suas prerrogativas para suavizar as pressões competitivas do mercado. Com vistas a sistematizar essas negociações, foram organizados importantes eventos do setor e que contaram com a presença de representantes das empresas aéreas e do regulador, o Departamento de Aviação Civil, DAC. Essas reuniões foram denominadas Conferências Nacionais de Aviação Comercial (CONAC). A 1ª CONAC foi realizada em 1961, a 2ª CONAC, em 1963 e a 3ª CONAC, em 1968. Ao final do processo, chegou-se ao consenso de que havia a necessidade de que se instaurasse uma política de estímulo à fusão de empresas, "*com o fim de reduzir o seu número a um máximo de duas na exploração do transporte internacional e três no transporte doméstico*" (Malagutti, 2001). Começava o que denominamos aqui de período de "Regulação Estrita", que equivale ao que se acostumou denominar no setor de "Período de Competição Controlada".

Durante esse período regulatório típico, variáveis como preços e freqüências de vôo passaram a ser ditadas pelas autoridades, a entrada de novas companhias aéreas foi banida, e o país foi dividido em cinco grandes áreas, monopólios especialmente desenhados para a operação das companhias aéreas regionais. Além disso, a competição entre companhias regionais e nacionais não era contemplada, visto que estas deveriam atuar apenas em ligações "tronco", em contraposição com as ligações alimentadoras (*feeder*) regionais. O regime de competição controlada foi ratificado com o Decreto 72.898, de 9 de outubro de 1973, onde às quatro grandes companhias aéreas de âmbito nacional do período era explicitamente atribuída toda a operação do sistema:

> "*A partir da publicação desde Decreto, fica concedido, pelo prazo de 15 (quinze) anos, às empresas de transporte aéreo Viação Aérea Riograndense S.A. (VARIG), Viação Aérea São Paulo S.A. (VASP), Serviços Aéreos Cruzeiro do Sul S.A. e Transbrasil S.A. Linhas Aéreas, o direito de executar o serviço aéreo de transporte regular de passageiro, carga e mala postal, independente de pedido*" (Decreto 72.898, de 9 de outubro de 1973, Disposições Transitórias, artigo n. 15).

Em paralelo à reestruturação regulatória do transporte aéreo naquele momento, um outro fator acabou por impulsionar as ações das autoridades, desta vez no sentido de fomentar o setor. Entre o final dos anos de 1960 e 1975, estabeleceu-se a introdução dos grandes jatos na frota das companhias aéreas. Esses novos equipamentos de vôo não operavam aeroportos de menor porte e muitos dos aeroportos regionais. Com essa substituição do *mix* da frota das grandes empresas, houve uma drástica redução do número de cidades atendidas pelo transporte aéreo – fenômeno que preocupou as autoridades. A Figura 28 permite evidenciar mostra essa tendência de queda na cobertura aérea no período:

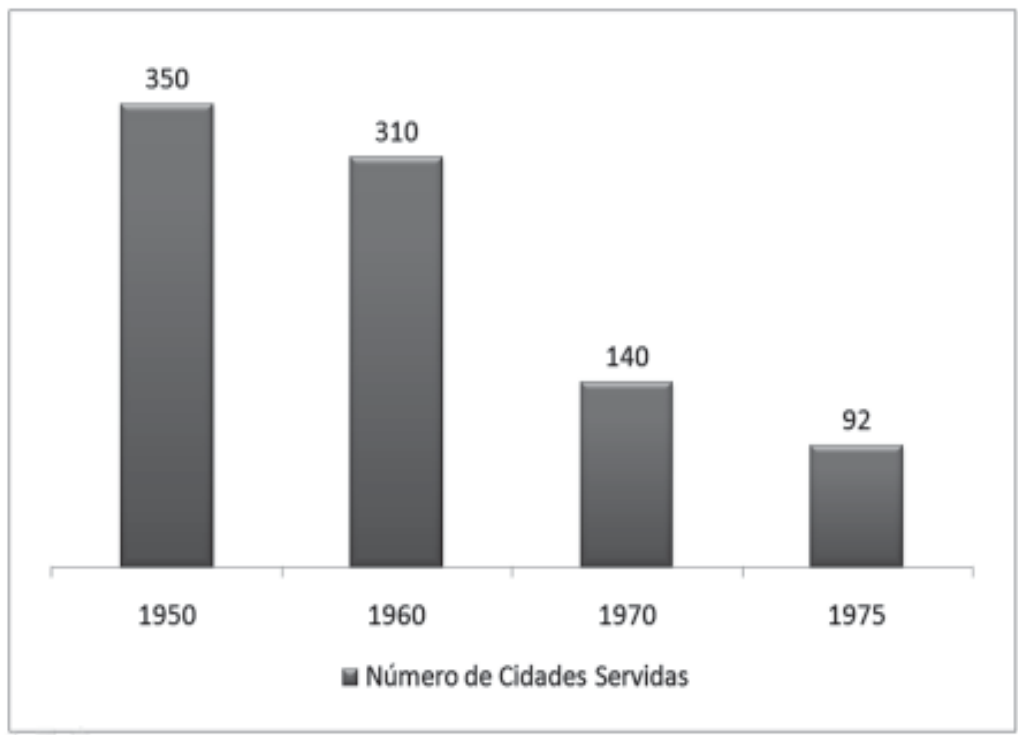

Figura 28 – Evolução da Cobertura Aérea no Pré-SITAR (Fonte: Gomes et al, 2002)

Dada a acentuada queda no número de cidades atendidas, o governo houve por bem fomentar uma nova modalidade de aviação, a aviação regional, e que contaria com subsídios para atender determinadas rotas de baixa viabilidade econômica. Dessa forma, foi criado em 1975, o Sistema Integrado de Transporte Aéreo Regional (SITAR), com o objetivo de atender localidades de médio e baixo potencial de tráfego. Assim, o território nacional foi dividido em cinco áreas, sendo cada uma delas atendido por uma empresa regional, que funcionava em regime de monopólio. Passaram a operar como empresas aéreas regionais oficialmente designadas: Nordeste Linhas Aéreas (Região Nordeste e parte dos estados de Minas Gerais e Espírito Santo), Rio-Sul Serviços Aéreos (Região Sul e parte dos estados do Rio de Janeiro, Espírito Santo e São Paulo); TABA Transportes Aéreos (Região Norte); TAM Transportes Aéreos (estado do Mato Grosso do Sul e parte dos estados do Mato Grosso e de São Paulo); e VOTEC Serviços Aéreos (Estado de Goiás, parte dos estados do Pará e do Maranhão, o Triângulo Mineiro e o Distrito Federal). Havia também o incentivo ao uso de equipamento da Embraer, mais precisamente o Bandeirante (EMB-110).

No âmbito do SITAR, estabeleceu-se o Adicional Tarifário, fixado em 3% sobre o valor dos bilhetes das linhas aéreas domésticas, sendo o produto dessa arrecadação destinado ao SITAR para suplementação tarifária de suas linhas. Esse adicional serviria como recurso para o fomento da atividade das operadoras regionais então criadas. O SITAR representava uma evolução do esquema de subsídios da RIN - Rede de Integração Nacional –, vigente entre 1963 e 1972, e que era baseada em recursos orçamentários.

> *O chamado "Adicional Tarifário" é um mecanismo de subsídio cruzado do transporte aéreo, criado pelo Decreto 76.590, de 11 de novembro de 1975 (o mesmo que criou o SITAR), na redação dada pelo Decreto no 98.996, de 2 de março de 1990. Na ocasião, previa-se um "adicional" de até três por cento a incidir sobre as tarifas de passagens aéreas das linhas domésticas. Os recursos provenientes da arrecadação do Adicional Tarifário seriam aplicados, exclusivamente, na Suplementação Tarifária de linhas aéreas regionais de passageiros, de baixo e médio potencial de tráfego, executadas dentro da Amazônia Legal e de baixo potencial de tráfego da região Nordeste e que atenderem aos requisitos exigidos.*

Concluindo, tem-se que, com a instituição do SITAR, alguns mecanismos relevantes de fomento a partir de um conceito de Política Industrial foram acrescentados ao marco regulatório da Regulação Estrita então vigente: criação de cinco monopólios regionais para a operação de companhias aéreas subsidiadas por suplementação tarifária, em rotas de baixa densidade, sobretudo alimentadoras de linhas-tronco, e com incentivo, por meio de linhas de crédito, ao uso de aeronave nacional da Embraer. Trata-se, portanto, de um período que representou a primeira e última tentativa do governo de estruturar, planejar e fomentar de maneira sistemática e global, o desenvolvimento desse setor, bem como de estabelecer políticas para a aviação regional.

## Fim da Utopia: Regulação com Política de Estabilização Ativa (1986-1992)

Com os anos 1980, o Estado brasileiro abandonou a utopia do planejamento do desenvolvimento industrial, sendo que o setor de transporte aéreo foi sendo paulatinamente afetado. Com os crescentes problemas de ordem macroeconômica e a necessidade de controle da inflação, inicia-se um período em que a política de estabilização — agora centro das atenções de governos e sociedade — passou a ser cada vez mais intrusiva nas questões setoriais. A regulação foi perdendo força ("Regulação Enfraquecida"), dado que um dos itens mais relevantes seria o controle de preços que, com os planos de estabilização, acabava por ficar totalmente a cargo dos gestores da economia.

Assim, dentre as políticas governamentais que tiveram impacto mais profundo no transporte aéreo, destacam-se: a) as seguidas medidas de desvalorização real da taxa de câmbio, visando o aumento da rentabilidade das atividades voltadas para exportação; e b) as interferências na precificação setorial de atividades orientadas para o mercado interno (setores com preços controlados). Por meio de a), foi promovida uma forte pressão do lado dos custos das companhias aéreas, visto ser esse um setor em que custos de alguns insumos relevantes apresentam alta correlação com a taxa de câmbio. Já através de b), afetou-se o mercado pelo lado da demanda (receitas), dado que setores voltados para o mercado doméstico acabaram por ter sua rentabilidade impactada pela defasagem dos reajustes de tarifas imposta pelo Conselho Interministerial de Preços (CIP) — o órgão de monitoramento e intervenção econômica dos mercados com a finalidade de controlar as pressões inflacionárias.

Sobretudo a partir de 1986 até o Plano Real, em 1994, as questões decorrentes das tentativas de estabilização econômica tornaram-se ainda mais prementes, dado que os seguidos planos tiveram impactos relevantes nas tarifas e na lucratividade dos setores de infra-estrutura — transporte aéreo incluso. Denominamos esse período "Regulação com Política de Estabilização Ativa", caracterizado pelo desgaste das políticas industriais vigentes e pela forte intervenção nas políticas de reajustes tarifários, levando a preços reais artificialmente baixos que causaram prejuízos ainda hoje contestados judicialmente pelas operadoras aéreas.

Importante enfatizar sobre esse período, que, em 1986, foi instituído o Código Brasileiro de Aeronáutica (Lei nº 7.565, de 19 de dezembro de 1986), que substituiu o Código Brasileiro do Ar (instituído pelo Decreto-lei número 32, de 18 de novembro de 1966).

## Ventos Liberalizantes: Liberalização com Política de Estabilização Inativa (1993-1997)

Este estágio representa o início da chamada "Política de Flexibilização da Aviação Comercial". Tendo início nos anos 1990, com um conjunto de portarias expedidas pelo Departamento de Aviação Civil (DAC), a liberalização do setor aconteceu de forma gradual e dentro das diretrizes do programa governamental de desregulamentação da economia do país no início daquela década.

A política de flexibilização do setor começou efetivamente a partir de 1992, dentro do chamado "Programa Federal de Desregulamentação" do Governo Collor (Decreto 99.179, de 15 de março de 1990), apesar de uma das mais relevantes medidas adotadas já estarem vigentes desde 1989 — as bandas tarifárias, conforme conta Rodrigues (2000). A desregulamentação do setor foi implementada em três rodadas, respectivamente, em 1992, 1998 e 2001, em uma trajetória muito semelhante aos "pacotes" de liberalização promovidos pela União Européia (UE). Com a abordagem européia, seguida pelo DAC, houve uma ênfase na graduação das políticas, de forma a se tentar evitar seus potenciais efeitos "danosos" de curto prazo, sobretudo em termos de um forte acirramento da competição no mercado, como aconteceu com a desregulamentação norte-americana, a partir de 1978.

A Primeira Rodada de Liberalização, (ou "Liberalização com Política de Estabilização Inativa"), foi efetivada sob a influência da V CONAC, realizada em 1991. A partir dela, os monopólios regionais, vigentes desde a época do SITAR, e que já se apresentavam distorcidos por conta da crescente competição entre as companhias nacionais e as companhias regionais, em busca de maior escala de operação, foram definitivamente abolidos (Portaria 075/GM5, de 6 de fevereiro de 1992 e Portarias 686 a 690 /GM5, de 15 de setembro de 1992).

As CONACs foram as Conferências Nacional de Aviação Comercial, eventos organizados com a finalidade de discutir e alterar os rumos da regulação do transporte aéreo. As três primeiras CONACs aconteceram na década de 1960, resultando no pleito por maior regulação e concentração do setor. A IV CONAC aconteceu em 1986, sem efeitos relevantes. A V CONAC foi realizada ao final de 1991, e forneceu bases e diretrizes para a formulação da "Política de Flexibilização da Aviação Comercial", a partir de 1992. Atualmente, a sigla CONAC se refere ao Conselho de Aviação Civil, o órgão de assessoramento do Presidente da República para a formulação da Política Nacional de Aviação Civil. Foi instituído pelo Decreto nº 3.564, de 17 de agosto de 2000.

Dessa forma, a política de "4 companhias nacionais e 5 companhias regionais" foi oficialmente extinta e, a partir de então, a entrada de novas operadoras passou a ser estimulada, o que resultou em uma onda de pequenas novas companhias aéreas entrantes no mercado (por exemplo Pantanal, Tavaj, Meta, Rico, etc), algumas oriundas de empresas de táxi aéreo. A única exceção com relação ao monopólio das regionais ficou por conta de alguns pares de aeroportos, ligando as cidades de São Paulo, Rio de Janeiro, Belo Horizonte e Brasília. Esses pares de aeroportos, em geral ligando os centros das cidades envolvidas e estes com a capital federal, eram conhecidos como "Vôos Direto ao Centro" (existentes desde 1986), e mais tarde, denominadas "Linhas Aéreas Especiais".

Outra medida adotada foi a introdução de preços de referência com novas bandas tarifárias, que agora variavam de – 50% a +32% do valor principal – sem dúvida, uma inovação diante do sistema de preços controlados do período de forte regulação. A competição em preços era agora vista como "saudável" para a indústria e passou a ser encorajada. Nesse sentido, as bandas tarifárias eram concebidas como instrumentos temporários para intensificar a rivalidade de preços. Contudo, os preços ainda eram, de certa forma, indexados, dado que as tarifas de referência em torno das quais os preços podiam flutuar eram, por definição, controladas e sujeitas às políticas de reajustes periódicos. A caracterização desse período como de política de controle de estabilização "inativa", remete ao fato de que não era preciso, no momento, que as autoridades macroeconômicas interferissem fortemente no mercado, visto não terem sido observadas maiores pressões em termos de aumento de preços (maior estabilidade de preços a partir de 1994). Adiciona-se a esse fato a ocorrência de baixa instabilidade dos custos, com taxas de câmbio (sobretudo o dólar) relativamente estável e favorável durante a maior parte do período.

## Segunda Rodada de Medidas: Liberalização com Restrição de Política de Estabilização (1998-2001)

No final dos anos 1990, as autoridades de aviação decidiram remover dois importantes dispositivos de controle da competição que ainda perduravam no setor: as bandas tarifárias e a exclusividade do direito de as regionais operarem as Linhas Aéreas Especiais. Isso gerou a Segunda Rodada de Liberalização (final de 1997 e início de 1998, com as Portarias 986 e 988/ DGAC, de 18 de dezembro de 1997, Portaria 05/GM5, de 9 janeiro de 1998 e Portaria 701/DGAC, de 30 de dezembro de 1998), que visava dar mais liberdade às companhias aéreas e que, na prática, teve o efeito de encadear o primeiro grande surto de competitividade desde o início da desregulamentação. De fato, em 1998, foram observados fenômenos de "guerras de preços" e "corridas por freqüências de vôo" – estratégia que os americanos denominam de "*cut fares and add flights*" –, muito divulgados pela mídia, que nada mais representavam que os efeitos de curto prazo das novas medidas implementadas, mas que geraram uma movimentação competitiva como não se via pelo menos desde a década de 1960.

Por outro lado, outra característica relevante no período foi o retorno da forte instabilidade da taxa de câmbio, especialmente após a mudança de regime cambial para livre flutuação, em janeiro de 1999. Igualmente importante foi a imediata desvalorização do real ante o dólar que se seguir. Isso representou uma fonte relevante de pressões nos custos operacionais das companhias aéreas, que serviu para arrefecer, de certa forma, o processo competitivo desencadeado em 1998. Assim, houve, em um primeiro momento, uma tendência generalizada de suspensão das tarifas promocionais vigentes desde o ano anterior e, depois, uma demanda por um realinhamento de preços em face do choque nos custos.

Com o cenário de aumentos de preços em diversos setores da economia em 1999, as autoridades macroeconômicas, e, mais precisamente, o Ministério da Fazenda, optaram por cercear a precificação da indústria quanto aos reajustes. Tal fato representou uma limitação às estratégias das firmas, dado que o controle dos reajustes de preços foi novamente posto em prática, necessitando de autorização prévia do DAC e do Ministério da Fazenda. Por isso, denominamos esse período "Liberalização com Restrição de Política de Estabilização".

É relevante mencionar que, a partir desse período, as autoridades antitruste passaram a também monitorar de perto o setor aéreo. Houve, inclusive, abertura de processo no Sistema Brasileiro de Defesa da Concorrência quando do reajuste em 10% nos preços das passagens, pelas companhias aéreas da Ponte Aérea Rio de Janeiro – São Paulo, em agosto de 1999.

## Quase-Desregulamentação (2001-2002) e o "Liberou Geral"

A meta de desregulação total dos preços das passagens aéreas domésticas no Brasil já havia sido estipulada desde 1998. Mas, como vimos, o grande empecilho à concretização dessa meta já em 1998 se deu com a mudança de regime cambial em janeiro de 1999. Visando segurar as pressões inflacionárias dos setores regulados, o Ministério da Fazenda atuou impossibilitando os reajustes naquele ano. A liberdade tarifária ainda era, portanto, assimétrica para cima, ou seja, era possível conceder descontos ou baixar preços livremente, mas não era permitido aumentar os preços de acordo com as novas condições de mercado.

Em 2001, um acordo entre o DAC e o Ministério da Fazenda permitiu que o mecanismo de regulação econômica que ainda persistia no setor finalmente fosse removido e a indústria ficou de uma vez por todas livre da interferência macroeconômica e do controle dos planos de estabilização. De fato, por meio de portarias paralelas dos dois órgãos governamentais, foi posta em prática uma total liberalização dos preços (Portarias 672/DGAC, de 16 de abril de 2001, e 1.213/DGAC, de 16 de agosto de 2001). Isso coincidiu com a flexibilização dos processos de entrada de novas firmas e de pedidos de novas linhas aéreas, freqüências de vôo e aviões (Terceira Rodada de Liberalização), em um processo que culminou com a entrada da Gol, em janeiro de 2001. Não se caminhou para uma total desregulação econômica, entretanto. Trata-se de uma "quase-desregulação" do setor, dado que ainda há questões importantes a serem equacionadas rumo a um mercado com maior liberdade de operação. Um importante exemplo disso é o acesso à infra-estrutura aeroportuária e a concorrência por horários de pouso e decolagem em aeroportos congestionados – os chamados *slots*. Além disso, não houve uma completa reformulação do aparato legal vigente, como, por exemplo, o Código Brasileiro de Aeronáutica (CBAer), que contém inúmeros dispositivos conflitantes com o processo de desregulação do setor posto em prática desde então. Um deles é a restrição de participação de capital estrangeiro em companhias aéreas brasileiras, que atualmente limitado a vinte por cento do capital.

Em suma, tem-se que, desde 2001 o que se observa no País é a completa liberdade de precificação por parte das companhias aéreas. O uso intensivo das práticas de *yield management* passou a ser uma constante desde então. Esse sistema permite às empresas oferecer diferentes tarifas e gerenciar a disponibilidade delas aos clientes no momento da compra ou reserva. Tanto que atualmente já faz parte da rotina do passageiro conviver com uma miríade de preços possíveis, que variam de acordo com o horário, o dia da semana, o tempo de antecedência da compra, o período da viagem e da reserva, o trecho e os aeroportos de origem e destino em um dado trecho. Esse amplo portfólio de oferta de tarifas contrasta em muito com o período regulatório, onde apenas uma tarifa era praticada por todas as companhias aéreas nacionais no mesmo trecho.

> *O yield management é a estratégia das companhias aéreas que visam maximizar as receitas de vôo com base em três instrumentos: 1. diferenciação de produtos com base na segmentação de consumidores - os chamados fare products; 2. discriminação de preços, com uma estrutura de tarifas disponíveis; 3. utilização de técnicas de alocação e controle do estoque de assentos (CEA) segundo as previsões de demanda para cada vôo, promovendo limites de passagens disponíveis aos segmentos de mercado. O yield management foi criado pela American Airlines e pela Delta Airlines, e foi fundamental na eliminação da concorrência proporcionada pela primeira onda de novas entrantes do pós-desregulação norte americano, nos anos 1980. Empresas como a People Express, sucumbiram nesse período.*

Pode-se dizer que o "liberou geral" em preços, permitido pelas Portarias 672 e 1.213 de 2001 permitiu inúmeros benefícios aos consumidores e empresas aéreas e, ao mesmo tempo, inúmeros desafios ao regulador. Os benefícios são nítidos, dado que empresas aéreas podem competir livremente e buscar maximizar suas receitas por meio dos métodos consagrados de *yield management*, e que os consumidores vêem ampliado o seu leque de possibilidades de compra. No caso do regulador, entretanto, os desafios são imensos, dada a dificuldade de observabilidade e de qualquer investigação, apuração e controle sobre as práticas tarifárias das companhias aéreas. Estas podem adotar práticas predatórias, de cartel ou mesmo guerras de preço sem sequer precisar alterar a estrutura de tarifas existente. Isso porque basta a elas alocar mais, ou menos, assentos nas classes tarifárias desejadas, para que as tarifas médias se desloquem para cima ou para baixo, conforme a estratégia anticompetitiva desejada. O regulador hoje em dia possui uma observabilidade muito ruim do que acontece com os preços das companhias aéreas e está longe de ter algum controle dessas práticas. É fundamental que ele invista em mecanismos de acompanhamento econômico de preços, visando embasar sua tomada de decisão e adequadamente pautar suas potenciais intervenções no mercado de forma a garantir o bem-estar econômico do setor. Essa responsabilidade aumentou significativamente após o "liberou geral" em preços de 2001 e tem que ser urgentemente endereçada.

## O Antitruste no Transporte Aéreo: Controle de Estruturas e Condutas do Setor

Uma temática que costuma dominar as discussões quanto aos efeitos das desregulações dos mercados aéreos diz respeito à preocupação com o aumento de poder de mercado das empresas incumbentes no ambiente competitivo liberalizado. Dentre as questões relevantes, encontram-se as causas e conseqüências da maior concentração do setor e a redução da contestabilidade pela existência de barreiras à entrada, ambos com ênfase nos riscos de exercício da dominância na indústria no pós-desregulamentação. Nessa mesma direção tem-se, como ilustração, a iniciativa do DAC, em seus últimos dias de existência enquanto regulador, de criar o instituto do "monitoramento" de certas ligações aéreas, mesmo após tendo liberalizado totalmente o processo de formação de preços. A idéia de monitorar por parte da autoridade, estava nitidamente ligada ao temor de que um mercado totalmente livre pudesse acarretar, em última instância, perdas de bem-estar econômico no setor.

Os economistas criaram o chamado paradigma "Estrutura-Conduta-Desempenho" – ou o "Modelo Bain-Mason" – para evocar a relação entre o grau de concentração em um mercado (estrutura) e o poder de oligopólio das empresas (conduta-desempenho). Esse paradigma clássico foi o mais utilizado por décadas em economia industrial, mas sofreu críticas importantes por conta do provável efeito de retro-alimentação do desempenho na própria estrutura, ou seja, por exemplo, mercados com maior lucratividade podem atrair novos entrantes que reduzem a concentração. O comportamento estratégico das firmas (conduta) também pode influenciar a estrutura, criando barreiras a entrada, por exemplo. A literatura do transporte aéreo tem sido cada vez mais enfática em apontar que estratégias cooperativas das empresas estabelecidas por todo o mundo, têm sido muito utilizadas nessa indústria, sobretudo dentre as companhias aéreas marcadas pela estrutura mais complexa de rede, como as "*Network Carriers*". Nitidamente, as evidências da formação de "alianças" e "acordos operacionais", têm se mostrado suficientemente concretas para se inferir que a potencialidade de exercício de poder de mercado na indústria tem crescido substancialmente, pela maior facilidade de coordenação estratégica decorrente daquelas práticas. Por outro lado, a contestabilidade do mercado por parte das chamadas companhias aéreas de Custo Baixo, Preço Baixo ("*Low Cost, Low Fare*") cresceu substancialmente por todo o mundo.

Dessa forma, pode-se dizer que o controle dos níveis de concentração e do poder de mercado em setores como o transporte aéreo, tornou-se a principal atribuição das autoridades no novo ambiente competitivo, visando a promoção do bem-estar. O conjunto de mecanismos legais e institucionais voltados a esse objetivo constitui a política antitruste, que objetiva aumentar a eficiência econômica por meio da promoção e estímulo à competição.

> *A política antitruste é realizada através de instrumentos de intervenção nos quais as autoridades podem atuar, tanto de maneira repressiva quanto preventiva, com relação aos atos considerados lesivos à competição (atos de concentração, como fusões e incorporações, ou práticas de articulação de mercado, como colusão tácita, coordenação de preços e quantidades), sem necessariamente impor condições a todos os participantes do mercado, como nos regimes regulatórios. São dois os seus tipos de processos: o controle da estrutura da indústria, voltado para o controle da formação de poder de mercado; e o controle das condutas, que se preocupa em coibir os abusos do poder de mercado que seja eventualmente detido por uma ou mais firmas da indústria. Esses processos podem envolver duas dimensões: a horizontal, quando as empresas objeto são competidoras em um mercado; e a vertical, que envolve empresas situadas em diferentes níveis da cadeia produtiva.*

O arcabouço institucional da política antitruste hoje praticada no Brasil foi estabelecido pela Lei de Defesa da Concorrência (Lei n. 8884, de 11 de junho de 1994). Os mecanismos de fomento à competição visam, nos termos da lei, prevenir as chamadas "infrações contra a ordem econômica", pautando-se pelos ditames da livre concorrência e da repressão ao abuso do poder econômico, dentre outros princípios constitucionais.

As instituições governamentais incumbidas da investigação antitruste no Brasil são a Secretaria de Direito Econômico (SDE) e o Conselho Administrativo de Defesa Econômica (CADE), vinculados ao Ministério da Justiça, e a Secretaria de Acompanhamento Econômico (SEAE), no Ministério da Fazenda. Enquanto a SDE tem um papel de acompanhamento e de instauração de processo administrativo para apuração de infrações, o CADE possui caráter judicante, de decisão, julgando os processos instaurados por aquela. Já à SEAE cumpre o papel de dar parecer econômico e proceder com investigações, em coordenação com os demais órgãos.

Um aspecto relevante da defesa da concorrência em setores regulados, mesmo que parcialmente desregulamentados, diz respeito à superposição de competências entre o órgão regulador, quando este existe, e os organismos antitruste. No caso do transporte aéreo brasileiro, a legislação atribuiu essa função à autoridade reguladora do setor, o Departamento de Aviação Civil (DAC). No entanto, essa função era de pouca importância em função do controle de preços e de entrada que esteve presente na indústria antes da liberalização. Somente com a "Flexibilização" da indústria é que emergiu uma maior preocupação com a defesa da concorrência no transporte aéreo, cuja condução coube aos órgãos antitruste.

Indubitavelmente, o antitruste, enquanto atuação governamental de maior impacto, tornou-se uma constante na indústria de transporte aéreo, de maneira mais nítida, a partir de 2000. Nessa ocasião, os três órgãos de defesa da concorrência, visando promover uma vigilância constante com relação às movimentações estratégicas das companhias aéreas que se seguiram à desvalorização cambial de 1999, promoveram a seguinte seqüência de ações no setor, apresentada na Tabela 4.

Tabela 4 – Investigações Antitruste do Transporte Aéreo no Começo dos Anos 2000

| Data do processo | Tipo de processo | Objeto | Companhias |
|---|---|---|---|
| Março de 2000 | Conduta | Aumento "coordenado e uniforme" de preços, na Ponte Aérea RJ - SP, após reunião entre representantes das empresas. Julgado pelo CADE em Setembro de 2004. | Varig, TAM, Vasp e Transbrasil |
| Março de 2000 | Conduta | Redução "coordenada e uniforme" da comissão das agências de viagem. | Varig, TAM, Vasp e Transbrasil |
| Abril de 2000 | Conduta | "Desequilíbrio na competição do setor" alegada pela TAM devido à inadimplência da Vasp. | Vasp e demais |
| Maio de 2000 | Estrutura e Conduta | Acordo operacional, seguido de redução de oferta de assentos e aumento de preços. | TAM e Transbrasil |
| Maio de 2000 | Estrutura e Conduta | Venda de aeronaves e transferência de linhas. | Varig e Vasp |
| Julho de 2000 | Estrutura | Assinatura de Carta de Intenções para a constituição da Plata, empresa de comercialização eletrônica de passagens aéreas e outros serviços relacionados a turismo. | Varig e TAM |
| Fevereiro de 2001 | Estrutura e Conduta | Denúncias de restrições à entrada: combinação de descontos no mês em que a Gol iniciou as operações; e lobby do Sindicato das Empresas Aeroviárias sobre o DAC para impedir aceitação de novas empresas. | Varig e TAM |
| Maio de 2001 | Estrutura | Aliança entre companhias internacionais em rotas entre Europa e América do Sul. | British Airways e Iberia |
| Fevereiro de 2003 | Estrutura | Assinatura de Acordo de Preservação de Reversibilidade da Operação (APRO) para a possível fusão entre as empresas, o primeiro na história do SBDC. | Varig e TAM |

Fonte: *webpage* de CADE, SEAE e SDE.

Como a Tabela 4 deixa claro, o antitruste tem sido um instrumento caracterizador da recente atuação governamental na indústria do transporte aéreo. A mesma Tabela evidencia que a política antitruste cuidou não apenas de processos envolvendo companhias aéreas nacionais, mas também de operações entre companhias internacionais com impacto no mercado doméstico. Como exemplo deste último caso, cite-se a aliança entre duas companhias internacionais que operavam vôos entre a Europa e a América Latina, com o possível objetivo de reforçar a posição destas na competição frente à Varig, empresa nacional que detinha participação expressiva nessas ligações e que mantinha aliança com outras companhias estrangeiras, a Star Alliance.

## Deu Revertério? O Breve Período de Re-regulação de 2003 e 2004

Em 2003, com o novo governo federal, e seguindo novas orientações de política setorial, o regulador voltou a implementar alguns procedimentos de interferência econômica no mercado. Em resposta à forte crise financeira das maiores empresas aéreas do País (TAM e Varig) em 2002, e ao processo falimentar que outras empresas importantes haviam alcançado (Transbrasil e Vasp), objetivou-se controlar o que foi chamado de "excesso de capacidade" e o acirramento da "competição ruinosa" no mercado.

Pelo texto das novas portarias, sobretudo a 243/GC5, de 13 de março de 2003 e a 731/GC5, de 11 de agosto de 2003, o DAC passa a exercer uma função moderadora, de "*adequar a oferta de transporte aéreo, feita pelas empresas aéreas, à evolução da demanda*", com a "*finalidade de impedir uma competição danosa e irracional, com práticas predatórias de conseqüências indesejáveis sobre todas as empresas*". Denominamos esse período de **"Re-regulação"**, uma fase onde pedidos de importação de novas aeronaves, novas linhas e mesmo de entrada de novas companhias aéreas, voltaram a exigir estudos de viabilidade econômica prévia, configurando-se uma situação semelhante ao do período regulatório típico; a grande diferença, nesse caso, foi que não houve interferência na precificação das companhias aéreas, ou seja, não houve re-regulação tarifária.

A idéia subjacente aos novos mecanismos era não apenas frear o processo de fragilidade financeira que havia feito Vasp e Transbrasil cessarem suas operações e que, no início de 2003, parecia ter afetado também TAM e Varig, as duas maiores empresas aéreas do País. A idéia de um subconjunto das autoridades da época era também estimular uma reformulação da estrutura do transporte aéreo doméstico, em prol de um fortalecimento dos agentes. A idéia da fusão TAM-Varig foi uma decorrência natural desse movimento e que foi implementada na prática por meio de um extensivo acordo de compartilhamento de aeronaves (*code share*) entre as operadoras. As medidas das portarias 243 e 731 visavam, portanto, sustentar esse arcabouço em prol de um arrefecimento da competição advinda sobretudo da nova entrante Gol e em prol da constituição da nova empresa – novo "campeão nacional", do tipo AMBEV – que emergiria com a fusão.

> *Um acordo "code share", ou de compartilhamento de códigos, é uma prática de mercado na qual duas companhias aéreas compartilham o mesmo código de dois dígitos usados na identificação das operadoras nos sistemas de reserva usados na comercialização dos vôos. Sob o acordo de compartilhamento, as firmas aliadas efetuam uma "fusão", parcial ou total, de seus sistemas de reservas e cada transportadora pode emitir bilhetes relativos a vôos operados pela(s) outra(s) empresa(s) participante(s) do acordo. Na prática, o acordo funciona como se fosse um acordo de compartilhamento de aeronaves, pois é como se uma empresa fizesse a extensão de sua capacidade de vôo utilizando parte da aeronave da empresa aliada.*

A união completa entre as duas grandes empresas não vingou, entretanto. O acordo de compartilhamento de aeronaves encerrou em abril de 2005, com o restabelecimento das operações independentes entre TAM e Varig. Independente disso, o breve período de Re-regulação efetivamente logrou segurar as pressões competitivas em 2003 e o primeiro semestre de 2004, bem como foi possível induzir um incremento da concentração do mercado, um duopólio TAM-Gol, surgido com a derrocada da Varig em 2006.

## Retomada da Desregulação com Redesenho Institucional (Desde 2005)

Seria possível argumentar que a re-regulação de 2003-2004 teria representado o fim do período da Política de Flexibilização da aviação comercial brasileira, dado que promoveu uma interrupção na trajetória de concessão de maiores graus de liberdade estratégica às companhias aéreas, e sinalizou ao mercado que o regulador teria a habilidade de intervir no mercado, de forma discricionária, quando julgasse necessário. Entretanto, um importante redesenho institucional veio recolocar o transporte aéreo na rota anteriormente delineada: a criação da Agência Nacional de Aviação Civil, ANAC, em substituição ao DAC.

*O DAC, Departamento de Aviação Civil, havia sido criado em 22 de abril de 1932, pelo Presidente Getúlio Vargas. Nasceu inserido na administração direta, subordinado ao então Ministério de Viação e Obras Públicas. Por mais de 70 anos, regulou, regulamentou, planejou e fiscalizou a aviação civil brasileira. Com o advento da Lei nº 11.182, de 27 de setembro de 2005, que estabelecia a criação da Agência Nacional de Aviação Civil, houve a troca oficial do regulador do transporte aéreo, que passou a ser constituído por uma agência fora do âmbito da administração pública direta e de cunho não-militar.*

Além do redesenho institucional permitido pela substituição do DAC pela ANAC, e com o intuito de promover uma maior coordenação entre as instituições da aviação civil, foi criada em 2007, (Decreto Nº 6.223, de 4 de outubro de 2007), a Secretaria de Aviação Civil (SAC). Cabe à SAC, além da função de coordenação e supervisão dos órgãos e das entidades do sistema, também assessorar o Ministro de Estado da Defesa em assuntos da aviação civil e na formulação das diretrizes da Política Nacional de Aviação Civil (PNAC), elaborar estudos, projeções e informações relativos aos assuntos de aviação civil, de infra-estrutura aeroportuária civil e de infra-estrutura de navegação aérea civil, dentre outros.

Com a Lei nº 11.182, houve uma clara retomada dos princípios liberalizantes da Política de Flexibilização dos anos 1990. Fruto de uma composição de forças dentro do governo federal, em que os princípios do Ministério da Fazenda foram os que prevaleceram em última instância, a Lei da ANAC proporcionou um restabelecimento da concepção de livre mercado para o setor. Importante salientar, entretanto, que aquela lei não revogou um conjunto de dispositivos de regulação mais estrita ainda em vigor no Código Brasileiro de Aeronáutica (CBAer) – Lei nº 7.565, de 19 de dezembro de 1986. De fato, o CBAer, na Seção IV, chamada "Do Controle e Fiscalização dos Serviços Aéreos Públicos", em seu artigo 193, reza que:

"Art 193. *Os serviços aéreos de transporte regular ficarão sujeitos*

*às normas que o Governo estabelecer para impedir a competição ruinosa e assegurar o seu melhor rendimento econômico podendo, para esse fim, a autoridade aeronáutica, a qualquer tempo, modificar freqüências, rotas, horários e tarifas de serviços e outras quaisquer condições da concessão ou autorização".*

Independentemente da questão jurídica acerca da possibilidade de intervenção governamental no funcionamento dos mercados aéreos, tem-se que, com a promulgação da Lei da ANAC e, sobretudo, após a posse da segunda diretoria da agência, constituída no final de 2007, tem-se que a orientação regulatória voltou a ser a mesma que conduziu o processo de liberalização do setor quinze anos antes.

No que diz respeito à legislação referente à concessão de linhas aéreas e seus impactos na tomada de decisão empresarial quanto à determinação da capacidade produtiva – freqüências de vôo e tipo e configuração de assentos das aeronaves –, vigora atualmente o regime de "Livre Mobilidade". Trata-se de um arcabouço mais liberal que visa dar agilidade e induzir eficiência no sistema de concessões de linhas aéreas para empresas regulares certificadas para atuar no segmento doméstico de passageiros. Por curiosidade, este regime foi implementado apenas nas disposições transitórias da lei nº 11.182, de criação da ANAC. Temos assim, no Capítulo VI, referente àquelas "Disposições Finais e Transitórias", a seguinte redação:

"Art. 48.§ 1º *Fica assegurada às empresas concessionárias de serviços aéreos domésticos a exploração de quaisquer linhas aéreas, mediante prévio registro na ANAC, observada exclusivamente a capacidade operacional de cada aeroporto e as normas regulamentares de prestação de serviço adequado expedidas pela ANAC".*

Caminhando na mesma direção, o Decreto nº 5.731, de 20 de março de 2006, que dispõe sobre a instalação, a estrutura organizacional da Agência Nacional de Aviação Civil - ANAC e aprova o seu regulamento, expressa que:

"Art. 10. *Na regulação dos serviços aéreos, a atuação da ANAC visará especialmente a: I - assegurar às empresas brasileiras de*

*transporte aéreo regular a exploração de quaisquer linhas aéreas domésticas, observadas, exclusivamente, as condicionantes do sistema de controle do espaço aéreo, a capacidade operacional de cada aeroporto e as normas regulamentares de prestação de serviço adequado*".

Dos conceitos acima encontrados, temos que apenas o de "prestação de serviço adequado" encontra definição explicitamente tratada no ornamento legal, mesmo que não específico do setor aéreo. De fato, a lei nº 8.987, de 13 de fevereiro de 1995, que a dispõe sobre o regime de concessão e permissão da prestação de serviços públicos previsto no art. 175 da Constituição Federal, trata, em seu Capítulo II, dessa relevante matéria no que tange os serviços regulados:

> "Art. 6º *Toda concessão ou permissão pressupõe a prestação de serviço adequado ao pleno atendimento dos usuários, conforme estabelecido nesta Lei, nas normas pertinentes e no respectivo contrato.*
>
> *§ 1º Serviço adequado é o que satisfaz as condições de regularidade, continuidade, eficiência, segurança, atualidade, generalidade, cortesia na sua prestação e modicidade das tarifas*".

A ANAC possui um sistema de mensuração da pontualidade, regularidade e eficiência operacional das companhias aéreas, herdada do extinto Departamento de Aviação Civil, e que auxilia o regulador no acompanhamento dos níveis de prestação de serviço adequado. Esse sistema vem sendo aperfeiçoado ao longo de 2008. Por outro lado, definições acima consideradas, como a de "condicionantes do sistema de tráfego aéreo" ou de "capacidade operacional de cada aeroporto" ainda carecerem de definição explícita no conjunto de normas que regem o setor.

No que tange especificamente ao controle feito pela autoridade em situações infra-estrutura aeroportuária e de controle de tráfego aéreo escassa, existe uma normatização infra-legal, emanada pela própria ANAC. Criada no sentido de preencher a lacuna quanto à alocação dos chamados "*slots*", isto é os horários de chegadas e partidas de aeronaves em aeroportos congestionados, esta normatização visou detalhar a forma de regulação nos casos considerados como exceção ao Regime de Livre Mobilidade consagrado pela Lei da ANAC.

> *Slot é o horário estabelecido para uma aeronave realizar uma operação de chegada ou uma operação de partida em um aeroporto congestionado ou saturado, e operado sob uma regra de restrição. Por definição, apenas os aeroportos ditos "administrados" ou "coordenados" pela regra restritiva das operações aéreas apresentam slots. E as regras restritivas costumam ser impostas apenas quando os aeroportos alcançam limites operacionais próximos de sua capacidade. Em casos de expansão da capacidade, toda a discussão da alocação de slots pode perder o sentido. Pelo mundo afora, tem-se como exemplo de aeroportos "eslotados": Boston/Logan, Nova York/ LaGuardia, Washington/National, Londres/Heathrow e São Paulo/ Congonhas.*

Assim, e após consulta e audiência públicas realizadas pela agência, expediu-se a Resolução n° 2, de 3 de Julho de 2006, que aprova o regulamento sobre a alocação de *slots* em linhas aéreas domésticas de transporte regular de passageiros, nos aeroportos que menciona, e dá outras providências. Os aeroportos que a Resolução menciona são aqueles que operarem no limite de sua capacidade operacional, como, por exemplo, o Aeroporto de Congonhas, em São Paulo. Um sistema de rodízio foi então desenvolvido no sentido de possibilitar a prestação do serviço pelas companhias aéreas regulares, sistematizando, em regra explícita, a configuração da alocação dos *slots* naquele aeroporto.

Por um lado, a normatização da importante questão regulatória possibilitou o início de uma maior compreensão, por parte da sociedade, de como funciona a distribuição da infra-estrutura escassa entre entes privados neste setor – fator que pode ser considerado extremamente benéfico dado que saímos da situação de completa ausência de regras claras para uma situação de regramento explícito. Por outro lado, entretanto, tem-se que a formatação da regra acabou por preservar as participações de mercado das companhias aéreas dominantes no Aeroporto de Congonhas (TAM, Gol e Varig), o que, na prática, apenas serviu como consolidação do sistema de *grandfather rights* que prevalecia até então[22].

> *Os chamados "grandfather rights" retratam uma situação típica do transporte aéreo mundial, onde a dominância histórica da(s) companhia(s) aérea(s) em um dado aeroporto se torna institucionalizada pelas próprias regras que governam aquele aeroporto, isto é, todo o arcabouço normatizador da rotina aeroportuária acaba sempre por consolidar a dominância do agente de operação aérea. Em outras palavras: quase que por tradição ou inércia, ficam com a maior quantidade e os melhores slots as empresas estabelecidas e que já possuem o maior número deles.*

As movimentações da ANAC, ao longo de 2008, foram no sentido de modernizar a legislação referente à alocação de *slots*, dado ser esse um mecanismo importante de incremento da contestabilidade do transporte aéreo e de indução de eficiências alocativas. De fato, ao final de 2008, a agência lançou uma consulta pública com uma proposta de resolução em substituição às regras estabelecidas na Resolução nº 2. A proposta visa criar um novo modelo de realocação de *slots*, de forma reduzir as barreiras à entrada de novas empresas no mercado e incrementar a competição. Adicionalmente, o novo modelo busca gerar mecanismos de incentivo à eficiência por parte das empresas já em operação nos aeroportos saturados, a partir do fato de que indicadores de desempenho (pontualidade, regularidade) passariam a ser utilizados como condição para a empresa continuar operando no mercado e para a realocação de *slots*. Assim, tem-se que a autoridade atualmente entende que é fundamental a redução do poder de monopólio das empresas instaladas, por meio da remoção de barreiras à entrada de novas operadoras – o que é salutar em se tratando de se buscar fomentar o bem-estar do consumidor.

### *Escala de Vôo: E o Transporte Aéreo Internacional?*

A partir de 2009, o transporte aéreo internacional brasileiro provavelmente passará por uma verdadeira revolução. É que desde essa data um processo gradativo de liberação de tarifas para todos os seus vôos internacionais será introduzido pela ANAC. Os preços de passagens aéreas de vôos saindo do Brasil para qualquer país poderão ter descontos sobre os valores mínimos obrigatórios. Gradualmente, teremos um percentual de descontos permitidos ampliado de 20% para 80%, de forma a atingir uma total liberação de tarifas em janeiro de 2010. A partir de 2010 as tarifas internacionais serão totalmente liberadas, se o cronograma da ANAC for implementado. Desde setembro de 2008 já está em vigor também a liberdade tarifária para os vôos saindo do Brasil para qualquer país da América do Sul. O instrumento legal utilizado pela ANAC para levar adiante a liberalização de preços internacionais foi a Resolução nº 61/2008, estabelecida após processo de consulta pública. O histórico da regulação desse importante segmento do transporte aéreo brasileiro sempre foi marcado pelas restrições dos Acordos Bilaterais. Na prática, as passagens internacionais no País tinham seus preços estritamente controlados. Os possíveis descontos sobre uma tabela de referência de valores eram restringidos por uma política de preços mínimos, que visava a evitar a competição entre as empresas atuantes no mercado. O maior receio das autoridades sempre foi a prática predatória de grandes companhias aéreas internacionais - sobretudo por parte das empresas norte-americanas. O consumidor não podia ser beneficiado com descontos por conta de um entendimento de que os mercados aéreos internacionais brasileiros funcionam de forma a levar à total monopolização por grupos de empresas internacionais. Esse entendimento mudou radicalmente nos últimos anos e as autoridades atualmente preconizam a liberalização do mercado.

A figura a seguir apresenta as tarifas médias de referência vigentes para o transporte aéreo internacional por continente. Cada país possui uma tarifa de referência própria, que poderá, com a liberalização, sofrer descontos progressivos até a total liberalização, em 2010.

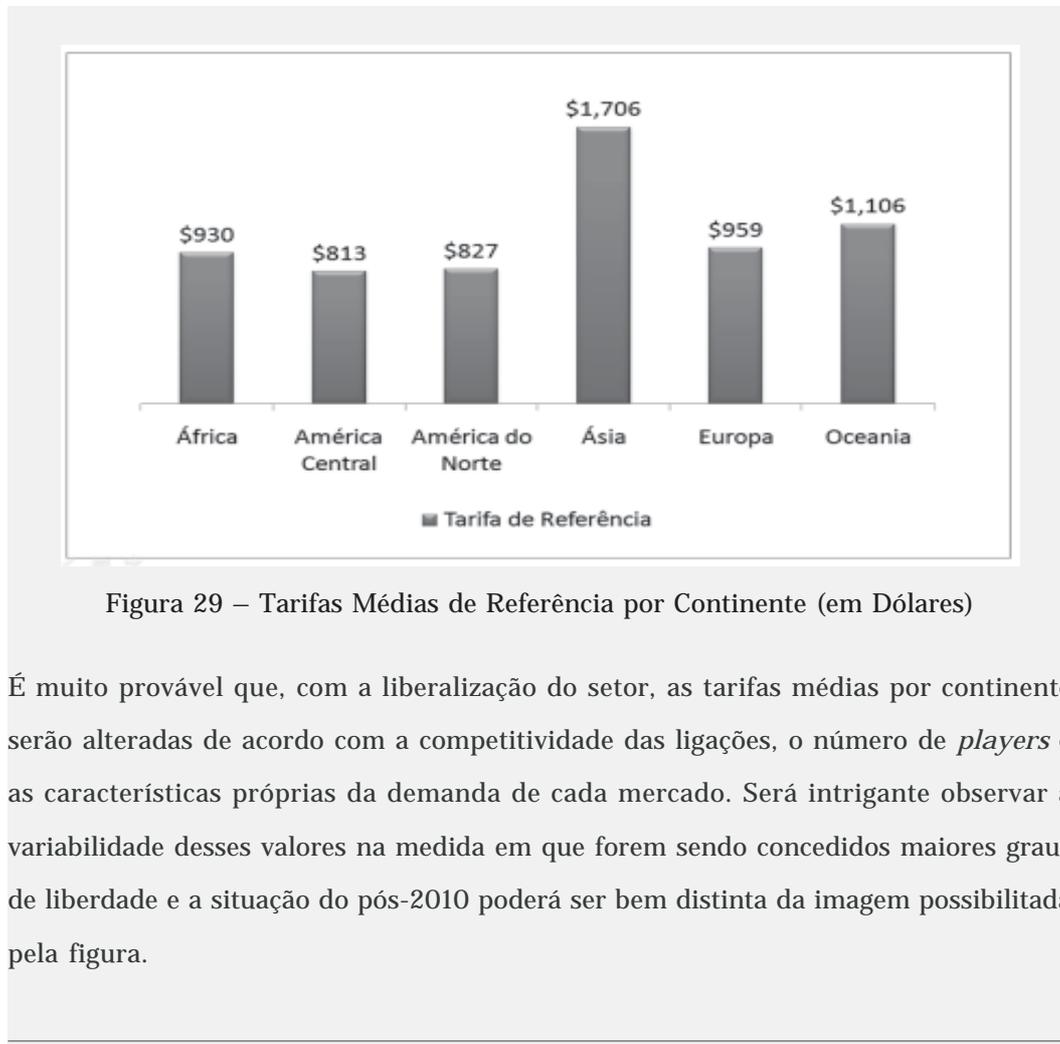

Figura 29 – Tarifas Médias de Referência por Continente (em Dólares)

É muito provável que, com a liberalização do setor, as tarifas médias por continente serão alteradas de acordo com a competitividade das ligações, o número de *players* e as características próprias da demanda de cada mercado. Será intrigante observar a variabilidade desses valores na medida em que forem sendo concedidos maiores graus de liberdade e a situação do pós-2010 poderá ser bem distinta da imagem possibilitada pela figura.

## Síntese da Dinâmica Regulatória: Para Onde Caminhamos?

A pergunta que fica, ao constatarmos a riqueza do processo evolutivo a que passou a regulação do transporte aéreo ao longo das últimas décadas, está voltada para o futuro. Afinal, para onde estamos caminhando? Quais os elementos são transitórios e estão fadados ao passado e quais são perenes e vieram para ficar? O que há de duradouro no debate regulatório? Como os "apagões" e as tragédias de 2006 e 2007 influenciarão essa trajetória? Como estará configurada a regulação da aviação brasileira ao longo das próximas décadas, por exemplo, em 2050? É possível aos *stackeholders* da aviação brasileira e, sobretudo, os investidores, anteciparem os cenários e fazerem projeções sobre as tendências quanto ao risco regulatório?

O conjunto de perguntas elencados acima certamente possui respostas não triviais. Elas requerem uma profunda reflexão sobre as tendências de postura não apenas da atual e das futuras diretorias da ANAC, mas também de todo o posicionamento do governo federal com relação a aviação civil, bem como das interações com a sociedade em geral, que se manifesta por meio de *lobbies* organizados e desorganizados, mobilizados ou não pelos próprios entes regulados, desejosos de capturar as ações do regulador em benefício próprio.

Para auxiliar no desenho dos cenários e alavancar o conhecimento sobre tudo o que já foi realizado até o presente momento em termos de configuração regulatória, propõe-se sistematizar a análise dos períodos. A seqüência de estágios evolutivos da regulação do transporte aéreo nas últimas décadas pode ser avaliada de forma mais adequada à luz de um conjunto de critérios de análise econômica. Os critérios aqui estudados foram os seguintes:

- ✦ Relação com a política econômica;

- ✦ Competitividade e concorrência;

- ✦ Entrada e saída de empresas (acesso e mobilidade);

- ✦ Precificação das empresas;

- ✦ Infra-estrutura aeroportuária.

O maior entendimento de como cada estágio de configuração regulatória lidou com esses itens ao longo do tempo proporciona uma melhor compreensão da atual problemática do transporte aéreo no País. Mais importante, permite entender as questões que ainda precisam ser abordadas na construção de um marco regulatório renovado e estável do setor.

As tabelas a seguir proporcionam uma breve descrição das características de cada uma das fases propostas, utilizando os critérios definidos de demarcação ("óticas"). Por meio delas, pode-se encontrar, mesmo que de forma resumida, uma boa gama das questões relevantes à constituição do novo marco regulatório e que nunca estiveram ausentes do debate em torno das políticas para esse setor. Questões importantes e atuais como a evolução do setor de controle do espaço aéreo, a privatização de aeroportos, a dominância de poucas empresas aéreas, a regulação dos atrasos de vôo, cancelamentos e *overbooking*, o acesso a localidades pelo interior do país, dentre muitas outras, podem ser melhor analisadas dentro dessa perspectiva histórica.

Em primeiro lugar, temos a Tabela 5. Ela retrata a relação entre a regulação econômica do setor e a política econômica do governo. Como vimos, pelas características do transporte aéreo na economia, tem-se que se trata de um setor de interesse para a autoridade econômica (Ministério da Fazenda), tanto no acompanhamento do seu crescimento quanto da sua competitividade. Isso porque o crescimento do transporte aéreo tem *spillovers* na indução de crescimento dos setores a montante e a justante, além do que apresenta contribuição importante nos indicadores de inflação. Historicamente, a intereferência macroeconômica oscilou entre "Ativa" e "Muito Ativa", por conta da necessidade de controle inflacionário. Desde 2001, com a Terceira Rodada de Liberalização do setor, observou-se que a participação do Ministério da Fazenda na rotina do setor ficou mais restrita à investigação antitruste – por meio da Secretaria de Acompanhamento Econômico, SEAE – e à participação nos debates quanto à configuração do marco regulatório e dispositivos de regulação.

Temos, assim, que, com o passar do tempo, a política macroeconômica evoluiu, passando de um caráter mais interventivo nos setores regulados para uma ênfase na liberalização com imposição de restrições de acompanhamento antitruste. Esse efeito pode ser considerado positivo, dado que contribui para a redução do risco regulatório. Igualmente importante, tende a se perpetuar no futuro.

Tabela 5 – Regulação do Setor e a Relação com a Política Econômica

| Estágio | Política Econômica | |
|---|---|---|
| | Regulação Econômica | Interferência Macroeconômica |
| 1. Regulação com Política Industrial | Presente, Estrita. | Ativa |
| 2. Regulação com Política de Estabilização Ativa | Presente, Estrita | Muito ativa |
| 3. Liberalização com Política de Estabilização Inativa | Parcialmente removida: Primeira Rodada da Liberalização | Possível, mas não ativa |
| 4. Liberalização com Restrição de Política de Estabilização | Parcialmente removida: Segunda Rodada da Liberalização | Ativa |
| 5. Quase- Desregulação | Removida: Terceira Rodada de Liberalização | Ausente |
| 6. Re-Regulação | Parcialmente Restabelecida | Ausente |
| 7. Retomada da Desregulação com Redesenho Institucional | Retorno dos Princípios Liberalizantes + ANAC | Ausente |

A Tabela 6 expõe a ótica da precificação das empresas. Primeiro, temos o problema do regulador em pleno período de regulação estrita. Nesse período, o regulador fixava uma tabela, chamada de "Curva Belga" do transporte aéreo, onde constavam os *yields* (preço por quilômetro transportado) a serem cobrados de acordo com a etapa de vôo das ligações concedidas para operação.

Era uma tabela onde constavam os índices tarifários a serem aplicados pelas companhias aéreas. Por meio dela, as operadoras obtinham os valores de *yield* permitidos para a formação dos preços nas respectivas ligações operadas.

Tabela 6 – Regulação do Setor e a Precificação das Empresas

| Estágio | Preços | | | |
|---|---|---|---|---|
| | Preço de Referência | Controle de Reajustes de Preço | Registro dos Preços | Banda tarifária |
| 1. Regulação com Política Industrial | Presente, Estrita. | Presente | Inexistente | Inexistente |
| 2. Regulação com Política de Estabilização Ativa | Imposto pelo DAC | Presente, com objetivos de política de estabilização | Inexistente | Ausente até 1988; [-25%, +10%] - 1989; [-50%, +32%] - 1990 em diante (só p/ tarifas com desconto); |
| 3. Liberalização com Política de Estabilização Inativa | Não imposto pelo DAC | Presente, mas associada à inflação do setor | ex-ante: 48 horas de antecedência, autom.aprovada se não houvesse resposta do DAC | [-50%, +32%], tanto para tarifas cheias quanto para tarifas com desconto |
| 4. Liberalização com Restrição de Política de Estabilização | Não imposto pelo DAC | Presente: mix de política estabilização c/ inflação do setor | ex-ante: somente em caso de mais de 65% de desconto | Inexistente |
| 5. Quase- Desregulação | Inexistente | Inexistente | ex-post: somente p/ acompanhamento do setor | Inexistente |
| 6. Re-Regulação | Inexistente | Inexistente | ex-post no primeiro ano e ex-ante in 2004 em diante | Inexistente |
| 7. Retomada da Desregulação com Redesenho Institucional | Inexistente | Inexistente | ex-post: somente p/ acompanhamento do setor | Inexistente |

A Tabela 7 a seguir ilustra a configuração da Curva Belga. Ela compõe a média dos coeficientes associados a um agrupamento de distâncias possíveis presentes na Portaria 1632/DGAC, de 9 de dezembro de 2003, que estabelecia "os índices tarifários de referência para o monitoramento das tarifas aéreas domésticas". Os valores monetários de yield (preço autorizado por

RPK) eram atribuídos de acordo com o tipo de companhia aérea "Empresa Nacional" (as principais) ou "Empresa Regional", conforme pode ser observado na Tabela. Os valores permitidos para empresas regionais eram sistematicamente superiores aos permitidos para empresas "nacionais", de forma que a relação entre os dois fosse constante ao longo da curva. Para efeitos da tabela, os valores monetários presentes na portaria foram trazidos a valor presente de janeiro de 2009.

Tabela 7 – Formatação da Curva Belga – Valores de *Yield* de Referência

| Distância | Empresas Nacionais Passageiros R$ Jan 2009 | Empresas Regionais Passageiros e Carga R$ Jan 2009 | Índice [até 500 km = 100] |
|---|---|---|---|
| Até 500 km | 1.17 | 1.52 | 100 |
| De 501 km a 1000 km | 0.73 | 0.94 | 62 |
| De 1001 km a 1500 km | 0.59 | 0.77 | 51 |
| De 1501 km a 2000 km | 0.52 | 0.67 | 44 |
| De 2001 km a 2500 km | 0.47 | 0.61 | 40 |
| De 2501 km a 3000 km | 0.43 | 0.56 | 37 |
| Mais de 3000 km | 0.39 | 0.51 | 33 |

Como forma de melhor entender a idéia subjacente à Curva Belga, tem-se o gráfico da Figura 30. Nele, pode-se visualizar a própria curva, que representa todo o arcabouço de regulação de preços utilizado no setor aéreo brasileiro ao longo de quase trinta anos. A relação não-linear com a etapa de vôo é bastante clara, sendo possível inclusive ajustar um polinômio de terceiro grau para sintetizar a forma funcional das curvas. Essa relação indica que o *yield* autorizado era decrescente com a distância percorrida, mas essa queda se efetuava a taxas decrescentes – o que é consistente com o que se espera em termos de custos unitários do transporte aéreo em relação à etapa de vôo.

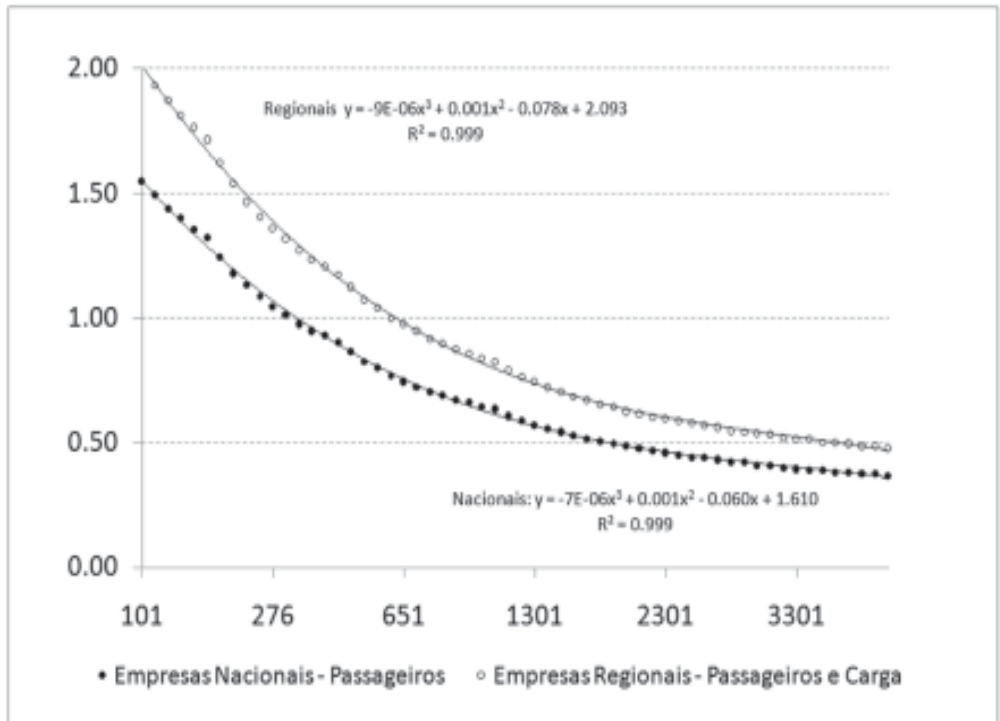

Figura 30 – Curva Belga para Empresas "Nacionais" e Empresas "Regionais"

Com a liberalização dos anos 1990, instituiu-se o instrumento das bandas tarifárias em adição ao instrumento da Curva Belga. Por meio das bandas, era possível conceder às companhias aéreas uma margem de flutuação ao redor do "preço de referência" representado pela Curva Belga. Os reajustes dos valores monetários representados pela Curva Belga continuavam sob controle do regulador que, periodicamente, revia os índices tarifários no sentido de ajustar pela inflação do período. Esse procedimento foi sendo aperfeiçoado ao longo do tempo, no sentido de se introduzir critérios relativos à inflação específica do setor aéreo.

Conta Rodrigues (2000), que a primeira versão das bandas tarifárias foi configurada de forma a permitir uma flutuação de até dez por cento para cima e vinte e cinco por cento para baixo da tarifa de referência - ou seja, liberdade de precificação dentro do intervalo [-25%,+10%]. Essa banda ampliada no início da década de 1990, para o intervalo [-50%, +32%]. Com a Segunda Rodada de Liberalização, ocorrida entre 1997 e 1998, com as Portarias 986 e 988 e 701, as bandas tarifárias foram extintas em prol do reforço de um mecanismo de registro de tarifas - seus "índices tarifários líquidos" - junto ao DAC. Esse mecanismo já existia desde a implantação das bandas tarifárias, e assegurava para o regulador a possibilidade de indeferir um pedido de tarifa promocional acima de 65% da tarifa de referência, por exemplo.

Desde a Terceira Rodada de Liberalização, a partir de 2001, o regulador não possui mais a problemática do registro de tarifas (*ex-ante* ou *ex-post*) ou do controle de reajuste de preços.  Com ela, a liberalização de tarifas aéreas no Brasil atingiu seu ápice, com a Portaria 1213: "*Os valores das tarifas aéreas aplicáveis às linhas aéreas domésticas serão estabelecidos livremente pelas empresas de transporte aéreo regular* (...)" (Artigo 3º). A autoridade, entretanto, continua buscando exercer algum acompanhamento econômico dos preços. As informações de preços são enviadas pelas companhias aéreas à autoridade regulatória, de acordo com legislação específica, sobretudo a Portaria DAC 447/DGAC, de 13 de maio de 2004, que estabelece as regras de funcionamento do sistema de tarifas aéreas domésticas.

Hoje em dia vigora o "regime de liberdade tarifária" foi consagrado com a Lei de Criação da ANAC[23]:

> "*Art. 49. Na prestação de serviços aéreos regulares, prevalecerá o regime de liberdade tarifária.*
>
> *§ 1o No regime de liberdade tarifária, as concessionárias ou permissionárias poderão determinar suas próprias tarifas, devendo comunicá-las à ANAC, em prazo por esta definido*".

A Tabela 8 apresenta a evolução da regulação do setor sob a ótica do acesso e da mobilidade das operadoras. O acesso diz respeito à possibilidade de certificação e entrada de novas operadoras no sistema e a mobilidade significa a possibilidade das operadoras já instaladas em fazer ajustes em suas malhas aéreas, entrando e saindo dos mercados por meio de inserção ou retirada de vôos em pares de aeroportos.

Tabela 8 – Regulação do Setor e a Entrada e Saída de Empresas

| Estágio | Entrada | |
|---|---|---|
| | Novas Firmas | Monopólios Regionais |
| 1. Regulação com Política Industrial | Não permitida: Política de "4 nacionais e 5 regionais" | Instituídos com o SITAR (5 regionais) |
| 2. Regulação com Política de Estabilização Ativa | Não permitida: Política de "4 nacionais e 5 regionais" | 5 regionais monopolistas |
| 3. Liberalização com Política de Estabilização Inativa | Permitida, tanto para o segmento nacional (linhas "tronco") como para regional | Ausente, com exceção das "Linhas Aéreas Especiais" |
| 4. Liberalização com Restrição de Política de Estabilização | Permitida | Inexistente |
| 5. Quase- Desregulação | Permitida, Remover Barreiras significa induzir as empresas à eficiência produtiva | Inexistente |
| 6. Re-Regulação | Permitida, mas com estímulo para aumentar a concentração | Inexistente |
| 7. Retomada da Desregulação com Redesenho Institucional | Permitida, Remover Barreiras significa induzir as empresas à eficiência alocativa | Inexistente, discussões sobre a mobilidade em mercados regionais |

Pode-se perceber que a regulação evoluiu de uma política que não permitia a entrada de novas firmas – o arcabouço de "4 companhias aéreas nacionais e 5 companhias aéreas regionais" dos anos 1970 – para uma política em que remover as barreiras regulatórias à entrada era vista como benéfica ao setor, e significava induzir as empresas à eficiência produtiva. A própria aquisição da Cruzeiro pela Varig, em 1975, e da VOTEC pela TAM, em 1986, serviram para desajustar a política inicialmente estabelecida. No início dos anos 1990, com a V CONAC, as companhias aéreas – e, sobretudo a TAM – mostravam-se insatisfeitas com os empecilhos que a regulação estrita causava no crescimento e na lucratividade do setor. Assim, da mesma forma que, ao final dos anos 1960, com as primeiras CONACs, as companhias aéreas lograram induzir a reforma regulatória que fosse consistente com os interesses privados da ocasião.

Com a Liberalização, ruíram oficialmente os esquemas de segregação de mercados e concessão de monopólios. Até 1992, a única exceção era dada pelas chamadas Linhas Aéreas Especiais – ligações entre os denominados aeroportos "centrais": Congonhas, Santos Dumont, Pampulha, entre si e com o Aeroporto de Brasília. Essas ligações eram operadas exclusivamente por empresas regionais, à exceção da Ponte Aérea Rio de Janeiro – São Paulo, operada por Varig, Vasp e Transbrasil em regime de *pool*. A Portaria nº 005/GM5, de 9 de janeiro de 1998 liberou qualquer empresa aérea brasileira para operar essas ligações. Em última instância, entrou também em caducidade o próprio conceito de "Empresa Regional", distinto do conceito de "Empresa Nacional".

Tabela 9 – Regulação do Setor e a Competitividade e Concorrência

| Estágio | Competição | |
|---|---|---|
| | Atitude das autoridades | Entre Nacionais e Regionais |
| 1. Regulação com Política Industrial | Presente, Estrita. | Inexistente em teoria, mas possível |
| 2. Regulação com Política de Estabilização Ativa | Inibir. ("Competição Controlada") | Inexistente em teoria, mas possível |
| 3. Liberalização com Política de Estabilização Inativa | Estimular | Permitida, com exceção das "Linhas Aéreas Especiais" |
| 4. Liberalização com Restrição de Política de Estabilização | Estimular, mas com controles antitruste | Permitida |
| 5. Quase- Desregulação | Estimular, mas com controles antitruste | Sem distinção |
| 6. Re-Regulação | Função "Moderadora", para evitar "excessos" de competição e de capacidade; uso de controles antitruste | Sem distinção |
| 7. Retomada da Desregulação com Redesenho Institucional | Princípios liberalizantes da lei da ANAC. Claro resgate durante a segunda diretoria. | Sem distinção |

A Tabela 9 mostra a evolução do setor sob a ótica dos anseios de competitividade e concorrência. De uma visão onde a concorrência era um mal a ser eliminado ("concorrência controlada") dos anos 1970, passou-se a uma visão na qual a concorrência era benéfica para o setor e que portanto deveria ser estimulada. Essa reviravolta no "consenso" entre os agentes do setor em pouco mais de duas décadas é indicativo do quanto os humores dos entes regulados variam ao sabor do ciclo de longo prazo do transporte aéreo. Variam também nitidamente ao sabor das oscilações de curto prazo do mercado, como visto com o período de "Re-Regulação" dos anos 2000. Se os lucros estão em queda, o clamor é por controle da competição; caso estejam em alta, pleiteia-se o livre mercado. A autoridade regulatória que tem por meta o bem-estar do consumidor deve estar atenta para evitar que esse comportamento seja fonte de captura por parte dos regulados. O setor deve possuir uma regulação pautada em um planejamento estratégico bem definido e que não flutue com as mudanças de humores dos agentes. Deve se flexível, contudo, a ponto de permitir que regras ruins não se mantenham apenas para manter a estabilidade. Mas deve-se sempre evitar a discricionariedade para acomodar interesses advindos da economia política típica do setor.

Por fim, a Tabela 10 mostra a evolução da regulação sob a ótica da sua relação com a infra-estrutura aeroportuária. A *interface* entre transporte aéreo e infra-estrutura aeroportuária é, em geral, planejada por meio da configuração de malha aérea das companhias operadoras. Em se tratando do transporte aéreo regular, essa *interface* é regulada por meio dos chamados HOTRANs (Horários de Transporte). O HOTRAN é a informação de um vôo regular solicitado pela companhia aérea ao órgão regulador. O conjunto de HOTRANs forma um sistema de informações dos vôos das companhias aéreas regulares, e contém o registro de todas as Linhas Aéreas Regulares aprovadas pela Comissão de Coordenação de Linhas Aéreas Regulares, COMCLAR, da ANAC. A Portaria 692/DGAC, de 20 de outubro de 1999, da época do DAC, dá instruções para o funcionamento deste órgão colegiado, de caráter consultivo, destinado originalmente a "*assessorar a Direção-Geral do Departamento de Aviação Civil no que tange a pedidos de aprovação de novas Linhas Aéreas Regulares e pedidos de alterações de Linhas Aéreas Regulares existentes*". No sistema BAV/HOTRAN são registradas todas alterações da oferta das companhias aéreas, como inclusões ou cancelamentos de escalas, alterações de freqüências de vôo, substituições de equipamento e ajustes de horários de vôo. O sistema HOTRAN, sob responsabilidade da Divisão de Estatística, representa, assim, o registro de toda a informação relativa aos vôos domésticos regulares autorizados pela COMCLAR.

Tabela 10 – Regulação do Setor e a Infra-Estrutura Aeroportuária

| Estágio | Capacidade e Infraestrutura | |
| --- | --- | --- |
| | Frequência, Rotas e Aeronaves | Aeroporto e Terminais |
| 1. Regulação com Política Industrial | Presente, Estrita. | Sob administração estatal: Infraero |
| 2. Regulação com Política de Estabilização Ativa | Controles econômicos, baseados nos fatores de aproveitamento; requeria autorização ex-ante da Comissão de Linhas Aéreas (CLA) | Sob administração estatal: Infraero |
| 3. Liberalização com Política de Estabilização Inativa | Autorização ex-ante (CLA); sem controle econômico; prioridade para companhias existentes | Sob administração estatal: Infraero |
| 4. Liberalização com Restrição de Política de Estabilização | Autorização ex-ante da Comissão de Coordenação de Linhas Aéreas Regulares (COMCLAR); sem controle econômico; processo mais ágil | Sob administração estatal: Infraero |
| 5. Quase- Desregulação | Autorização ex-ante da Comissão de Coordenação de Linhas Aéreas Regulares (COMCLAR); sem controle econômico; processo mais ágil | Sob administração estatal: Infraero; alguns aeroportos congestionados causando problemas de acesso e entrada. |
| 6. Re-Regulação | Restabelecimento dos controles econômicos para a autorização ex-ante | Sob administração estatal: Infraero; alguns aeroportos congestionados causando problemas de acesso e entrada. |
| 7. Retomada da Desregulação com Redesenho Institucional | Apagões de 2006 e 2007; regras mais transparentes e estritas de slots, concatenadas com regularidade e pontualidade | Rumo a uma inserção privada: ainda Infraero, mas discussões sobre o modelo de concessão, privatização de aeroportos. |

Com a Política de Flexibilização, e, sobretudo a partir da Quase-Desregulação de 2001 e da Lei da ANAC de 2005, maior pressão vem sendo exercida pelas companhias aéreas sobre a questão da aprovação de linhas aéreas. As empresas buscam exercer sua influência sobre todo o processo decisório dado que se trata de uma questão crucial do seu planejamento de malha, alocação de aeronave e tripulação e sobre a lucratividade em geral. Ou seja, o exame dos pedidos de novas linhas passou a ser elemento fundamental de disputa entre empresas, dado o ambiente de livre competição instaurado desde 2001. Com a saturação de alguns aeroportos-chave da malha aeroportuária brasileira – sobretudo o Aeroporto de Congonhas, em São Paulo – a tendência é caminharmos para uma sistemática mais estruturada e racional de análise de pedidos de novos HOTRAN, em paralelo à questão da alocação de horários em aeroportos "eslotados", isto é, sob regra de concessão de horários de pouso e decolagem por conta de estarem operando próximo ao limite de sua capacidade.

> *Um dos participantes da COMCLAR é a Empresa Brasileira de Infraestrutura Aeroportuária – ou INFRAERO, como é mais conhecida. A INFRAERO é um dos agentes que compuseram a rotina do setor aéreo ao longo das últimas décadas. Empresa pública federal brasileira, pertencente à administração indireta vinculada ao Ministério da Defesa, a INFRAERO foi criada pela Lei nº 5862, em 12 de dezembro de 1972, sendo responsável pela administração dos principais aeroportos do país. Atualmente são 67 aeroportos sob a administração da INFRAERO, o que compõe a maior parte do tráfego aéreo nacional e internacional. Na medida em que outros operadores privados começam a entrar no mercado brasileiro, será preciso criar um marco regulatório do setor aeroportuário, em que empresas como a INFRAERO, sejam vistas como entes regulados, e não como autoridades do setor.*

A questão da infra-estrutura aeroportuária no Brasil passa atualmente por intensos debates. Com os "apagões" de 2006 e 2007, as autoridades governamentais atentaram para os problemas de gargalo da capacidade do sistema, sobretudo na Região Metropolitana de São Paulo, que envolve os importantes aeroportos de Congonhas e Guarulhos. Um novo aeroporto deve ser construído na capital paulista, dado que nenhum daqueles aeroportos apresenta condições de expansão a ponto de aumentar consideravelmente a capacidade do sistema e resolver o problema de saturação. Adicionalmente, medidas como a abertura de capital da INFRAERO e a privatização dos aeroportos de Viracopos/Campinas e Galeão-Tom Jobim/Rio de Janeiro (ambos administrados pela INFRAERO), estão definitivamente presentes no cenário nacional.

A conclusão que podemos extrair da análise evolutiva da regulação, sob as diversas óticas aqui levantadas, é a seguinte: o setor está caminhando para um ambiente cada vez mais competitivo, onde prevalecerá os operadores com maior eficiência e que sejam os mais efetivos e bem-sucedidos na conquista do passageiro e na geração de receitas. Os movimentos de regulação discricionária, como os episódios de "Re-Regulação" de 2003 e 2004, ainda devem permanecer por um bom tempo adormecidos, mas deve-se estar atento para os riscos de captura em momentos de crise e potencial falência de operadoras. A questão aeroportuária, sobretudo no que tange a saturação de certos aeroportos, é de difícil equacionamento no curto prazo e tende a gerar preços mais elevados e necessidade de regulação de *slots* e tarifação a contento da infra-estrutura aeroportuária. O teor de previsibilidade do setor aéreo ainda é baixo, dado que há muitos mecanismos regulatórios ultrapassados ainda em vigor e que provavelmente serão alvo de consultas públicas e reformas no futuro próximo. Os investidores terão que ponderar entre as perdas possíveis advindas do risco regulatório e o elevado potencial de crescimento do mercado aéreo brasileiro.

### Pouso: Participação Estrangeira no Capital de Companhias Aéreas

O atual Código Brasileiro de Aeronáutica (Lei 7.565 de 19 de dezembro de 1986) impõe uma restrição máxima de 20% de capital estrangeiro na composição do capital votante de uma empresa aérea. A experiência internacional mostra que os Países tentam manter suas razões para não flexibilizar as restrições de controle e capital na incerteza de que as companhias estrangeiras sirvam aos interesses e propósitos da própria nação. De acordo com o relatório da ICAO chamado "*Survey of Contracting States*", de maio de 2001, o desenvolvimento econômico é citado como a razão mais importante para a manutenção das restrições atualmente vigentes. Conformidade com acordos internacionais é a segunda. Outros argumentos comumente citados são o interesse econômico das companhias nacionais, o turismo e o comércio internacional, a segurança de vôo, a manutenção e criação de empregos, a segurança nacional e a balança comercial. A figura a seguir mostra como essa restrição varia de país para país.

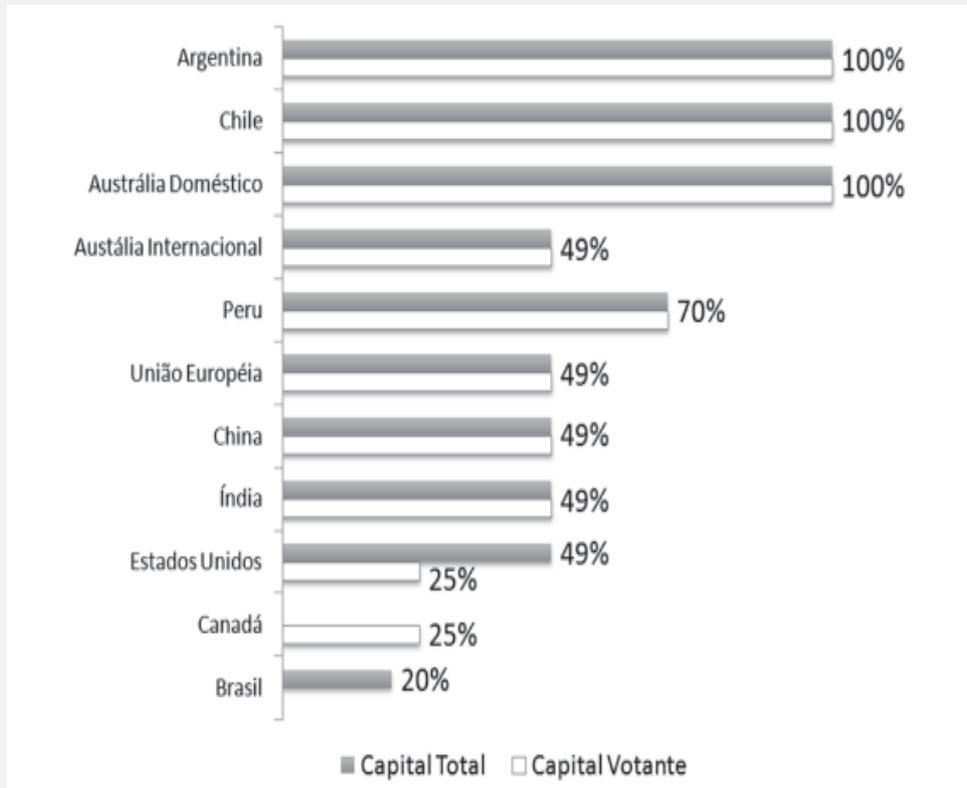

Figura 31 – Restrição de Participação de Capital Estrangeiro ao Redor do Mundo

No estudo "Estudo da Mudança de Restrição de Participação de Capital Estrangeiro nas Companhias Aéreas Brasileiras" (Kretzmann, 2007), realizado no âmbito do NECTAR, o autor efetuou uma avaliação dos impactos financeiros da flexibilização da atual restrição de capital estrangeiro em companhias aéreas brasileiras. Para isso, desenvolve um modelo de simulação que incorporou variáveis quanto a cenários pós-mudança da regra. O modelo de cálculo baseia-se em prováveis efeitos financeiros da mudança de restrição de capital, que levaria a 1. Nova oferta de ações ordinárias das companhias aéreas listadas em bolsa; 2. Investimento estrangeiro em ações ordinárias; 3. Variação do preço das ações ordinárias; 4. Variação do preço das ações preferenciais; e 5. Nova oferta de ações ordinárias. Os cenários analisados mostram a possibilidade de injeção de volumosos recursos financeiros como conseqüência de uma eventual mudança de restrição de capital estrangeiro, podendo chegar a 19% de aumento do valor de mercado para a Gol e 22% para a TAM.

# Desempenho nos Mercados Aéreos: Concentração e Competitividade

**Decolagem: Estrutura e Contestabilidade de Mercados**

Uma vez estudada a evolução da regulação do setor aéreo ao longo das últimas décadas, cumpre entender os efeitos das medidas e ações de políticas implementadas nos efetivos resultados de mercado. Buscamos aqui entender a relação entre a evolução da regulação do setor, e fatores-chave do mercado, como a estrutura (número de *players* e a concentração industrial), conduta (práticas concorrenciais ou anticoncorrenciais) e desempenho de mercado das empresas reguladas (eficiência, ineficiência). É a partir do entendimento dessas relações que se pode melhor visualizar a dinâmica competitiva do setor aéreo e as necessidades de inserção do regulador e das políticas públicas setoriais.

Em particular, o regulador tem papel fundamental na configuração de barreiras regulatórias à entrada de novas firmas. Essas barreiras seriam fruto de regras ou mecanismos regulatórios que dificultariam ou mesmo inviabilizariam a entrada de novos operadores. No limite, a própria contestabilidade dos mercados – a capacidade de novos *players* surgirem para competirem com as atuais empresas instaladas, chamadas de incumbentes – e, em última instância, o bem-estar do consumidor, seriam potencialmente afetados por barreiras regulatórias desse tipo. Em regime de barreiras à entrada e ausência de controles antitruste, as firmas estabelecidas podem ter incentivos para exercer seu poder de mercado, quer seja na forma de práticas colusivas tácitas, quer seja na forma de cartelização do mercado, o que é extremamente danoso ao consumidor.

> *A Contestabilidade dos mercados é a capacidade de oferta de um bem ou serviço por fornecedores alternativos, novos entrantes. Um mercado é perfeitamente contestável se novos participantes podem entrar (e sair) com facilidade, derrubando assim os lucros dos oligopolistas ou monopolista. Processos colusivos ou cartéis são mais difíceis de serem montados quando a contestabilidade é alta.*

Cabe única e exclusivamente ao regulador definir a magnitude e a abrangência das barreiras regulatórias existentes. Muitas vezes, entretanto, o regulador não se dá conta de que certas regras podem ter como conseqüência não antecipada a imposição de barreiras com danos à concorrência. Por exemplo, um regulador bem-intencionado e preocupado com o melhor uso de um aeroporto, pode exigir que somente empresas com balanço patrimonial anual publicado sejam habilitadas a participar de processos de alocação de *slots* em aeroportos administrados com essas regras – aeroportos "eslotados". A justificativa para isso poderia ser que empresas sem balanço ainda não comprovaram a viabilidade da própria operação e que conceder *slots* a elas seria arriscado no sentido de aumentar o risco de ruptura da prestação do serviço caso a mesma entre em regime falimentar. Por outro lado, essa regra se configura em uma barreira à entrada econômica nos aeroportos congestionados, dado que, por não possuírem balanço anual, novas entrantes recém-constituídas estariam excluídas *ex-ante* da concorrência. E sua própria viabilidade financeira poderia não ser garantida caso a mesma não possuísse algum acesso ao aeroporto congestionado, mas altamente lucrativo. Temos assim que a falta de acesso ao aeroporto congestionado poderia se configurar em um sério problema de contestabilidade dos mercados, dado que as novas entrantes sempre estariam fragilizadas financeiramente, por estarem confinadas a aeroportos com operação menos lucrativa.

Lógicas como a perseguida no parágrafo anterior devem estar na rotina dos processos de formação de regras regulatórias no setor aéreo. Sempre existirão efeitos colaterais importantes de regras interventivas, e esses efeitos devem ser antecipados pelos reguladores. Regras ruins são aquelas que promovem danos à concorrência. Entretanto, e como visto, historicamente as regras regulatórias vêm evoluindo no sentido de buscar fomentar a maior competitividade do setor aéreo. Vejamos como esse processo afetou os elementos do mercado aéreo e sua competitividade ao longo das décadas.

## Evolução da Estrutura do Mercado Doméstico

A indústria do transporte aéreo brasileiro foi marcada por pouquíssimos episódios de entrada nos últimos 35 anos. De fato, foi apenas com a reforma regulatória dos anos 1990 é que o setor teve sua estrutura de mercado oficialmente aberta para novas entrantes, quando se rompeu com o arcabouço institucional de 4 companhias aéreas de âmbito nacional (linhas-tronco) e 5 companhias aéreas de âmbito regional que caracterizou a fase regulatória mais estrita, de "competição controlada". Como visto, a operação das companhias aéreas regionais iniciou-se com o esquema de subvenções do SITAR (Sistemas Integrados de Transporte Aéreo Regional Regional, a partir de 1976), onde foram designados monopólios geográficos para TAM, Taba, VOTEC, Rio-Sul e Nordeste, com vistas a se alavancar o transporte aéreo em rotas menos densas do sistema. Antes disso, havia apenas a operação do triopólio Varig-Cruzeiro, Vasp e Transbrasil, que operavam sobretudo nas linhas troncos do segmento doméstico.

A Figura 32 abaixo, exibe a evolução do número de companhias aéreas nacionais certificadas, que tinham seus dados operacionais e econômicos registrados nos Anuários do DAC no período de 1970 e 2008.

o

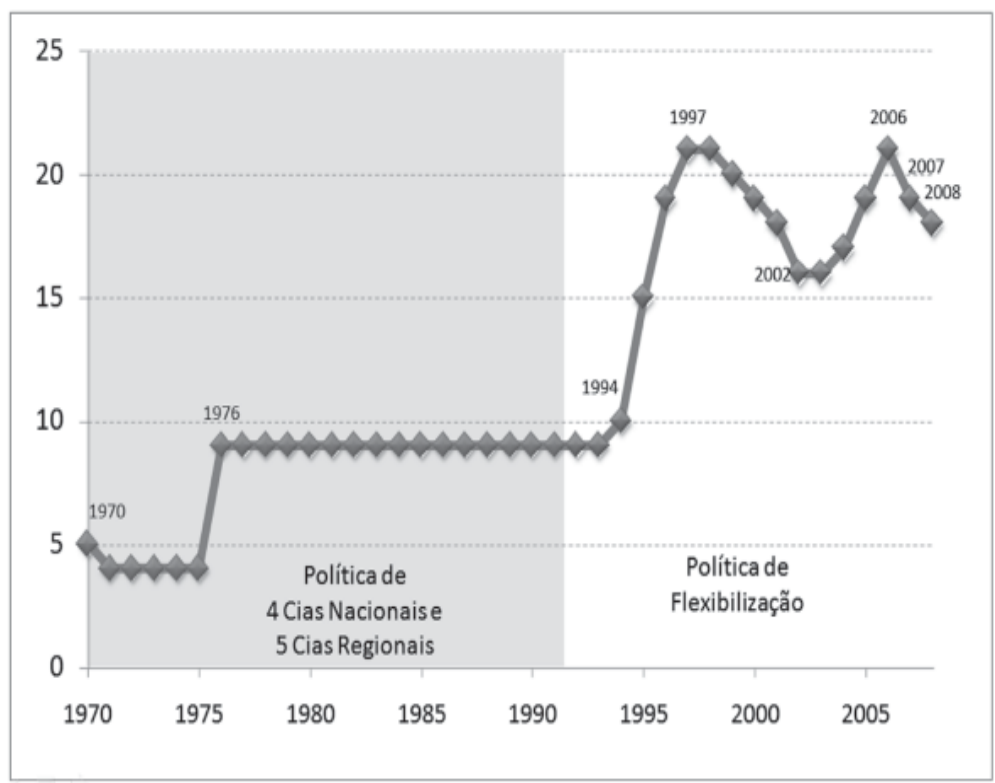

Figura 32 – Número de Companhias Aéreas Regulares no Transporte Aéreo[24]

Como pode ser analisado por meio da Figura 32, afora o episódio de entrada das companhias aéreas regionais quando da implementação do SITAR, observou-se apenas dois outros eventos de nova entrada em massa de operadoras novatas – o primeiro, que se seguiu à Primeira Rodada de Liberalização, de 1992, e o segundo, já nos anos 2000, até 2006. O mais expressivo deles foi o observado em meados da década de 1990. De fato, naquele período, o número de companhias mais que dobrou, passando de 9 para 21 – fenômeno que pode ser caracterizado como uma "onda de novas entrantes" seguida à liberalização da oferta caracterizada pelo fim dos monopólios regionais. O choque cambial de 1999 e o conseqüente choque de custos operacionais, contudo, representaram um retrocesso neste quadro, dado que o número de novatas na aviação começou a declinar desde então, vindo a se recuperar, ainda que de forma lenta, apenas recentemente (2004 e 2005).

*A saída de empresas, com encerramento total das operações e dos vôos, faz parte da rotina de mercados aéreo livres. Nos Estados Unidos, por exemplo, a desregulamentação fez com que encerrassem suas operações as três principais companhias aéreas do país: Panam, TWA e Eastern. Esses episódios geram inúmeros problemas para o consumidor, entretanto. Em geral, uma companhia aérea quando encerra suas operações, é porque não tem viabilidade e nem saúde financeira para proceder. Em geral possui dívidas de curto prazo junto a fornecedores como empresas de arrendamento de aeronaves e distribuidores de combustível e que inviabilizam totalmente o prosseguimento das operações. Essas circunstâncias sempre geram problemas de curto prazo, dado que haverá um estoque de passageiros com bilhetes comprados e que não poderão viajar e que pleitearão o reembolso dos valores comprados. Uma das soluções é a incorporação por uma outra empresa, interessada na marca da empresa em dificuldades. Uma outra solução, de cunho mais regulatório, é o endosso das passagens. Essa a solução parcial encontrada durante os episódios de saída da Transbrasil (2001), Vasp (2004), BRA (2007) e mesmo da crise da Varig (2006). A autoridade sempre possui o dilema entre "intervir agora e incorrer no risco de prejudicar totalmente as chances da empresa" e "intervir depois e incorrer no risco da empresa quebrar". Quanto mais as autoridades investirem na criação de mecanismos claros e transparentes de atuação, menos traumático será o encerramento das operações das empresas aéreas.*

A Figura 33 apresenta o número de episódios de entrada no segmento doméstico desde 1994. Nela é possível constatar um caráter cíclico da indústria, onde seqüencias de entradas são seguidas por ondas de saídas de empresas. Não obstante o aumento expressivo no número de companhias aéreas, caracterizado pelo período posterior à Primeira Rodada de Liberalização, este acréscimo não se traduziu em uma participação de mercado que possa ser considerada relevante. Pelo contrário, essa entrada é melhor caracterizada como variações na "franja" do mercado – ainda predominantemente marcada pela operação de empresas regionais – do que no mercado típico das chamadas companhias nacionais, que, em geral operam tanto em linhas-tronco quanto em linhas regionais.

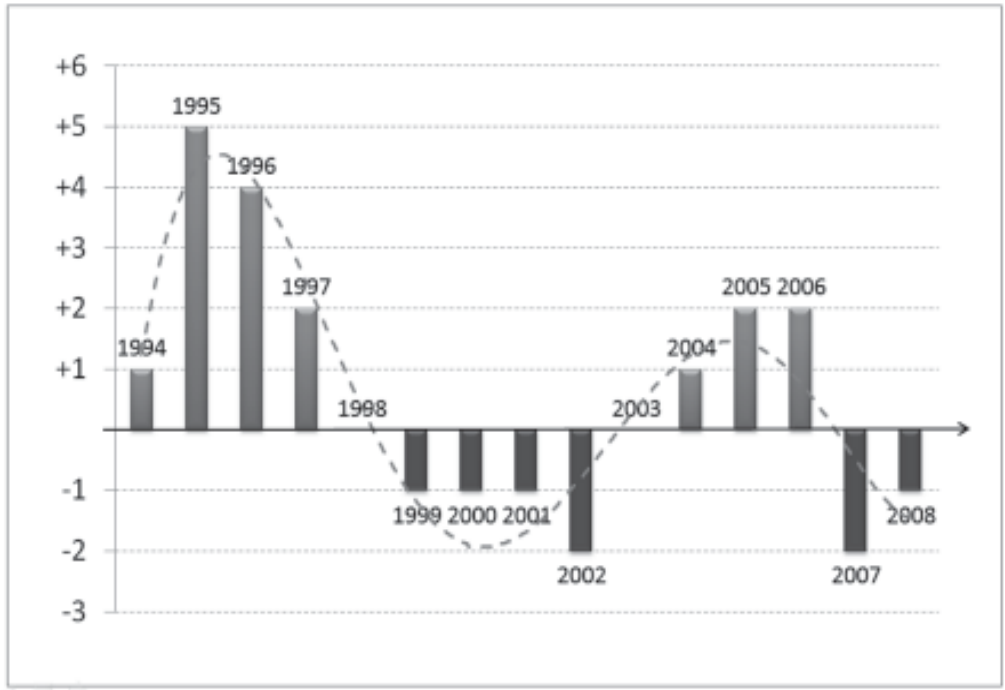

Figura 33 – Episódios de Entrada no Transporte Aéreo[25]

Pela Figura 34 é possível analisar a evolução da franja de mercado do transporte aéreo brasileiro – em termos de participação no total de RPK (passageiros-quilômetro transportados pagos) – nos últimos 30 anos:

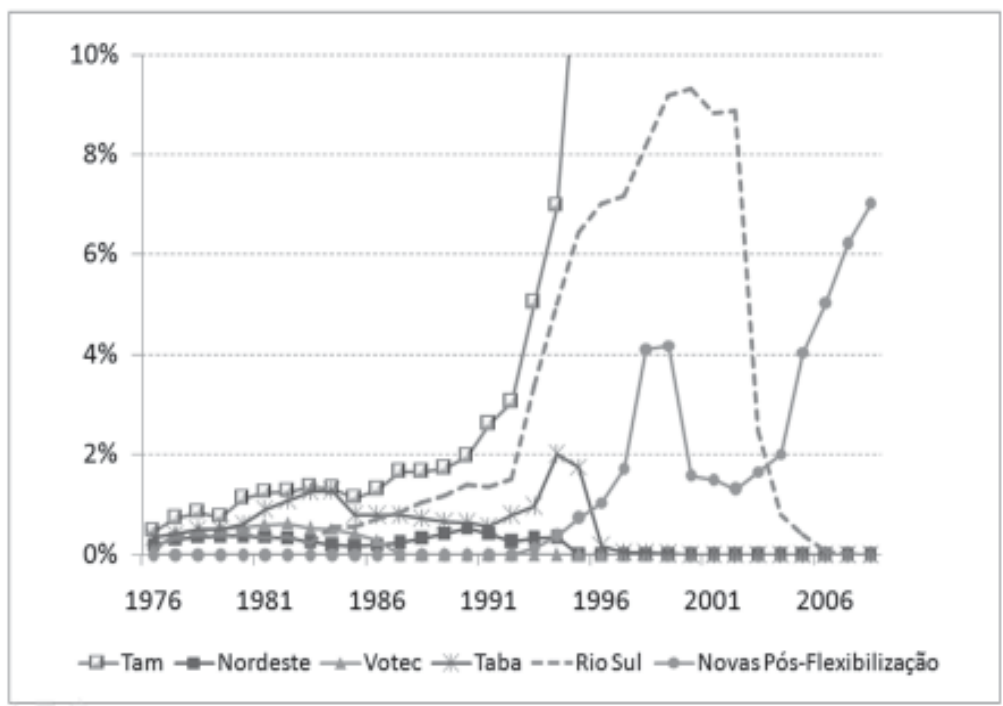

Figura 34 – Participação das Regionais do SITAR e Novatas do Pós-Flexibilização[26]

"Franja de mercado" é aqui definida aqui como aquele conjunto de firmas com participação reduzida no setor – um critério certamente mais qualitativo do que quantitativo. Companhias regionais típicas, como Taba e VOTEC, sempre fizeram parte da franja, bem como as novatas entrantes no período imediato à Primeira Rodada de Liberalização. Entretanto, há que se mencionar que tanto a Rio-Sul quanto a Nordeste, por serem, na época, subsidiárias de grupo empresarial, e, pelo seu porte, a TAM (e outras companhias do mesmo grupo), desde o início dos anos 1990, podem ser consideradas como fora da franja.

Esta definição é suficientemente geral para englobar todas as firmas fora do conjunto restrito formado por Varig (grupo), Vasp, Transbrasil (grupo), TAM (grupo) e Gol. Todas as demais se encontram dentro da franja. O gráfico da Figura 35 mostra a evolução do tamanho do total da franja (em participação no RPK total da indústria doméstica). Note-se que a TAM somente está presente neste cômputo até 1991, sendo considerada fora da franja a partir da Primeira Rodada de Liberalização.

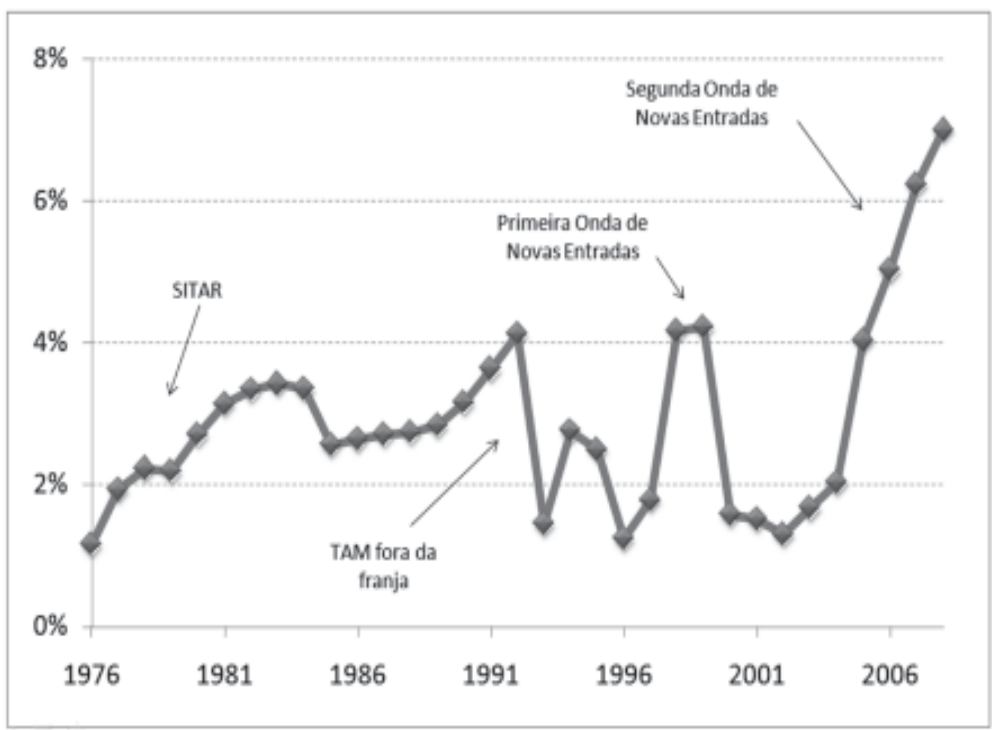

Figura 35 – Tamanho da Franja de Mercado[27]

A onda de novas entrantes que se seguiu à Primeira Rodada de Liberalização – sobretudo com o fim dos monopólios regionais – certamente significou um relevante impacto na contestabilidade do setor aéreo, dado que, a partir de então, um número maior de companhias certificadas e em operação estava presente no mercado. Apesar desse conjunto de novas entrantes estar confinado em determinadas regiões do País, é sempre importante salientar que, se garantida a mobilidade com acesso a aeroportos eslotados, as mesmas podem sempre ser consideradas potenciais entrantes nos mercados das companhias incumbentes e podem atuar de forma benéfica no sentido do incremento da contestabilidade aos mercados. A ascensão da TAM enquanto "*major player*" no mercado brasileiro pode ser considerado uma decorrência disso, e ilustra o fato de que, uma vez as novas entrantes presentes no sistema aéreo – isto é, certificadas e em operação – elas já incorreram em muitos dos custos enterrados do setor e têm, provavelmente, mais facilidade de entrar em outras regiões do que teriam outras futuras novatas.

### *Escala de Vôo: A Entrada da Azul Linhas Aéreas*

A mais nova empresa aérea do Brasil – Azul Linhas Aéreas, que iniciou suas operações no mercado doméstico em dezembro de 2008 – nasceu a partir do interesse do empresário David Neeleman em investir na aviação do País. Neelman ficou famoso mundialmente por ter fundado duas das maiores e mais lucrativas empresas aéreas seguidoras do conceito *Low Cost, Low Fare* na América do Norte – a americana jetBlue e a canadense Westjet. Uma particularidade do empresário é a sua peculiar combinação de experiência na indústria e isenção legal no que se refere à propriedade de empresas aéreas no Brasil: cidadão brasileiro e americano, David Neeleman pode contornar os limites de propriedade de capital estabelecidos na legislação e aplicados a não-brasileiros pela Agência Nacional de Aviação Civil – órgão regulador da aviação civil brasileira. De fato, estrangeiros não podem deter mais que vinte por cento do capital de empresas aéreas nacionais. As primeiras veiculações públicas do interesse de Neeleman em se estabelecer no mercado brasileiro foram veiculadas pela mídia em janeiro de 2008. Já naquele instante, aspectos-chave do plano estratégico da nova empresa foram anunciados ostensivamente, como a data de início previsto de operações, tamanho da frota e das aeronaves que a comporiam e a morfologia da rede. Desde então, o cronograma de eventos que antecedem ao lançamento da nova empresa manteve-se em ritmo contínuo e intenso, o que conferiu plena credibilidade à entrada de Neeleman no mercado brasileiro. Dois aspectos incrementaram em muito a visibilidade da nova empresa aérea. Primeiramente, menos de dois meses após seu anúncio de lançamento, a nova empresa conduziu uma campanha de alcance nacional que decidiria seu nome por meio de votação pública. Após contar com quase 160 mil votos provenientes de mais de 100 mil cadastros, a escolha do nome ocorreu em duas fases: a primeira de caráter sugestivo; a segunda por meio de votação a partir dos 10 nomes mais freqüentes. Em segundo lugar, concomitante ao anúncio do nome escolhido – Azul –, oficializou-se a compra de 36 aeronaves da Embraer, mantendo-se fiel ao plano estratégico de trabalhar com apenas um tipo de aeronave e com tamanho menor que as operadas pelas principais empresas estabelecidas. Para se ter uma idéia do que isso representa, tem-se que, da frota ativa de 315 aeronaves comercial de transporte de passageiros no Brasil ao final de 2007, 222 aviões possuíam mais de 118 assentos, ou seja, uma proporção de 7 em cada 10 aeronaves apresentava configuração superior à escolhida pela entrante. Levandose em conta que, das 93 aeronaves com capacidade igual ou inferior a 118 assentos, apenas 28 eram jatos, todos de geração anterior, tem-se que o plano de frota da Azul pode representar um marco importante de uma nova etapa da nossa aviação comercial. Um aspecto muito marcante no processo de lançamento da Azul Linhas Aéreas foi, em um primeiro momento, a reiteração do modelo de negócios que pretendia seguir: diferenciação de produto como vantagem competitiva, malha de vôos que aposta na ligação direta entre mercados que hoje requerem escalas ou conexões, especialmente entre capitais de Estados, e prática de "preços não-predatórios", ou seja, alinhados com o mercado e respeitando flutuações na cotação do preço do petróleo. Ou seja, afirmações condizentes com um comportamento de baixa agressividade e de "contestação responsável" a um mercado desde 2006 caracterizado por um quase-duopólio formado por Gol e TAM, que, juntas, detinham mais que 90% do total das viagens domésticas (2007). Mais tarde, às vésperas da entrada, o anúncio de tarifas promocionais, cerca de 35% inferiores às da concorrência e próximas às das operadoras de ônibus, deixava claro que havia a necessidade de se fazer um *marketing* mais agressivo para a campanha de inserção inicial no mercado.

## Concentração e Competição Oligopolística

A competição no mercado de transporte aéreo brasileiro é marcada pela estrutura de oligopólio parcialmente desregulado, onde vigora forte concentração. Em geral, pode-se dizer que é um mercado que tem, nas últimas quatro décadas, em geral comportado 3 *players* efetivos e uma franja de pequenas empresas. De fato, o triunvirato dos anos 1970-80 (Varig-Cruzeiro, Vasp e Transbrasil) evoluiu para o triunvirato dos anos 2000 (TAM, Varig e Gol), sendo que, das grandes incumbentes do período regulatório (*legacy*), apenas a Varig continua em operação em 2009. A partir de 2006, chega-se à situação de quase-duopólio TAM-Gol.

A Figura 36 abaixo mostra a evolução das participações de mercado das maiores companhias aéreas brasileiras nos últimos quarenta anos, o que permite tecer considerações acerca da estrutura do mercado de aviação comercial doméstico no País. O gráfico foi construído agrupando-se a participação de todas as companhias aéreas pertencentes a um mesmo grupo empresarial (por exemplo, Varig, Rio-Sul e Nordeste), mas, entretanto, sem agrupar firmas participantes de acordos code share:

Pode-se notar, pela Figura 5, que, mais relevante do que a onda de novas entrantes do pós-Primeira Rodada de Liberalização, foi a ascensão da TAM enquanto companhia aérea de âmbito nacional no mercado, e também a entrada da Gol (janeiro de 2001), já durante a Segunda Rodada de Liberalização. Em ambas as ocasiões, o efeito sobre a participação de mercado das companhias *legacy* brasileiras instaladas (ou "incumbentes") foi extremamente influente, alterando efetivamente a composição das participações no mercado.

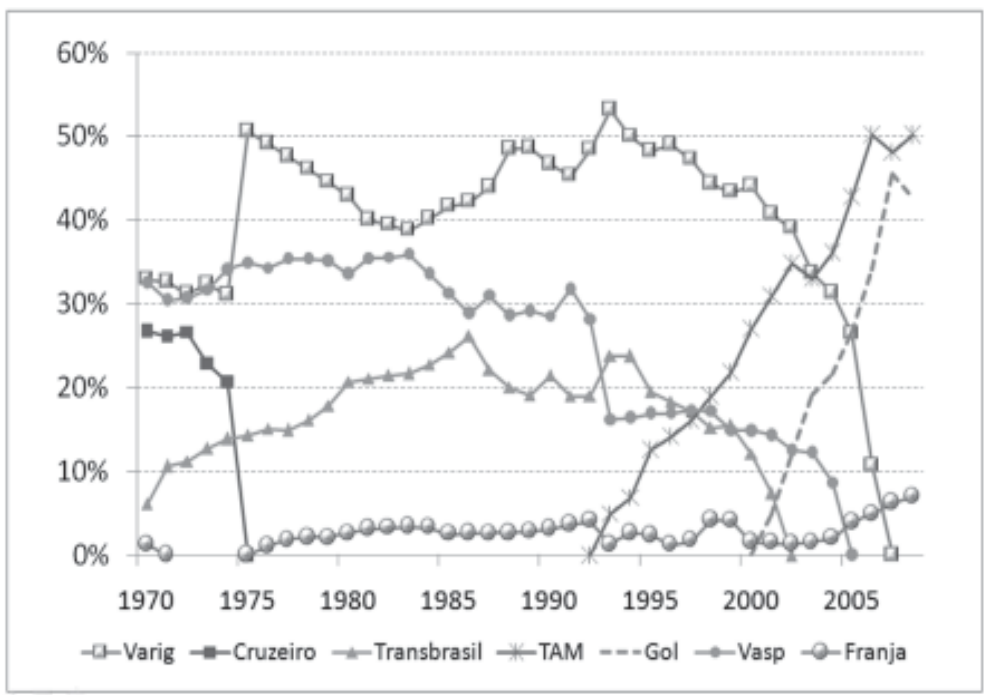

Figura 36 – Participações de Mercado – Transporte Aéreo Doméstico[28]

Um maior entendimento da evolução da estrutura de mercado na indústria do transporte aéreo pode ser realizado por meio do cômputo do índice de concentração denominado de Índice Herfindhal-Hirschman ou Índice HHI.

*O HHI, Índice de Herfindhal-Hirschmann, mede o grau de concentração em um dado mercado. Varia entre zero e um, sendo quanto maior quanto for o total do mercado em poder de poucas empresas. Ele é calculado por meio da soma dos market share das firmas instaladas elevados ao quadrado. É como se o analista extraísse uma média ponderada dos market shares das empresas, onde os pesos seriam os próprios market shares. Isso significa que participações maiores, de empresas dominantes por exemplo, entram com peso maior no cômputo do índice. O Departamento de Justiça dos Estados Unidos utiliza o HHI multiplicado por 10 mil e manifesta preocupação quanto à estrutura de mercado sempre que o índice for superior a 1800, qualificando o mesmo de "altamente concentrado".*

Por meio da extração do Índice HHI de concentração para o mercado doméstico regular de passageiros no Brasil, ao longo de todo o período analisado, tem-se a Figura 37 a seguir:

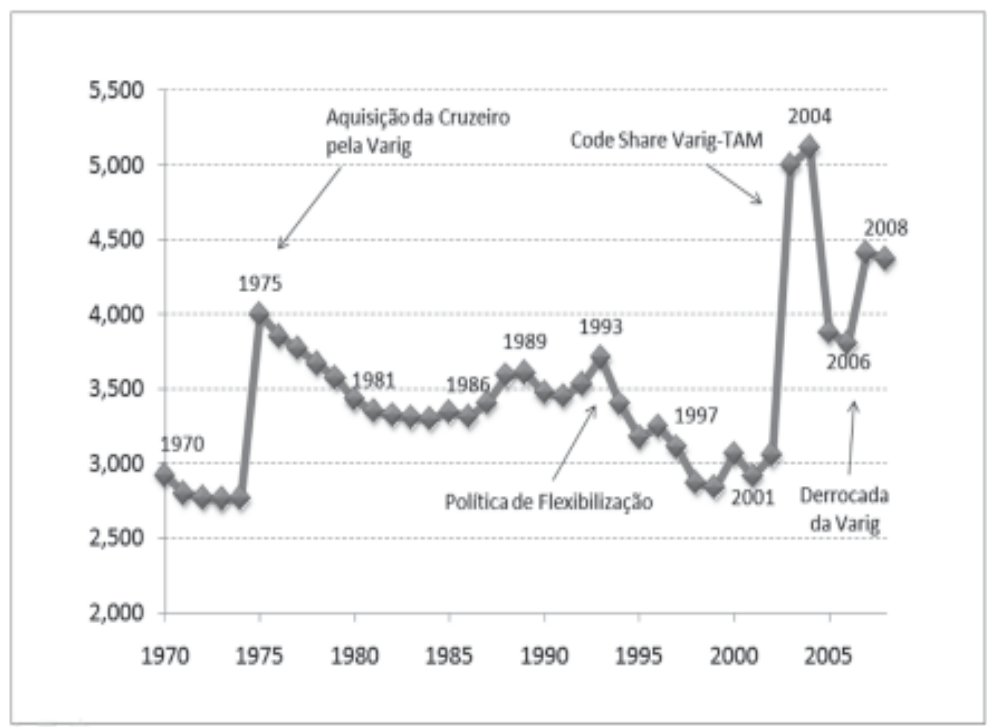

Figura 37 – Índice Herfindhal Global (HHI) – RPK Doméstico[29]

A Figura 37 pode ser visualizada de uma forma alternativa, utilizando-se uma outra métrica de estrutura de mercado, o chamado "Número de Competidores Efetivos". O Número de Competidores Efetivos é calculado a partir da extração do inverso do HHI (ou seja, 1/HHI). Esse indicador oferece um cálculo do número de empresas participantes com efetivo poder e influência no mercado, e contrasta com a simples contagem do número de participantes, que em geral inclui empresas com participações de mercado muito reduzidas. Assim, no caso do indicador de Número de Competidores Efetivos, tem-se uma métrica que apresenta as mesmas vantagens do HHI, sendo definida a partir do *market share* de cada participante. A série histórica de "Número de Competidores Efetivos" pode ser vista na Figura 38:

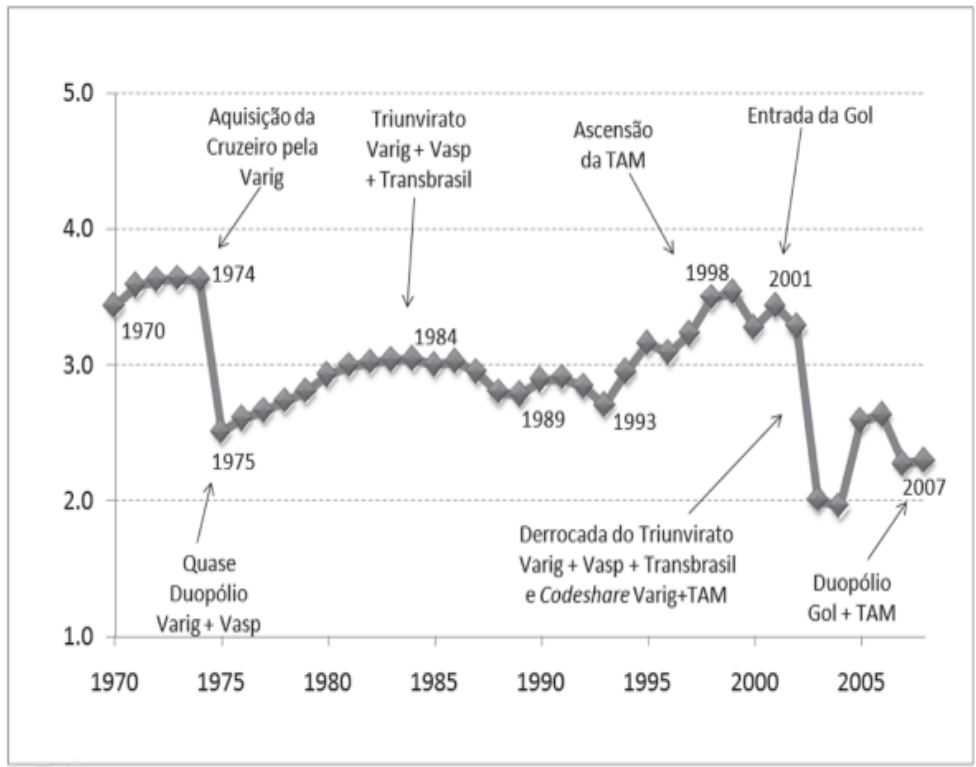

Figura 38 – Número de Competidores Efetivos (1/HHI) [30]

Pode-se perceber que os picos históricos de número de competidores efetivos ocorreram pouco antes da aquisição da Cruzeiro pela Varig e a partir da Segunda Rodada de Liberalização (1998 em diante), tendo ocorrido valores acima dos 3,5 tanto em 1999 quanto 2003. São exemplos de anos onde coexistiam no mercado quatro *major players* no mercado (Varig-Vasp-Transbrasil-Cruzeiro, em 1974, Varig-TAM-Transbrasil-Vasp, em 1999, e Varig-TAM-Vasp-Gol, em 2001).

## Conduta e Desempenho: Influência de Fatores Macroeconômicos

A análise da conduta competitiva das companhias aéreas no mercado doméstico permite importantes ilustrações sobre desempenho do setor ao longo das últimas décadas. Um adequado entendimento da evolução de indicadores de conduta e desempenho, e, sobretudo, de seus determinantes, proporciona subsídios para uma análise da eficácia das reformas regulatórias do setor, e em especial, da política de Flexibilização.

Como indicador principal do desempenho, pode ser utilizada a margem operacional, o *markup* preço-custo, MPC, definido como MPC = (preço médio – custo médio)/ custo médio. O MPC é uma variável *proxy* para as margens de lucro das empresas e para o chamado índice de Lerner – uma métrica que os economistas criaram para medir o poder de mercado das empresas. O uso do MPC é um procedimento razoável sob a suposição de custo marginal constante e lado dos custos perfeitamente observável pelo analista. De fato, os custos médios ao nível da rede (sistema) são observáveis, pois os dados econômicos globais das companhias aéreas estão presentes nos anuários da ANAC. Deve-se resguardar, contudo, que a observabilidade do lado dos custos pelo analista não é perfeita, tanto por fatores de complexidade do processo produtivo no setor de transporte aéreo, como por questões de assimetria informacional entre DAC (quem coleta os dados de custos) e firmas reguladas (quem repassa a informação de custos).

Qualquer análise acerca dos impactos de medidas de reformas regulatórias na conduta competitiva e desempenho das firmas reguladas deve passar por uma decomposição de diversos fatores que são potencialmente explicativos do comportamento empresarial. Assim, deve-se primeiro apontar o que é variação devido a fatores estruturais de deslocamentos da demanda e dos custos, para então se identificar se a reorientação das políticas setoriais é fator explicativo dos indicadores de desempenho das empresas. No caso do transporte aéreo, duas variáveis-chave, de âmbito macroeconômico, têm papel crucial na conduta das companhias operadoras: o PIB e a taxa de câmbio.

A Figura 39 apresenta a evolução do MPC do mercado doméstico, no período 1972-2007. Uma análise pormenorizada deste gráfico permite visualizar o comportamento cíclico deste indicador de performance setorial. Uma curva polinomial foi ajustada aos dados, como forma de melhor identificar este comportamento:

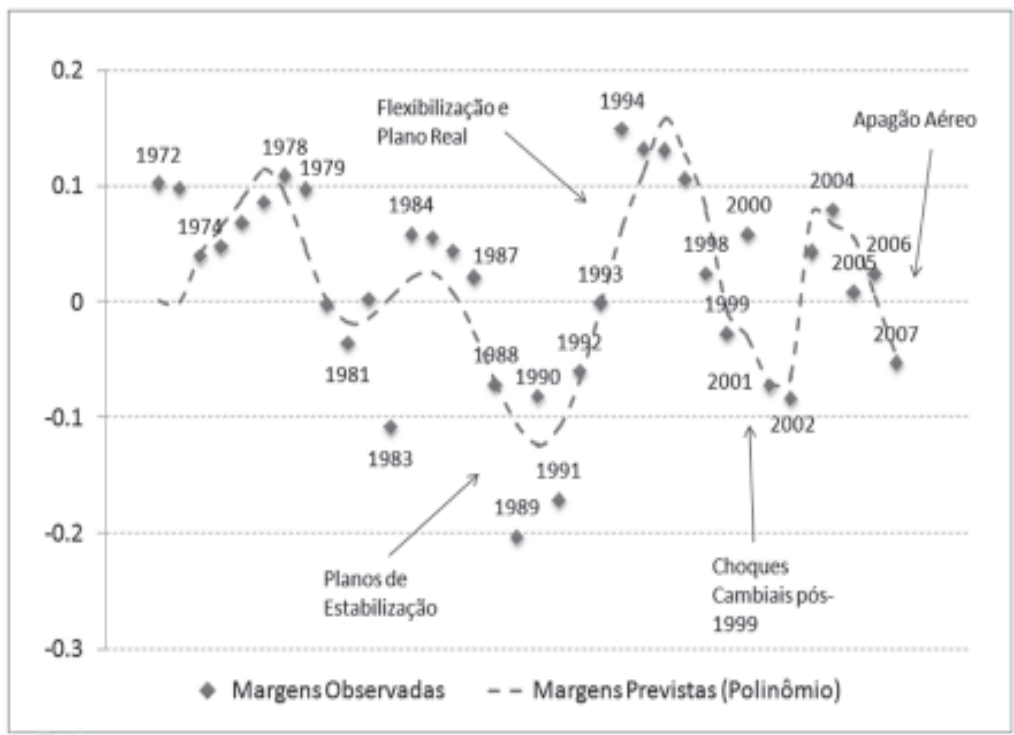

Figura 39 – Os Ciclos do MPC no Transporte Aéreo Doméstico[31]

Ao longo do período de análise, foram observados alguns picos e vales do MPC, sendo os anos de 1973, 1979, 1987, 1997 e 2004, os derradeiros das fases de maior lucratividade; não por coincidência, a maioria desses anos foi caracterizada por choques macroeconômicos de alta relevância, e exógenos ao setor. Os anos de 1983, 1991, 1999, 2001 e 2002, anos de choques cambiais, são marcados por lucratividade altamente negativa. 2007 também foi um ano de lucratividade ruim, provavelmente devido ao apagão aéreo, o trágico acidente de julho daquele ano e as restrições subseqüentes sobre o aeroporto de Congonhas. Também, percebe-se que entre 1988 e 1992 constata-se prejuízos seguidos, contendo, inclusive, os dois menores valores da série (1989 e 1991). Este período foi inclusive alvo de disputas na justiça com relação às perdas acarretadas pela política macroeconômica na lucratividade deste setor. Por fim, tem-se que as maiores lucratividade estão associadas a períodos áureos de crescimento do PIB e do consumo, como do "Milagre Econômico", do pós-Plano Real, e da expansão da renda, do crédito e dos programas sociais do governo Lula.

**Evolução da Oferta e da Capacidade Ociosa**

Associadas às variações no PIB (e à formação de expectativas com relação a essas variações) também estão as variações na oferta do setor, usualmente medidas por meio de variações na quantidade de assentos-quilômetros oferecidos (ASK). A Figura 40 abaixo apresenta a evolução das quantidades no setor, exibindo os indicadores ASK (bilhões de assento-quilômetro oferecidos), RPK (bilhões de passageiros-quilômetros transportados pagos) e o fator percentual de aproveitamento de vôo (*load factor*) entre 2000 e 2007:

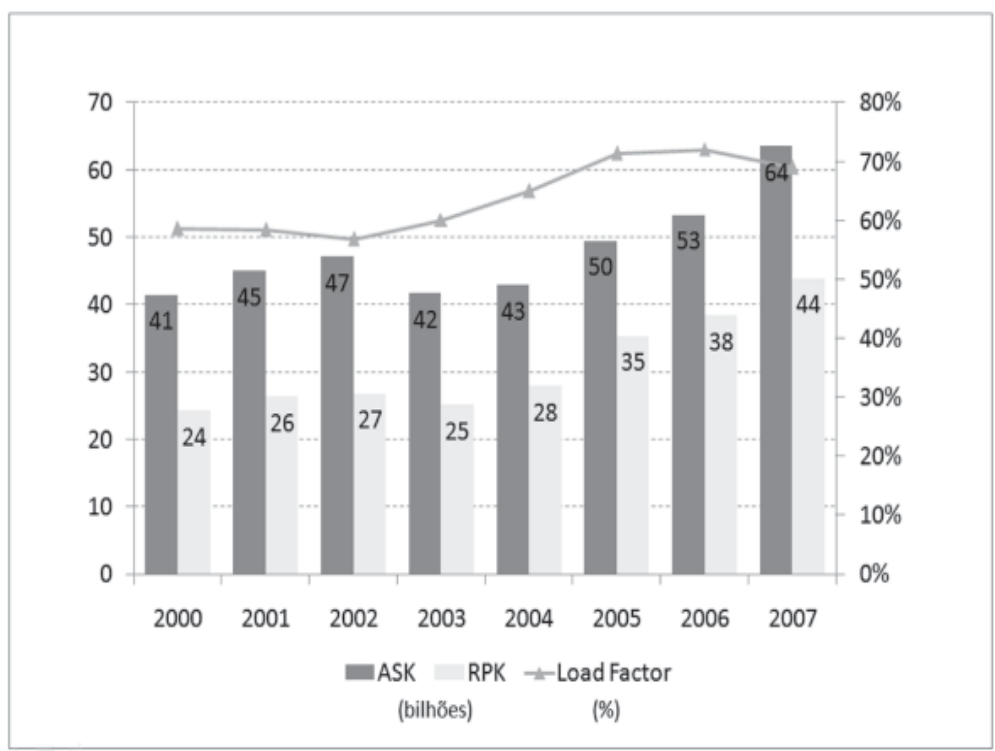

Figura 40 – Evolução de ASK, RPK e Load Factor[32]

Pode-se notar que uma constante do transporte aéreo é, em média, sempre apresenta alguma ociosidade dos vôos. Isso é facilmente constatável dado que RPK é sempre menor que ASK e, por decorrência, o fator de aproveitamento, ou de *load factor*, é sempre menor do que a unidade, um percentual menor que 100. Na Figura 40, constata-se que o excesso de capacidade passou de 17 (= 41-24) em 2000, para 20 ( = 64-44) em 2007. Essa é uma métrica absoluta de excesso de capacidade.

Uma das maiores preocupações das autoridades regulatórias com todo o processo de desregulação do setor dos anos 1990, dizia respeito a um possível descolamento entre oferta e demanda. Essa preocupação está externada, por exemplo, no artigo 193 do Código Brasileiro de Aeronáutica (CBAer), Lei nº 7.565, de 19 de dezembro de 1986. Como visto, entretanto, em regime de operações normais, tem-se que, na situação típica, sempre haverá alguns assentos em cada vôo que partirão vazios. Para se ter uma idéia do que um dado nível absoluto de ociosidade pode significar na operação de uma companhia aérea, pode-se somar todos os assentos-quilômetros (ASK) ociosos em todo o ano de operação. Se dividirmos esses ASK ociosos pelo tamanho médio das aeronaves e pela etapa de vôo média, obteremos um indicador do número de vôos que partiriam vazios em um ano caso todos os assentos ociosos pudessem ser neles agrupados. Essa conta resultaria, somente para o transporte aéreo doméstico no Brasil, em centenas de milhares de vôos vazios por ano! A Figura 41 permite visualizar esses números e inferir que o setor apresentou uma queda expressiva em sua ociosidade a partir de 2003 – ano do *code share* Varig-TAM.

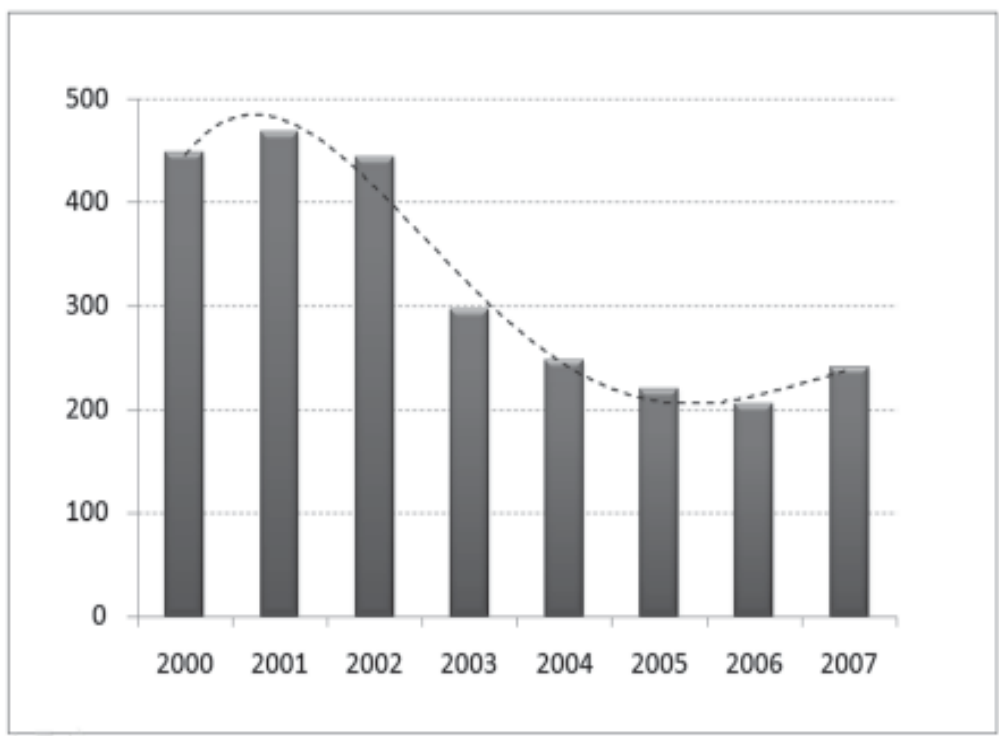

Figura 41 – Evolução do Número de Assentos Vazios
(em Milhares de Vôos de Tamanho Médio na Etapa Média) [33]

O excesso absoluto de capacidade tende a ser enganoso, entretanto. O número de assentos vazios tende a crescer proporcionalmente com o tamanho do mercado, o que dificulta análise do excesso absoluto de capacidade. Um indicador mais adequado de ociosidade de assentos é o excesso de capacidade relativa, ECR, que nada mais é do que o complemento do fator de aproveitamento médio (1 – aproveitamento), e que tende a ser igualmente ilustrativo. A evolução deste indicador está apresentada no gráfico abaixo:

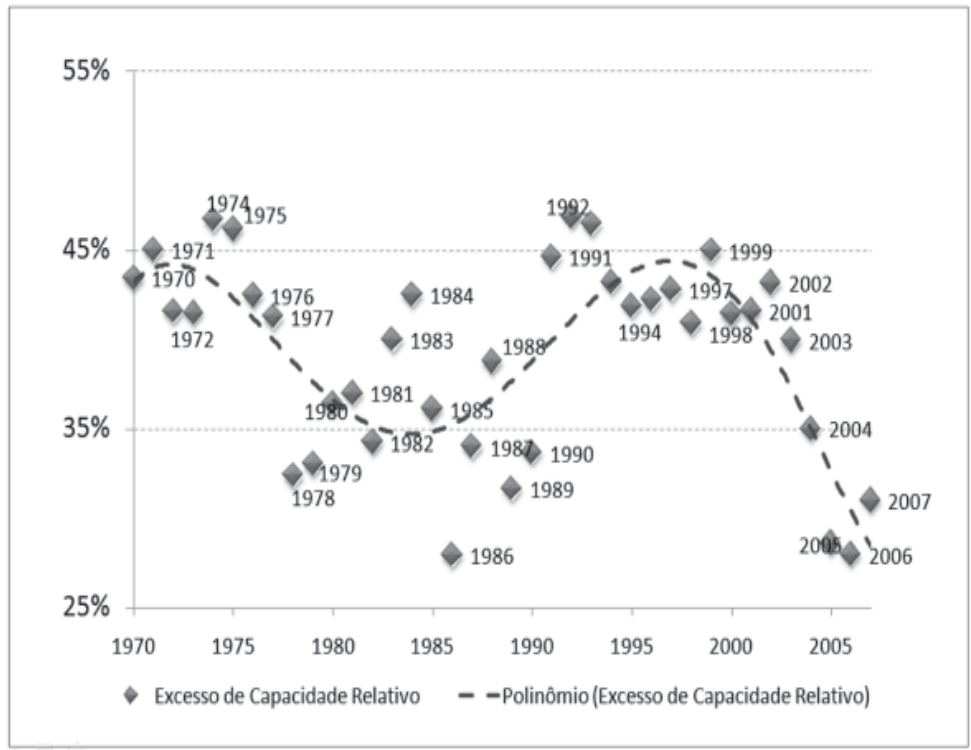

Figura 42 – Evolução do Excesso de Capacidade Relativa (ECR) da Indústria[34]

Como a Figura 42 permite observar, com a Flexibilização do setor, houve um incremento no ECR, se compararmos com os níveis de ECR que vigiam desde o final da década de 1970; de fato, os níveis de ECR, que estavam em torno de 35% nesse período, passaram a ser superiores a 40% no pós-Flexibilização. Este aumento, entretanto, tornou os níveis de ECR comparáveis à situação que vigia no período do "milagre econômico", e, assim, não há bases para se inferir que o mesmo não seja apenas uma decorrência do próprio ciclo econômico setorial. Assim, tem-se que, não obstante a constatação de que o excesso de capacidade, medido em termos absolutos, tenha sofrido acréscimo relevante na última década, este acréscimo não foi capaz de produzir efeitos relativos, isto é, normalizados pelo tamanho do mercado (ECR), que sejam além daqueles acarretados pela própria dinâmica do crescimento do setor.

### *Pouso: Lado dos Custos – O Querosene de Aviação (QAV)*

O aumento dos preços dos insumos é recorrentemente citado pelas empresas como um fator fundamental no acréscimo de custos e conseqüente perda de competitividade e repasse em preços. Muitos desses aumentos podem não estar ligados diretamente ao insumo em si, mas sim, ao seu transporte até o local da produção. Fontes de insumos importantes podem se situar em alguns poucos pontos específicos do país, dificultando seu transporte para outras regiões e/ou encarecendo-o bastante. O caso do querosene de aviação (QAV) no Brasil é um exemplo típico que custos logísticos podem influenciar decisivamente na competitividade de empresas e no bem-estar dos consumidores. O combustível representa uma parcela significativa dos custos das companhias de aviação e os reajustes afetam principalmente as companhias de pequeno porte. De acordo com recentes estudos realizados pela IATA, o consumo de combustível representa o segundo maior componente dos custos diretos operacionais das empresas aéreas, atrás apenas dos custos referentes à mão de obra. Estima-se que a participação do consumo de combustível esteja na faixa de 20% a 40% dos custos diretos operacionais totais nas empresas aéreas. Os centros de distribuição do combustível no País existem em número pequeno e são localizados de forma muito concentrada ao longo do território nacional. Isso acaba agravando os impactos dos aumentos de preços, fazendo com que empresas de aviação regional de diferentes regiões tenham sensibilidades a variações no custo desse insumo distintas.

No estudo de "Análise de Custos Logísticos do Transporte Aéreo Regional" (Fregnani, Ferreira, Griebeler e Oliveira, 2007), os autores realizam, por meio de modelos econométricos, a estimação de uma função custo Cobb-Douglas, para verificar se empresas brasileiras de aviação regional que atuam em regiões diversas, algumas com maior proximidades dos centros de distribuição de QAV em relação a outras, têm diferentes sensibilidades nos seus custos a variações no preço desse insumo.

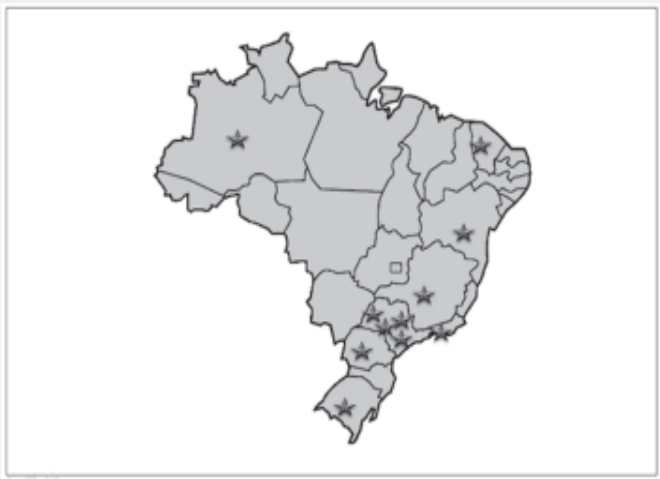

Figura 43 – Refinarias Produtoras de QAV no País (Fonte: Petrobrás)

O que se buscou testar no trabalho é se uma companhia que atua apenas em uma região e, por isso, tem poucas opções de fornecedores de QAV, é mais sensível ao preço quando comparada a uma outra que atua em diversas localidades e, por conseguinte, tem maiores opções de compra. Para tal, um par de empresas foi analisado: uma que tem atuação restrita à região Norte, a Rico; e outra de atuação mais ampla, englobando cidades das cinco regiões nacionais, a Passaredo. Os resultados comprovam a expectativa inicial. A primeira apresentou uma sensibilidade QAV-Custo de 0,26, valor 63% superior ao encontrado para a Passaredo (0,16). Um resultado condizente com a realidade da região Norte do Brasil, onde há apenas uma refinaria de QAV. O fato de a Petrobrás ser responsável por 85% da produção do QAV comercializado no país e ter de importar os 15% restantes e destiná-los especialmente às regiões Norte e Nordeste somente vem a embasar os resultados encontrados.

# Demanda e Oferta de Transporte Aéreo: Análise de Três Estudos de Caso

## Decolagem: Compreendendo o Setor por Meio de Estudos

A proposta deste capítulo é proporcionar um maior entendimento dos problemas relacionados às políticas públicas para o transporte aéreo a partir da análise de três Estudos de Caso. Esses estudos retratam pesquisas realizadas no âmbito do Núcleo de Economia dos Transportes, Antitruste e Regulação (NECTAR), do Instituto Tecnológico de Aeronáutica. Os Estudos serão apresentados de maneira sintética, sendo que aspectos específicos da metodologia utilizada podem ser encontrados com detalhamento nos respectivos textos originais.

No primeiro Estudo de Caso, será abordado o tema da relação entre transporte aéreo e turismo, com uma análise do perfil dos turistas que viajam pelo Brasil. No segundo Estudo de Caso, será realizado um diagnóstico da cobertura do transporte aéreo regular doméstico ao longo do território nacional, sendo analisados indicadores sintéticos inéditos de microrregiões cobertas, e as respectivas coberturas de PIB por setor da economia. No terceiro e último Estudo de Caso, serão avaliados os determinantes dos preços das passagens aéreas no País. Para conferir um maior dinamismo à leitura do capítulo, nos três estudos apresentados, serão dadas respostas às seguintes questões: "*O que foi feito no Estudo?*", "*Qual a relevância do Estudo?*", "*Quais os principais resultados do Estudo?*" e, por fim, "*Quais as recomendações de políticas públicas obteníveis a partir deste Estudo?*".

## Estudo de Caso 1: Turistas que Viajam pelo Brasil

Durante o ano de 2006, uma equipe de pesquisadores coordenada pela Professora Lucia Helena Salgado, do IPEA, realizou uma série de estudos sobre oferta e demanda de bens turísticos – denominada "Organização Industrial do Turismo"[35]. Esses estudos buscaram fazer um apanhado da teoria e metodologia econômicas ao objeto Indústria do Turismo. Um desses estudos foi realizado no âmbito do Núcleo de Economia dos Transportes, Antitruste e Regulação (NECTAR), e resultou no artigo "Determinantes da Escolha por Tipo de Viagem e por Destinos de Viagens Turísticas no Brasil" (Oliveira e Vassallo, 2007), que constituirá o estudo de caso desta seção.

## O que foi feito no Estudo

Os pesquisadores do NECTAR estudaram o comportamento de uma amostra de viajantes no País, tendo por base de dados de recente pesquisa com questionários junto a famílias realizada pela Fundação Instituto de Pesquisas Econômicas (FIPE) da Universidade de São Paulo[36]. Utilizaram uma metodologia com uso de modelos econométricos de escolha discreta com o objetivo de identificar nos dados os determinantes da escolha por tipo de viagem, classificadas como *viagens rotineiras*, *viagens de excursão* (ou *viagens sem pernoite*), *internacionais* e *domésticas* (com pelo menos um pernoite no destino). As informações de viagens foram cruzadas nos modelos com informações dos entrevistados referentes a aspectos como renda, hábitos de viagem, estilo de vida, de forma a se apontar os fatores dos indivíduos que mais alavancam turismo. Adicionalmente, realizaram um estudo de apontamento dos fatores de competitividade das localidades turísticas, em termos de captura de *market share* do total de turistas entrevistados. Utilizaram, para isso, atributos turísticos das cidades encontrados no Guia 4 Rodas *On Line*: o número de atrações, restaurantes e hotéis, de forma a identificar estatisticamente os determinantes da atratividade de um município em face à competição com os demais municípios brasileiros.

## Relevância do Estudo

Com base nos modelos estimados por meio de uma abordagem Teorométrica – isto é, de modelos econométricos aplicados ao Turismo –, foi possível melhor compreender o perfil do turista brasileiro e antecipar qual o impacto que políticas públicas visando a indução do setor teriam na demanda por turismo no País. Quanto maiores forem as habilidades das autoridades em prever com acurácia e agilidade os efeitos de medidas de políticas na demanda por turismo, maior a probabilidade de que medidas eficazes e bem-sucedidas serão implementadas a contento. Adicionalmente, os diversos municípios com potencial turístico ao longo do território nacional têm, neste estudo, um instrumento efetivo de mapeamento de seu poder competitivo na atração de turistas domésticos. Análises dessa natureza são importantes no sentido de propiciar a compreensão das vantagens e desvantagens competitivas das localidades, na disputa pelos turistas domésticos, bem como possibilita projeções de necessidades de investimentos para suprir carências nessa disputa.

## Principais Resultados Deste Estudo

Em primeiro lugar, o estudo chegou à conclusão de que quanto maior a idade do indivíduo, menor a propensão a viajar. Estes resultados são observados sistematicamente tanto quando se analisa as viagens em geral, quando se enfoca as viagens sem pernoite e as viagens domésticas. Eles indicam que *as viagens turísticas (as "viagens domésticas") são mais intensas dentre os mais jovens*, independentemente das demais variáveis relevantes, como renda, posição familiar, sexo, hábitos, etc. Esses resultados são bastante intuitivos, dado que os jovens, em geral, estão à procura de experiências, além de estarem mais abertos a novas descobertas – o que é natural do processo de amadurecimento e aprendizagem do ser humano. Adicionalmente, tem-se como possível explicação que os mais jovens são muito mais influenciados pelos atrativos do "Novo Turismo" – aquele não considerado de massa ou rotineiro – e que melhor caracteriza a realidade atual do setor; em oposição, as gerações anteriores formaram-se sob a cultura de formas anteriores de organização turística – sobretudo o turismo de massa – que se encontra em trajetória declinante no mundo. É, portanto, de certa forma, natural observar as novas gerações tendo maior propensão a consumir um bem (turismo) que é comercializado de uma forma muito distinta de décadas atrás.

Outro resultado foi que as viagens sem pernoite são também mais fortemente preferidas pelas novas gerações – o que pode ser explicado pela maior energia e disposição a enfrentar percursos até os destinos. As exceções a esse fenômeno de maior propensão a viajar dos mais jovens são as viagens internacionais e as rotineiras com pernoite. Nestas, a idade em geral mostrou não ser atributo determinante na decisão de viajar. Entretanto, a faixa etária entre 51-65 anos mostrou-se mais propensa a viagens internacionais que as demais faixas etárias. Ou seja, se compararmos uma pessoa de 20, 30 ou 40 anos com uma de 60 anos, de mesma classe social (faixa de renda), tem-se que, estatisticamente, a segunda apresentará maior propensão a fazer viagens internacionais.

Em terceiro lugar, temos os resultados com relação à renda. Considerando como *caso base os indivíduos de classe de renda mais alta, pode-se notar que o fator renda exerce forte influência sobre a propensão a viajar, sobretudo com relação* às viagens internacionais. Isto acontece por conta do maior valor monetário associado a estas viagens, fruto da restrição orçamentária do consumidor. Em todos os tipos de viagens, percebe-se que, quanto maior a renda, maior a propensão a viajar e, portanto, tem-se que são todos considerados, do ponto de vista da teoria microeconômica, como bens normais (não-inferiores). Nota-se também que quanto maior a parcela de gastos destinada ao lazer pelo indivíduo/família, maior a propensão a viajar.

*"Bem Normal" é um conceito inventado pelos economistas para descrever um produto ou serviço que tem a sua demanda aumentada quando a renda de um indivíduo (ou da economia como um todo) aumenta. Ou seja, ele é mais procurado por indivíduos ou países com renda superior que a média. O transporte aéreo é nitidamente um "Bem Normal" dado que o fator renda é, na maioria dos casos, um fator restritivo nas viagens dos indivíduos e famílias, e é de se esperar que países com rendas per capta maiores apresentem maior pujância do setor aéreo do que países com renda per capta menor. Em contraste, os "bens inferiores" são aqueles preteridos na medida em que a renda cresce. Por exemplo, na medida em que a renda de um indivíduo ou família cresce, a demanda pelo sistema público de transporte (ônibus urbano, por exemplo), cai.*

Os resultados referentes às experiências dos turistas com outros tipos de viagens mostram que *experiências com outros tipos de viagens são fortes indutores de viagens*; isto acontece fortemente com as viagens domésticas ("turísticas"): tanto viagens sem pernoite como viagens ao exterior são indutores de viagens domésticas, o que demonstra que os viajantes são impulsionados pela experiência adquirida em outras viagens para realizar turismo doméstico; Isso é indicativo de um processo de aprendizado do viajante, que vai aumentando sua propensão ao consumo do bem turístico na medida em que exerce o consumo de outros tipos de viagens. Essa constatação vai de acordo com o que seria esperado, uma vez que a realização de viagens longas está muito atrelada à superação do medo do desconhecido. Neste sentido uma pessoa com constantes experiências de viagens, principalmente internacionais, se sentirá mais propensa a realizar viagens domésticas.

Com relação ao estilo de vida do viajante, o estudo mostra que alguns fatores condicionam fortemente a sua propensão a viajar. Em geral, pessoas cujo primeiro lazer, declarado na pesquisa, é "descansar" ou "assistir TV", têm menos propensão a viajar do que pessoas cujo primeiro lazer é "teatro ou cinema" e "esportes". Pessoas que têm na *internet* como fonte primeira de lazer não estão nem mais nem menos propensas a viajar. *Este resultado indica como as pessoas com maior vida social e menos introvertidas*

*(esportes e cinema) têm maior propensão a realizar viagens.*

Adicionalmente, em uma análise desagregada por faixa etária, visando distinguir o comportamento do "Novo Turista" do comportamento do "Velho Turista", foi realizada. *Obteve-se o resultado de que os "Novos Turistas" (aqueles com faixa etária mais baixa) têm maior propensão a viajar, tudo o mais mantido constante.* Aqueles turistas são menos influenciados por restrições relacionadas com a renda, e são fortemente impulsionados às viagens quando o primeiro lazer é "teatro e cinema". Inclusive, tem-se que estes veículos de lazer podem ser, assim, considerados alvos potenciais para propaganda de destinos turísticos. Já os "Velhos Turistas", ou aqueles que tiveram, no passado, alguma influência do Turismo de Massa, tem menor propensão a viajar, são mais influenciados pela restrição orçamentária, e são menos influenciados por formas de lazer cultural, como "teatro e/ou cinema".

> *Os "Novos Turistas" são segmentos de consumidores com papel muito importante para o transporte aéreo e para o turismo. Eles representam as novas gerações de consumidores, apresentam maior familiaridade com a internet, estão acostumados a pesquisar preços nos sites das companhias aéreas e são naturalmente abertos a novas propostas de modelos de negócios das empresas. É da maior disposição a viajar dos "Novos Turistas" que dependerá o crescimento do transporte aéreo acima do seu crescimento histórico.*

No estudo referente à competição por *market share* entre as localidades turísticas, chegou-se a um conjunto de resultados importantes. *Tanto a variável "Número de Atrações" quanto a de "Número de Hotéis", que são contabilizadas pelo Guia 4 Rodas, mostram-se significativos instrumentos para alavancar a fatia de mercado turístico dos municípios brasileiros.* Igualmente importante, tem-se que o modelo permite inferir que o efeito da presença de um hotel adicional é aproximadamente 15% superior ao efeito da adição de uma atração adicional em uma dada localidade, independente da região. Esse fenômeno é indicativo de que a maior carência na alavancagem do turismo no País ainda está relacionada à infra-estrutura disponível ao turista – nesse caso representada pela rede hoteleira.

Para municípios da região Sudeste, os dados mostram que o que importa na atração de turistas, é o número de atrações; por outro lado, para municípios da região Nordeste, a variável fundamental em termos de atratividade é a relativa a Hotéis. Esse fato parece ser novamente indicativo de que a infra-estrutura da região Sudeste já é suficiente, e que o principal elemento alavancador de turistas domésticos é o número de atrações da localidade; por outro lado, no que diz respeito à região Nordeste, tem-se que, dado o alto Número de Atrações por localidade, tem-se que o fator diferenciador na atração de turistas é a infra-estrutura sendo, por exemplo, o diferencial alavancador a infra-estrutura hoteleira do local. Isto pode ser indicativo de que mais investimentos em infra-estrutura turísticas são necessários para as localidades do Nordeste se tornarem melhor posicionadas na concorrência por turistas domésticos. Para as localidades do Centro-Oeste e Sul, tem-se que investimentos em infra-estrutura de hotéis também têm sido importantes para elevar a posição dos seus destinos no mercado nacional.

## Recomendações de Políticas Públicas a Partir Deste Estudo

Dado que há maior propensão a viagens turísticas por parte das novas gerações, tem-se, como recomendação, o fomento do turismo nas gerações anteriores a estas. Por exemplo, um maior investimento em informação a esse tipo de consumidor representativo do "Velho Turismo" poderia ser implementado, dado que os aspectos tecnológicos do "Novo Turismo" ainda são, em parte, desconhecidos nessas faixas etárias.

Recomendou-se às autoridades e mesmo ao *trade*, um foco especial no chamado viajante "*day tripper*", que não é usualmente considerado dentre aqueles nos quais pode-se gerar demanda turística no Brasil. Em outros Países, esse tipo de viajante pode ser origem de recursos importantes para a economia de certas localidades, sobretudo se próximas a grandes centros urbanos. Como recomendação de políticas, tem-se aqui que o investimento em maior preparo das localidades para criar e investir na qualidade de suas atrações, preparar uma infra-estrutura e um conjunto de serviços de apoio ao turista, e, sobretudo, divulgar adequadamente a informação com relação a suas atrações nos centros de maior densidade populacional mais próximos.

A inferência com relação à maior intensidade das viagens turísticas dos mais jovens é extremamente importante do ponto de vista das políticas de fomento do turismo, e, sobretudo, da sua potencial interação com as políticas educacionais do País; isto porque, dada esta constatação, pode-se idealizar mecanismos de integração do turismo com as atividades educacionais (incentivando-se, por exemplo, viagens escolares, excursões, congressos da juventude universitária, dentre outros), dado que esta é a parcela da população mais propensa ao turismo. O resultado de políticas facilitadoras do turismo nessa faixa etária será o de melhor formação do jovem em idade escolar, com ampliação de sua bagagem cultural, enriquecimento da qualidade de suas experiências de vida que resulte em maior amadurecimento para o próprio mercado de trabalho.

A constatação de que a faixa etária a partir dos 50 anos mostrou-se mais propensa a viagens internacionais, permite um melhor entendimento do comportamento da balança de serviços (viagens internacionais), na medida em que aumenta a percentagem da população que se enquadra nas faixas etárias mais elevadas. Adicionalmente, oferece a importante indicação de que, se o governo deseja ampliar a inserção internacional do País, precisa estabelecer políticas de incentivo às viagens internacionais dos mais jovens, como forma dos mesmos adquirirem experiências, culturas e formas de pensar de outros países. A Inglaterra é um exemplo de país onde a população jovem (sobretudo a masculina) tem elevada propensão a viagens internacionais, o que contribui acentuadamente com a capacidade daquele país em formar mão-de-obra apta a trabalhar em um ambiente competitivo globalizado.

Sendo, do ponto de vista microeconômico, um "bem normal", tem-se que o crescimento econômico que seja indutor de renda e de uma melhor distribuição da mesma, será altamente benéfico para a indução do turismo. A recomendação de políticas públicas é a de que as autoridades devem ter um posicionamento atuante com relação aos aspectos da política macroeconômica, de forma a pleitear que políticas comprometidas com um ambiente de estabilidade de regras e dos negócios de forma a propiciar investimentos públicos e privados, bem como políticas fiscal e monetária adequadas sejam implementadas.

Dado que existe um processo de aprendizado do viajante, tem-se que é o acúmulo de experiências que irá aumentar ou diminuir a propensão ao turismo na economia. Isto significa que políticas corporativas, por parte das operadoras turísticas, hotéis, companhias aéreas, etc., visando o atendimento ao cliente no pós-vendas, com o desenvolvimento de questionários de qualidade e medidas de compensação no caso de insatisfação, são extremamente importantes. Essas políticas também deveriam ser perseguidas pelos municípios com intenção de ampliar seu *market share* turístico.

Dado que cidadãos com constantes experiências de viagens, principalmente internacionais, sentem-se mais propensos a realizar viagens domésticas, tem-se, como recomendação, que o País precisa, portanto, desenvolver um ambiente propício às viagens, sejam elas turísticas ou não, de longo ou curto percurso, etc. Ao se promover um estímulo às viagens, as autoridades estarão, ao mesmo tempo, alavancando nos indivíduos-cidadãos o gosto pelo consumo do bem turístico, na medida em que há um processo de aprendizado em cada viagem realizada. Condições para que um ambiente como este se efetive passam pela melhoria de estradas, terminais rodoviários, ampliação de aeroportos, fortalecimento das companhias aéreas, operadoras turísticas, etc. Muitas dessas condições estão nas mãos das autoridades responsáveis, sendo passíveis de políticas públicas;

Tanto as autoridades quanto o *trade* turístico devem investir na realização de um *marketing* mais agressivo tentando mudar o hábito de pessoas que têm na televisão e na *internet* suas principais atividades de lazer, dados que estes são os consumidores potenciais que ainda apresentam baixa demanda pelo turismo. Por outro lado, tem-se ser mais eficaz definir campanhas publicitárias voltadas para um marketing mais orientado aos atributos dos destinos (e, portanto, permitindo a comparabilidade e a competição entre os mesmos) aos consumidores que tem um estilo de vida mais ativo;

Investimentos em campanhas publicitárias e *marketing* devem ser cuidadosamente orientados de acordo com o segmento de consumidores a ser considerado, dado que as especificidades dos mesmos irá com certeza condicionar os resultados em termos de alavancagem da demanda tanto por viagens, quanto pelos destinos em si.

Constatou-se que a maior carência na alavancagem do turismo no País ainda está relacionada à infra-estrutura disponível ao turista – no caso do estudo, a rede hoteleira. Ou seja, um incremento na magnitude da rede hoteleira de um município tem um efeito relativamente maior sobre o seu poder de captura da massa de turistas domésticos, dada a relativa escassez de infra-estrutura pela grande maioria dos municípios ao longo do território nacional.

## Estudo de Caso 2: Cobertura do Território Nacional

Uma análise da malha aeroportuária brasileira no período recente é de extrema relevância dado que pode proporcionar análises quanto à necessidade de se ter políticas públicas para o transporte aéreo regional. O presente Estudo de Caso foi retirado do texto "Constituição do Marco Regulatório Para o Mercado Brasileiro de Aviação Regional" (Oliveira e Salgado, 2008), produzido no âmbito do NECTAR, por solicitação da Associação Brasileira de Empresas de Transporte Aéreo Regional, ABETAR.

## O que foi feito no Estudo

Esse estudo buscou elaborar um diagnóstico, um "raio-x" da situação geográfica atual da aviação regular no país, analisando os aspectos de qualidade de cobertura ao longo do território nacional. Por cobertura aérea entende-se o provimento do serviço de forma a "atender" o espaço geográfico brasileiro, isto é, no sentido de disponibilizar os serviços de transporte aéreo para o máximo de localidades possível, dadas as restrições de demanda, oferta e tecnologia. Foram compiladas estatísticas públicas de provimento de serviços aéreos – basicamente, freqüências de vôos regulares a partir do Relatório HOTRAN, da Agência Nacional de Aviação Civil. Essas estatísticas de número de aeroportos operados, obtidas nos registros do HOTRAN, foram utilizadas na criação de indicadores sintéticos inéditos de cobertura como o número de microrregiões e microrregiões cobertas, o PIB coberto dos setores da agricultura, da indústria, dos serviços e PIB total, bem como a população coberta.

**Relevância do Estudo**

Essa análise visa identificar com maior clareza os problemas estruturais no setor no que tange à cobertura do transporte aéreo ao longo do território nacional e, mais especificamente, como ele vem desempenhando o papel de desenvolvimento das localidades pelo País afora. Permite, portanto, um mapeamento das oportunidades da aviação regional brasileira.

**Principais Resultados Deste Estudo**

A Figura 44 a seguir apresenta uma das imagens produzidas pelo Estudo de Caso, acerca da malha aeroportuária brasileira no período entre 1998 e 2008. Nela, é apresentado o mapa do Brasil e a localização dos aeroportos que tiveram alguma movimentação com vôos regulares no período.

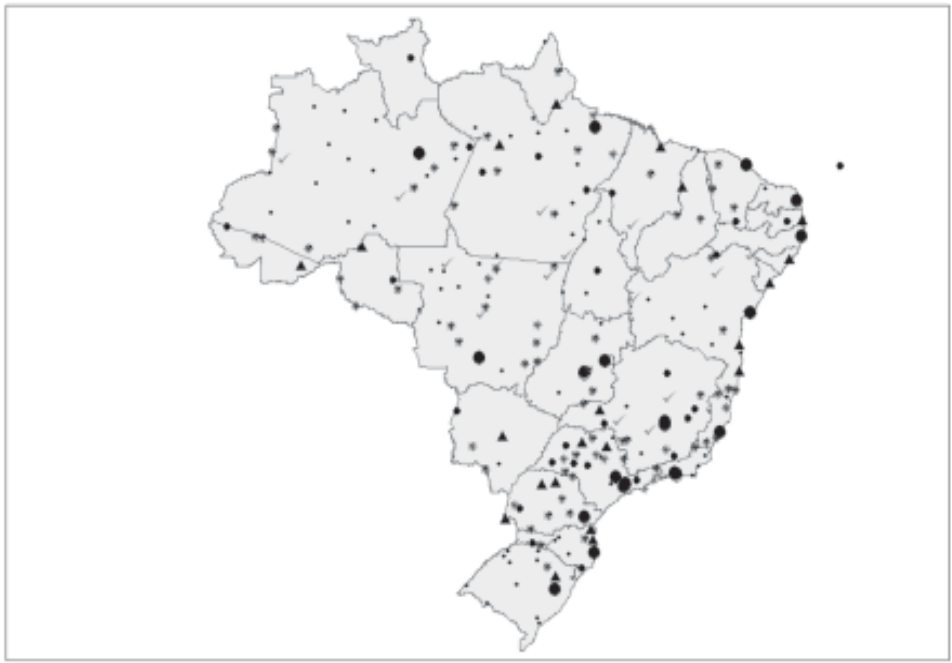

Legenda: ● = Hub Grande; ▲ = Hub Médio; ◉ = Hub Pequeno; • = Aeroporto Local
* = Operações Encerradas ou Frustradas ✓ = Operações Iniciadas
Figura 44 – Operações Aeroportuárias 1998-2008: Brasil

Para efeitos desta análise, foi necessário efetuar uma padronização na classificação de aeroportos, a partir da constituição de um *ranking* dos mesmos. Foi adotada, com essa finalidade, as definições de aeroportos "hub grande", "hub médio" e "hub pequeno" adotada pela *Federal Aviation Administration*, FAA, dos Estados Unidos.

> *Segundo a Federal Aviation Administration (FAA), para um aeroporto ser considerado "hub grande", ele deve movimentar mais do que 1% dos embarques e desembarques de passageiros de um País. Os aeroportos "hub médios" da FAA são aqueles que movimentam entre 0,25% e 1% do total dos passageiros, enquanto os "hubs pequenos" são aqueles que movimentam entre 0,05% e 0,25%. Abaixo dos 0,05% de participação da movimentação de passageiros no total do sistema aéreo, o aeroporto é considerado "não-hub"; para efeito das análises aqui empreendidas, denominaremos esses aeroportos de "locais". Essa classificação de hub pode ser encontrada no documento "Introduction to the Airport Improvement Program" da FAA, de novembro de 2002. Ela se baseia na característica dos aeroportos hubs de serem "grandes terminais", com grande movimento de embarque e desembarque de passageiros. Isso nem sempre é verdade, dado que o conceito de hub está mais relacionado com o fato de ser um concentrador de conexões do que com o seu tamanho e market share de movimentação propriamente ditos.*

Utilizou-se, para efeito do mapa da Figura 44, as seguintes as legendas: ● para aeroportos classificados como "hubs grandes", ▲ para "hubs médios", ● para "hubs pequenos" e • para "aeroportos locais". A Figura 44 também apresenta os aeroportos que apresentaram início de operações (✓), aqueles que tiveram operação encerrada bem como aqueles que tiveram operação regular "frustrada", ou seja, iniciada mas encerrada no período (*).

Pode-se perceber, por meio da Figura 44, como o país apresentou queda na cobertura aérea no período entre 1998 e 2008. De fato, houve encerramento de operações em todas as regiões do país, e em número superior ao número de operações iniciadas.

A Tabela 11 a seguir apresenta uma síntese das informações de cobertura ao longo do território nacional. Nela, é possível visualizar não apenas a variação no número de aeroportos operados, mas também de importantes indicadores sintéticos construídos a partir da metodologia de divisão territorial proposta pelo Instituto Brasileiro de Geografia e Estatística, IBGE: número de microrregiões, de municípios, PIB (agricultura, indústria e serviços) e população coberta pelo transporte aéreo regular.

Tabela 11 – Resumo das Operações Aeroportuárias 1998-2008: Brasil[37]

| Cobertura | Pre-Liberalização | Pos-Liberalização | Variação | Var% |
|---|---|---|---|---|
| Aeroportos Operados | 199 | 155 | -44 | -22.1% |
| Microrregiões Cobertas | 166 | 131 | -35 | -21.1% |
| Municípios Cobertos | 1,821 | 1,437 | -384 | -21.1% |
| PIB Agricultura Coberto (R$ bilhões 2005) | 44.6 | 36.7 | -7.9 | -17.7% |
| PIB Indústria Coberto (R$ bilhões 2005) | 391.1 | 365.6 | -25.5 | -6.5% |
| PIB Serviços Coberto (R$ bilhões 2005) | 893.0 | 851.9 | -41.1 | -4.6% |
| PIB Total Coberto (R$ bilhões 2005) | 1,570.8 | 1,486.9 | -83.8 | -5.3% |
| População Coberta (milhões 2005) | 113.3 | 104.7 | -8.6 | -7.6% |

A metodologia de cálculo dos indicadores presentes na Tabela 11 pode ser descrita pelos seguintes passos:

> ✈ Os aeroportos operados pela aviação regular, constantes do sistema HOTRAN do antigo Departamento de Aviação Civil, DAC, e do atual regulador, a Agência Nacional de Aviação Civil, ANAC, são identificados a partir de extrações mensais de HOTRAN desde 1998;

> ✈ Para cada aeroporto constante no HOTRAN é atribuída uma microrregião do IBGE a que pertence, de acordo com sua localização territorial;

✦ O número de microrregiões "cobertas" pelo transporte aéreo regular é igual ao número de microrregiões em que foram observados aeroportos com operação regular;

✦ O número de municípios "cobertos" pelo transporte aéreo regular é igual ao número de municípios pertencentes à microrregião onde foram observados aeroportos com operação regular;

✦ O PIB "coberto" (agricultura, indústria e serviços) e a população "coberta" pelo transporte aéreo regular é igual ao PIB e a população das microrregiões onde foram observados aeroportos com operação regular.

Na Tabela 11, é possível, portanto, observar a variação de diversos indicadores sintéticos produzidos para medir as diferenças de cobertura entre o "Pré-Liberalização" – período aqui definido como os anos anteriores à Terceira Rodada de Liberalização de 2001 (janeiro de 1998 a janeiro de 2000) – e o "Pós-Liberalização" – os anos mais recentes, englobando o período entre janeiro de 2006 e janeiro de 2008. Essa contraposição entre pré e pós-Liberalização será mantida também nas análises das figuras a seguir, e é de fundamental importância para um maior entendimento de como se processou o ajuste das malhas das companhias aéreas a partir da concessão de maiores graus de liberdade no mercado por parte das autoridades regulatórias brasileiras desde os anos 1990 – como vimos, a "Política de Flexibilização da Aviação Comercial.

As figuras a seguir apresentam a síntese da análise de cobertura geográfica e dos indicadores sócio-econômicos que configuraram o diagnóstico da aviação regular ao longo do transporte aéreo regional.

A Figura 45 apresenta um gráfico com o número de aeroportos operados no pré e pós-liberalização. De fato, houve nítida perda na qualidade da cobertura com a liberalização do setor, conforme apontam quase todos os indicadores levantados. Houve queda no número de aeroportos operados em todas as regiões do país, sendo que a região Norte foi a que perdeu mais cobertura, em termos absolutos      (-13). Ao todo, foram menos quarenta e quatro aeroportos operados em todo o Brasil (queda de 22%).

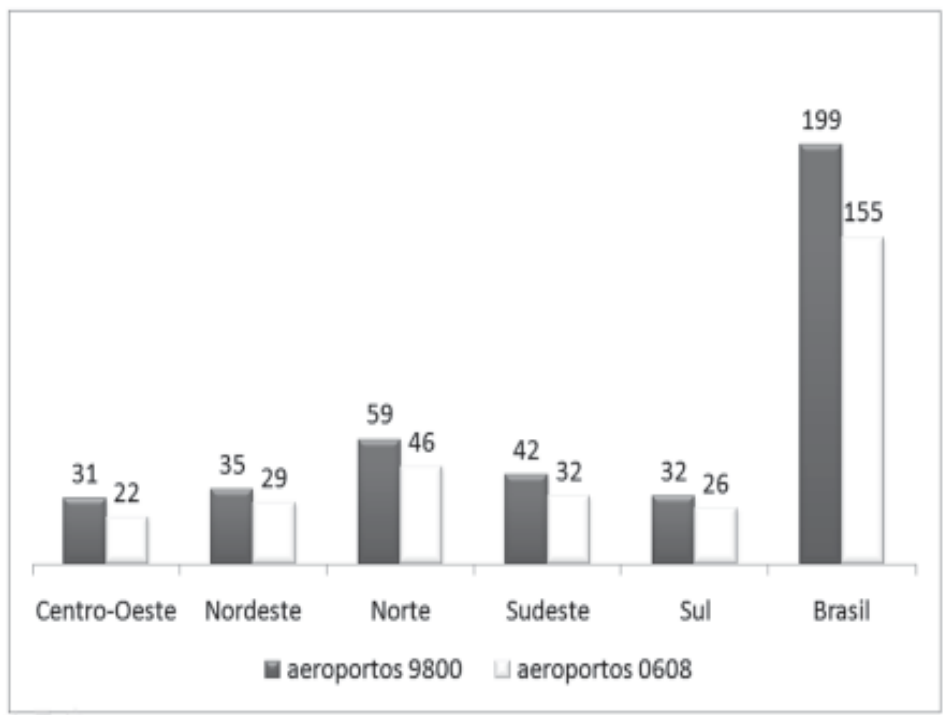

Figura 45 – Aeroportos Operados

A Figura 46 e a Figura 47 a seguir permitem um entendimento de como se efetivou a queda na cobertura do país, ocasionando concentração das operações aéreas nos "hubs grandes" e "hubs médios", em detrimento dos aeroportos locais. Dos quarenta e quatro aeroportos com operações encerradas, temos que a grande maioria (72%) é constituída de "aeroportos locais". De fato, o número de aeroportos locais caiu de 137 para 105 (-32 aeroportos), conforme mostra a Figura 46.

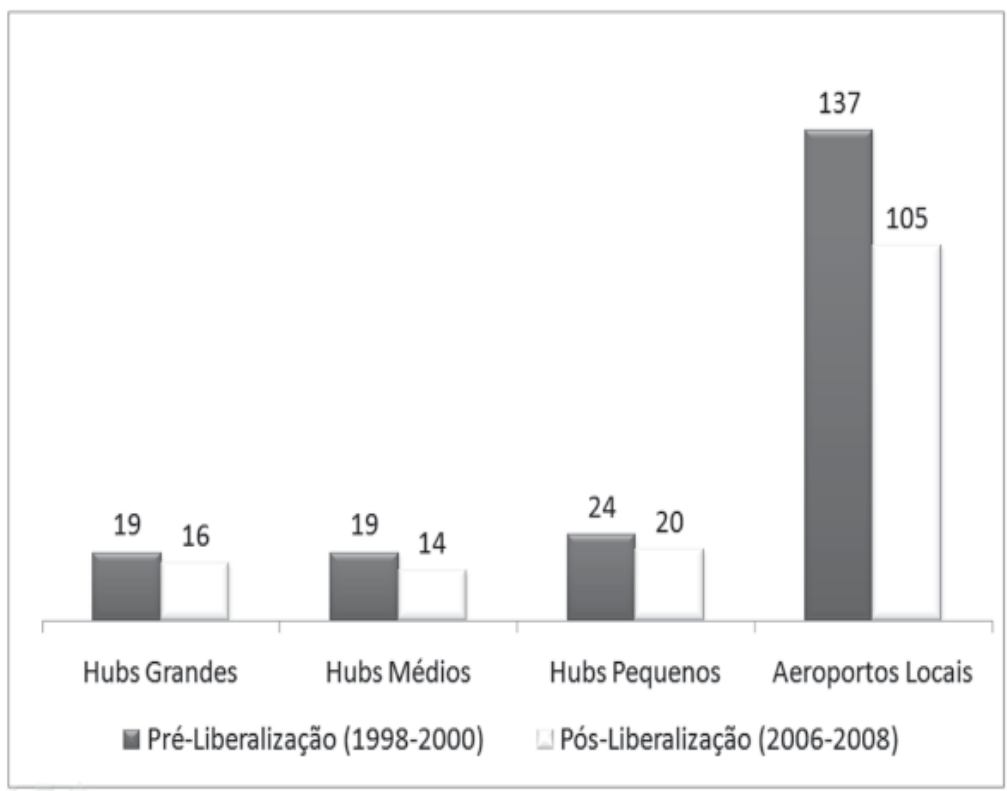

Figura 46 – Aeroportos Operados

A operação dos demais tipos de aeroportos também caiu porque o número de aeroportos com fatia de mercado acima de 1% ("*hub* grande"), de 0,25% ("*hub* médio") e 0,05% ("*hub* pequeno") caiu, dado que os principais aeroportos do País (sobretudo Aeroporto de Congonhas, em São Paulo, e Aeroporto Juscelino Kubitschek, em Brasília) passaram a concentrar maior participação no sistema. De qualquer forma, a queda no número de aeroportos locais foi muito mais expressiva que as demais quedas, e esse fator pode ser atestado pela Figura 47, que apresenta a participação de mercado de cada um dos tipos de aeroportos:

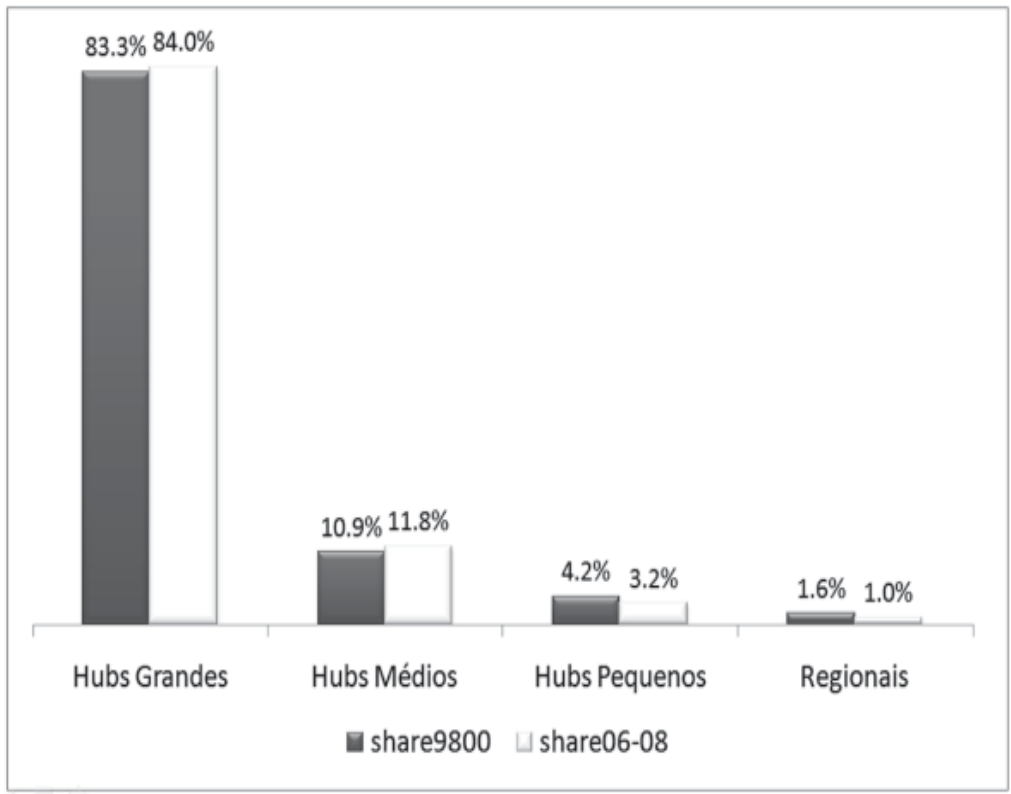

Figura 47 – Aeroportos Operados

Como pode ser observado, enquanto a fatia de mercado de "*hubs* grandes" e "*hubs* médios" aumentou com a liberalização do setor aéreo, a fatia de mercado dos "*hubs* pequenos" e "locais" caiu expressivamente. Os "aeroportos locais" tiveram participação em queda de 37%, caindo de 1,6% para apenas 1% dos movimentos de passageiros no sistema aéreo doméstico brasileiro.

A Figura 48 e a Figura 49 apresentam gráficos com o número de microrregiões cobertos pela aviação regular e o número de municípios dessas regiões, outro tipo de indicador que apresentou queda considerável ao longo de todas as regiões do País.

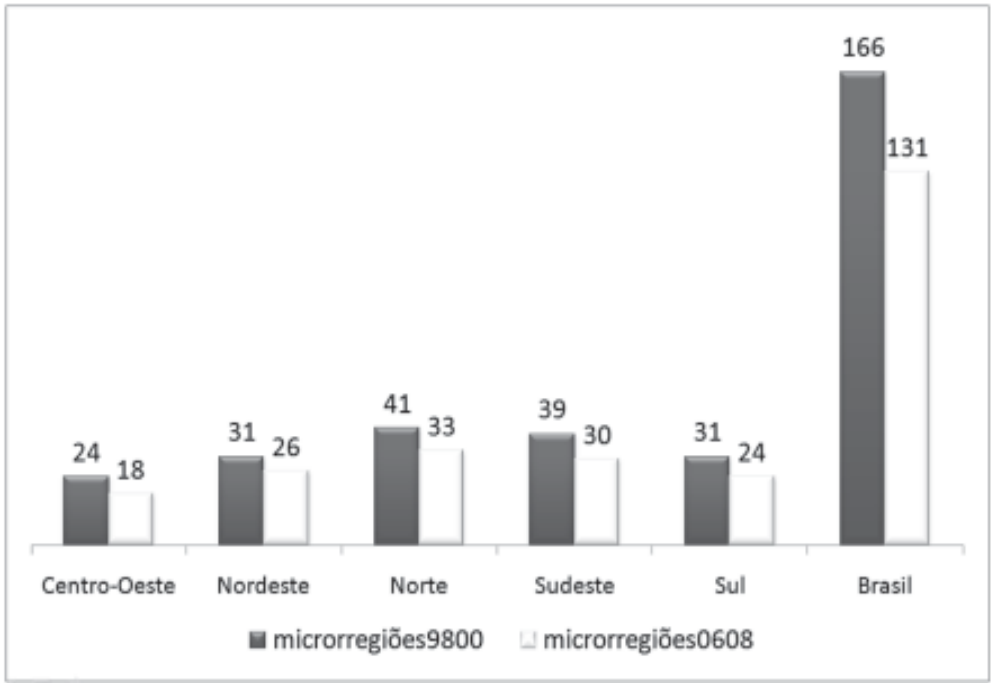

Figura 48 – Microrregiões Cobertas

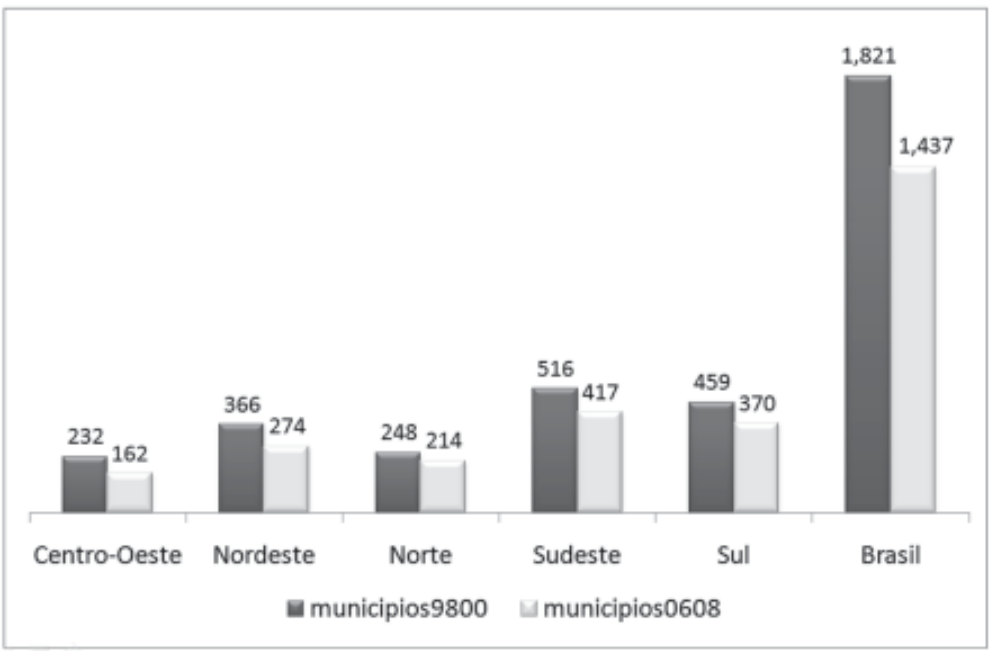

Figura 49 – Municípios Cobertos

## Recomendações de Políticas Públicas a Partir Deste Estudo

Todos os fatores acima apontados no diagnóstico da cobertura aérea da aviação regular no Brasil são conseqüência de uma maior busca pela eficiência das companhias aéreas brasileiras por meio do enxugamento de suas operações. De fato, com a liberalização do setor, as empresas passaram a competir de forma mais intensa, e tiveram que explorar de forma mais eficiente as operações ao longo de suas malhas aéreas. A decorrência mais visível desse fenômeno é a queda no nível e na qualidade da cobertura do transporte aéreo ao longo do território nacional, dado que localidades com operações menos lucrativas são preteridas em prol de localidades com evidências de maior lucratividade.

Por outro lado, entretanto, a queda na cobertura é um indicador de que o transporte aéreo teve seus potenciais de alavancagem do desenvolvimento sócio, econômico e turístico comprometidos, com potencial perda da geração de encadeamentos e *spillovers* advindos da maior cobertura aérea (desenvolvimento regional, crescimento sustentável do país, etc). Do ponto de vista das políticas públicas, a queda na cobertura aérea do país é preocupante, e, uma vez diagnosticada, deve ser tratada com mecanismos de incentivo à operação regional, como concessão de espaço em aeroportos congestionados, vigilância antitruste da concorrência predatória de grandes companhias aéreas e incentivo às alianças entre as grandes companhias aéreas e as empresas regionais.

### Escala de Vôo: O que Determina a Disponibilidade de Assentos e de Vôos?

A demanda por transporte aéreo é bastante sensível a flutuações econômicas. Esse fenômeno também é observado pelo lado da oferta de vôos e do número de assentos disponíveis nos *websites* das companhias aéreas. Variáveis econômicas exercem influência direta sobre as decisões de oferta das empresas aéreas. No curto prazo, choques de custos ou da atividade econômica podem ser respondidos com alterações na alocação de frota, atribuindo-se aeronaves em rotas diferentes, cidades e ligações podem ser acrescentadas ou retiradas numa readequação de malha de vôo, o número de horas voadas por dia pode ser mais ou menos intensivo e até mesmo a velocidade média de cruzeiro pode ser alterada, em função de cálculos referentes ao custo horário de operação, função do custo com combustível e da taxa de depreciação da aeronave, ambos cotados em dólar. No médio e longo prazos, todo o planejamento de frota da empresa aérea pode ser revisto: aeronaves arrendadas podem ter seu contrato renovado ou então serem devolvidas; a encomenda e entrega de novas aeronaves pode ser antecipada, postergada ou cancelada. Em certa medida, o início da operação da Azul Linhas Aéreas em 2008 pôde ser antecipado devido ao aluguel de aeronaves da empresa aérea americana jetBlue que se tornaram excedentes devido ao decréscimo econômico nos EUA.

Um artigo publicado recentemente no *Journal of Air Transport Management* (Bettini e Oliveira, 2008) e que colocou sob análise os 30 aeroportos brasileiros de maior oferta de assentos entre 1997 e 2006 revelou que, no curto prazo, o câmbio impacta a oferta de transporte aéreo de acordo com uma elasticidade de - 0,04, efeito negativo que se explica pela exposição que as empresas aéreas têm a insumos cotados na moeda americana. Por sua vez, por deslocar a demanda por vôos, o PIB acaba por impactar a oferta com elasticidade positiva de 0,47. Ou seja, se o dólar desvaloriza em 10%, a oferta de assentos das companhias aéreas reduz em quase meio ponto percentual. Ao contrário, se o PIB aumenta 10%, a oferta de assentos aumenta 4,7%. Além destes efeitos, nitidamente macroeconômicos, este mesmo estudo obteve estimativas para a elasticidade de fatores concorrenciais: a presença de uma empresa *low cost* em uma ligação estimula a oferta de assentos em 7%, enquanto o acordo de compartilhamento de assentos (*code share*), em vigor entre as duas maiores empresas aéreas nacionais durante parte do período, aparece associado a um decréscimo de 5% na oferta de assentos. Outro fator associado à variabilidade da oferta de assentos é o estado regulatório vigente. No estudo citado, o período que contou com medidas que restringiu a liberdade de empresas aéreas para a importação de novas aeronaves e a expansão desimpedida de rotas, referido como "Re-regulação", surge também associado a uma elasticidade negativa, porém de pequena magnitude, 2%.

A sensibilidade da oferta de transporte aéreo às flutuações econômicas não é privilégio brasileiro. Recentemente, a Northwest Airlines, maior empresa aérea americana desde que anunciou sua fusão com a Delta Air Lines, cancelou o único vôo regular transatlântico que partia do aeroporto de Hartford, Connecticut, ligando esta comunidade à Holanda. Citando os preços exorbitantes do petróleo e do combustível de aviação, a empresa decidiu pela suspensão dos vôos, a despeito da oferta de US$ 650.000 que a empresa aeroportuária e organismos locais ofereceram à aérea como recompensa caso mantivesse suas operações. Segundo estas autoridades, os impactos do fechamento desta ligação são enormes para a economia local.

## Estudo de Caso 3: Preços das Passagens Aéreas

Um dos assuntos mais estudados e discutidos no âmbito do Núcleo de Economia dos Transportes, Antitruste e Regulação (NECTAR) é a precificação das companhias aéreas brasileiras. O Estudo de Caso a seguir aborda essa temática, e também foi extraído do relatório da pesquisa coordenada pela Professora Lucia Helena Salgado, do IPEA, da "Organização Industrial do Turismo" (vide Estudo de Caso 1).

## O que foi feito no Estudo

Uma das principais perguntas a que o presente estudo se propõe a responder diz respeito ao padrão de precificação das companhias aéreas brasileiras no segmento doméstico regular de passageiros. Qual a influência de variáveis macroeconômicas (PIB e taxa de câmbio) nos preços das passagens? Existe diferencial de preços causado por diferenciação de produtos? Existem economias de densidade de tráfego (quanto maior o tráfego, menores os custos operacionais e menores preços)? A precificação varia de acordo com a região do País? Qual a influência da presença de uma companhia aérea baseada no modelo "*Low Cost, Low Fare*"? Fatores turísticos influenciam na precificação? Como os preços se comportam ao longo do ano, ou seja, qual a influência da alta e baixa estação turísticas? Este é o conjunto de perguntas foi respondido à luz de um modelo econométrico de preços, aplicado a dados reais do setor aéreo. O modelo estimado é bastante simples, e o leitor interessado é convidado a se remeter a textos do NECTAR com abordagem mais aprofundada sobre o tema e que chegaram a resultados semelhantes (ex. Oliveira e Huse, 2008).

Com vistas a estudar a precificação no setor aéreo, um conjunto de dados foi obtido junto ao Departamento de Aviação Civil, DAC, e à Agência Nacional de Aviação Civil, ANAC. A maior parte da base contém informações mantidas e gerenciadas pela Divisão de Assuntos Econômicos (SA3) e Divisão de Estatística e Gestão do Sistema BAV/HOTRAN (SA5) do DAC. Estas informações são enviadas pelas companhias aéreas à autoridade regulatória, de acordo com legislação específica, sobretudo a Portaria DAC 447/DGAC, de 13 de maio de 2004, que estabelece as regras de funcionamento do sistema de tarifas aéreas domésticas, e a Norma (Instrução de Aviação Civil) IAC 1505, de 30 de abril de 2000, que tem por objetivo traçar diretrizes para normatizar o envio dos dados estatísticos do tráfego aéreo, prestado pelas empresas brasileiras.

A relação de ligações aéreas sujeitas a acompanhamento econômico, e prevista pela Portaria 447/DGAC, compreende um subconjunto dos possíveis pares de aeroportos entre 29 cidades do território nacional. A relação de cidades é a seguinte: Araçatuba, Bauru, Belém, Belo Horizonte, Brasília, Campinas, Campo Grande, Cuiabá, Curitiba, Florianópolis, Fortaleza, Foz do Iguaçu, Goiânia, Joinville, Londrina, Macapá, Manaus, Marília, Navegantes, Porto Alegre, Porto Seguro, Recife, Ribeirão Preto, Rio de Janeiro, Salvador, São José do Rio Preto, São Paulo, Uberlândia e Vitória. Ao todo são 32 aeroportos com transporte aéreo regular. Com vistas a reduzir o conjunto de mercados passíveis de monitoramento, além de excluir aqueles cuja relevância em termos de densidade de tráfego é reduzida, a portaria explicitamente elenca os pares de aeroportos envolvidos, relacionando, assim, 67 ligações.

A amostra de ligações representada pelos Relatórios Mensais de *Yield* é bastante significativa: apesar de envolver apenas 134 dos 1878 pares de aeroportos direcionais listados no Anuário Estatístico do DAC, Volume I – Dados Estatísticos, em 2001, esta amostra representa o tráfego, para aquele ano, de 21 milhões de passageiros, aproximadamente setenta por cento do total transportado no Brasil. Se considerarmos apenas as cem ligações aéreas domésticas mais densas do País, temos que 97 dessas ligações estão incluídas na amostra, abrangendo 98,8% do tráfego de passageiros destas ligações.

O estudo econométrico envolveu modelar o determinantes da variável *yield*, que representa a receita média por RPK (passageiro-pagante transportado vezes a etapa percorria). Essa variável é a melhor aproximação para o preço médio pago pelos passageiros e é bastante usada em transporte aéreo. Variáveis utilizadas na decomposição do *yield*: distância entre a cidade de origem e destino em quilômetros e seu termo quadrátrico, da densidade de tráfego medida pelo número de passageiros transportados na rota e seu termo quadrático, o PIB real e a taxa de câmbio real, o número de incumbentes, ou seja, o número de companhias aéreas "major" presentes na rota (exceto a Gol), a presença da empresa *low cost*, ou seja, uma variável *dummy* (binária) indicativa da presença da Gol na rota, o share de assentos *nonstop* (fatia de mercado da companhia aérea, no mercado de vôos *nonstop*), o share de assentos durante o pico (fatia de mercado da companhia aérea, no mercado de vôos durante horários de pico), o *share* na cidade de origem/destino (fatia de mercado da companhia aérea, na cidade de origem e destino), o tamanho médio da aeronave, dentre outras.

**Relevância do Estudo**

A competição em preços das companhias aéreas é tema de extrema importância para o setor, dado que envolve não apenas a rentabilidade da operação, mas também o comportamento dos agentes de oferta (operadoras) e de demanda (passageiros). Proporciona bases para um adequado entendimento do funcionamento da indústria e propiciando formas de acompanhamento econômico, antitruste e regulatório.

Por exemplo, um duopólio de duas companhias aéreas pode não ser, necessariamente, um mal à economia e ao consumidor. Essa tem sido uma situação observada no Brasil desde a derrocada da Varig, em 2006, quando, a partir de então, TAM e Gol detêm, juntas, mais do que 90% do mercado doméstico regular de passageiros. Um duopólio pode ser, sob certas circunstâncias, bastante competitivo. Em outras circustâncias, que, em geral são ditadas pela própria autoridade regulatória, pode ser bastante cartelizado ou apresentar comportamentos de colusão tácita e de divisão de mercado. A melhor forma para as autoridades antitruste (CADE, SEAE e SDE) e a autoridade regulatória (ANAC) acompanharem e investigarem essas circunstâncias é por meio de uma análise dos preços *vis-a-vis* os custos e as condições de demanda.

**Principais Resultados Deste Estudo**

Os principais resultados da modelagem econométrica realizada pelo NECTAR podem ser resumidos pela tabela a seguir:

Tabela 12 – Decomposição dos Preços das Passagens Aéreas Domésticas

| Variável | Efeito | Estatisticamente Significante? |
|---|---|---|
| PIB | Negativo | Sim |
| Taxa de Câmbio | positivo | Sim |
| Número de Incumbentes | Negativo | Não |
| Presença de Low Cost | Negativo | Sim |
| Share de Assentos Nonstop | positivo | Não |
| Share de Assentos no pico | positivo | Sim |
| Share na Cidade de Origem/Destino | positivo | Sim |
| Tamanho Médio da Aeronave | Negativo | Sim |

O cerne do presente estudo foi uma identificação empírica dos padrões de precificação das companhias aéreas nacionais, utilizando-se dados disponíveis para o setor. Os resultados apontaram para a relevância de variáveis macroeconômicas, como PIB e taxa de câmbio, na precificação das empresas. Identificou-se que, tudo o mais constante, o setor é mais competitivo nos períodos de maior atividade econômica (isto é, ao longo do ano), em contraposição aos períodos de alta estação turística, quando a competição é menos acirrada. Identificou-se também que o setor aéreo apresenta alta vulnerabilidade a choques cambiais, dado que boa parte dos custos operacionais está atrelada ao dólar (combustível, manutenção, *leasing*), que gera pressões altistas nos preços das viagens domésticas quando ocorrem desvalorizações da taxa de câmbio.

Outras conclusões do estudo de padrões de formação de preços do setor foram: 1. a entrada da companhia *Low Cost* Gol acarretou uma pressão baixista nos preços no período analisado; 2. as companhias aéreas possuem atributos que tornam o seu produto heterogêneo do ponto de vista do consumidor, atributos estes que conferem vantagens competitivas às empresas, resultando, em última instância, em um maior poder de precificação; 3. do lado dos custos, alguma evidência é encontrada quanto à existência de economias de densidade de tráfego, dado que os preços caem na medida em que aumenta o tamanho médio das aeronaves; e 4. o segundo semestre é o período onde os preços dos bilhetes aéreos costumam estar mais altos; isto se dá provavelmente devido à aproximação da alta estação turística (férias de final de ano).

## Recomendações de Políticas Públicas a Partir Deste Estudo

As autoridades devem incentivar a entrada de empresas eficientes, ainda mais se tiverem seu modelo de negócios do tipo "*Low Cost, Low Fare*", como era a Gol no início de suas operações. Empresas como essas induzem fortemente o tráfego por meio de estímulo em preço da demanda, além de promoverem a maior competitividade do setor. Em última instância, o consumidor pode ser extremamente beneficiado. Adicionalmente, tem-se que as autoridades devem estar alertas com relação os mecanismos de concessão de direitos de pouso e decolagem nos aeroportos congestionados e/ou sob estrita administração – os chamados "*slots*". Como foi visto com a significância estatística da variável "*Share* de Assentos no pico" e "*Share* na Cidade de Origem/Destino", temos que conceder slots é o mesmo que conceder direitos de usufruto de poder de mercado no transporte aéreo. O desenho correto de mecanismos de concessão de slots é fundamental para se garantir a geração de bem-estar econômico para o passageiro.

### *Pouso: O Mercado é Livre. Acompanhar preços para quê?*

Em notícia no Jornal do Brasil de 11 de novembro de 2007, intitulada "Governo quer fim do duopólio TAM e Gol", lia-se "*O governo prepara medidas para acabar com o duopólio da TAM e da Gol, detentoras de 86% do mercado doméstico de aviação comercial. Capitaneada pelo ministro da Defesa, Nelson Jobim, a missão envolve a Agência Nacional de Aviação Civil (Anac) e a Secretaria de Acompanhamento Econômico do Ministério da Fazenda (Seae). (...) A idéia do governo é aumentar a concorrência no setor a fim de reduzir o preço das passagens e aumentar o número de aeroportos atendidos pelas companhias aéreas.*". Argumentava na ocasião o presidente da Associação Nacional em Defesa dos Direitos dos Passageiros do Transporte Aéreo (Andep): "*O duopólio de TAM e Gol é altamente prejudicial, pois há a quebra do princípio da concorrência*".

Poucos meses depois, sem que ainda houvesse uma concretização das medidas anunciadas, em 07 de Julho de 2008, uma notícia do Jornal Estado de São Paulo divulgava que "*Passagens aéreas subiram 61,24% nos últimos 12 meses*", mostrando que a inflação nos preços das passagens aéreas no varejo alcançou, em junho, 61,24% no índice acumulado em 12 meses - o maior maior patamar em 11 anos. O levantamento foi realizado pela Fundação Getúlio Vargas. A pesquisa tomou por base dados do Índice Geral de Preços - Mercado (IGP-M) do mês anterior. A pergunta de analistas e sociedade como um todo, é se esse aumento de preços foi exclusivamente devido ao choque de custos dos preços do combustível de aviação, ou se há um componente disso devido ao duopólio TAM/Gol-Varig. Segundo as estatísticas da ANAC (Dados Comparativos Avançados, no acumulado de janeiro a outubro de 2008), as participações de mercado das empresas apontavam para um das maiores concentrações do setor em todos os tempos no Brasil:

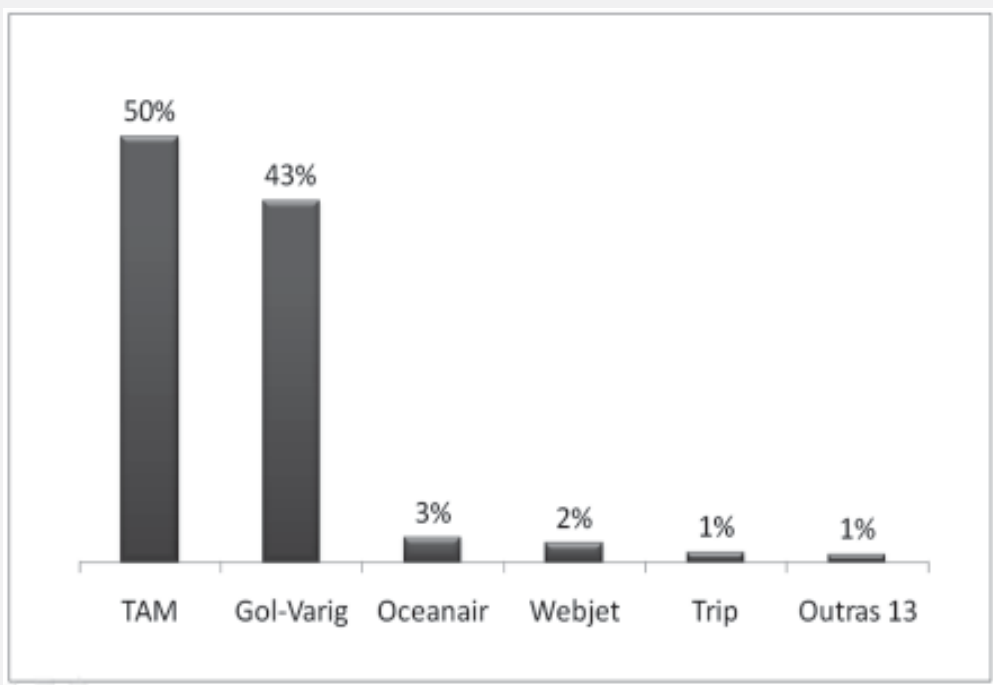

Figura 50 – Market Share no Mercado Doméstico de Passageiros (Jan-Out 2008)

Essa composição aponta para uma concentração acima de 4 mil pontos na escala do índice de concentração HHI. Para entender se esse nível de concentração é prejudicial, é preciso acompanhar os preços das companhias aéreas e efetuar estudos sistemáticos de decomposição desses preços entre fatores de custos, qualidade e efeitos colusivos. Essa tarefa é fundamental para o bem-estar econômico no setor, e deve ser realizada pela autoridade regulatória.

**5**

# Princípios Balizadores da Regulação e das Políticas Públicas do Transporte Aéreo

## Decolagem: A Economia Normativa do Transporte Aéreo

Há algum tempo que os economistas convivem com a diferença entre os conceitos de "Economia Positiva" e "Economia Normativa". Enquanto a Economia Positiva busca descrever a economia como ela *é*, a Economia Normativa busca tecer considerações sobre como a economia *deveria ser*. A primeira tem um cunho mais *descritivo* e a segunda, nitidamente mais *prescritivo*. Noventa e nove por cento do tempo de um bom economista é consumido analisando a economia e seus fenômenos, e, portanto, produzindo avaliações e estudos no campo da Economia Positiva. Por vezes, é chamado a opinar sobre elementos das políticas econômicas e acaba entrando na seara da Economia Normativa. Por exemplo, se diz que "*Uma regra regulatória que fixa os preços da infra-estrutura aeroportuária pode causar ociosidade de um aeroporto*", está nitidamente fazendo uma afirmação de Economia Positiva. Ao contrário, se disser que "*O regulador deveria baixar a tarifa dos aeroportos que apresentam ociosidade*", aí estará efetuando uma afirmação de Economia Normativa. É claro que podemos sempre pensar que a Economia Normativa pode e deve ser construída à luz de análises embasadas na Economia Positiva.

O presente trabalho visou analisar a indústria de transporte aéreo, com ênfase no seu segmento doméstico, ao longo das últimas décadas. Buscou produzir um conjunto de inferências de Economia Positiva acerca da evolução das políticas regulatórias e comportamento das companhias aéreas operadoras no setor. Por outro lado, objetivou-se também com o presente texto, efetuar algumas considerações de Economia Normativa. Essas acabam por se configurar em um conjunto de aspirações para o setor, ou sobre quais caminhos deveriam ser seguidos neste momento único por que passa a indústria, de reformulação do arcabouço regulatório rumo a um marco estável e de qualidade. Abaixo segue um conjunto de princípios sugeridos para a Reforma Regulatória do Transporte Aéreo.

## Princípio 1 – O Passageiro Sempre em Primeiro Lugar

As maiores e melhores empresas de inúmeros setores no País e pelo mundo afora já sabem disso há tempos, mas parece que esse princípio básico é, de tempos em tempos, esquecido tanto por operadores do transporte aéreo e pelo próprio regulador no Brasil. Não custa nada reafirmar em alto e bom tom: a regulação tem que ser totalmente voltada para o cliente, pois é no seu bem-estar que está a chave para a eficácia do regulador e para o desempenho das companhias aéreas.

Em suma, o regulador deve sempre procurar induzir as situações competitivas, nunca proteger empresas. A competição vai sempre gerar os resultados que são mais pró-consumidor do que qualquer coisa que ele possa fazer ou do que qualquer regra que ele possa inventivamente elaborar. As empresas devem cortejar o cliente, pois o passageiro satisfeito é quem gera os lucros. Nunca devem esquecer dele, mas também nunca negligenciem a própria eficiência produtiva: ganhar em qualidade mas perder em preços pode também ser fatal. Não devemos nunca negligenciar a sensibilidade a preço do passageiro em potencial, cada vez mais acostumado a nagevar pela *internet* e comparar ofertas, etc. Vide o exemplo do comportamento do "Novo Turista" em nosso Estudo de Caso 1 do capítulo anterior. Avião está longe de ser um restaurante, mas também não pode ser um purgatório, uma sala dos horrores. Os passageiros lembrarão de suas boas experiências a bordo quando saírem a procura do próximo vôo nos *sites* das empresas.

### Escala de Vôo: O Caso do "Overbooking" da Tam no Natal de 2006

Em uma notícia do Jornal Valor Online do dia 03/04/2007, continha a seguinte matéria: "*Anac descarta overbooking da TAM durante o Natal*". Tratava-se do resultado de uma investigação feita pelo regulador sobre o conjunto expressivo de atrasos e cancelamentos de vôos da TAM nas vésperas do Natal de 2006. Esse episódio foi um dos mais marcantes dos chamados "apagões aéreos" de 2006 e 2007. O relatório final da ANAC concluía que os problemas ocorridos naqueles dias foram o resultado de uma conjunção de fatores excepcionais, mas chamava à atenção para o fato de que a empresa assumiu riscos além de sua capacidade. De fato, nas 58 principais rotas da companhia, ficou claro, no entendimento da agência, que houve uma saturação nos vôos da companhia. Entre os dias 18 e 24 de dezembro de 2006, quando a crise ocorreu, foram realizadas 664 reservas acima da capacidade operacional da companhia, em 65 rotas. Essas reservas além da capacidade representaram, entretanto, apenas 0,13% do total das realizadas pela TAM no período. Argumentou a agência: "*Casos de overbooking certamente contribuíram para o agravamento da situação de crise. Entretanto, dada a quantidade de ocorrências do tipo (...), pode-se concluir que as mesmas não se revelaram como causa primária dos problemas.*" De qualquer maneira, a ANAC julgou que a decisão da companhia de atuar próxima do limite de sua capacidade, particularmente nas rotas mais importantes, foi importante para a detonação dos problemas daquela semana de dezembro de 2006.

A prática de *overbooking* tem sido amplamente utilizada por empresas aéreas por todo o mundo como uma maneira para lidar com os problemas de gerenciamento de capacidade em suas malhas de vôos. A prática é fundamentada em questões de custos das empresas e do comportamento do passageiro que efetua reserva, mas não comparece e nem efetua o cancelamento – chamado de passageiro "*no show*". O *overbooking* implica no aumento virtual da capacidade de uma aeronave, no âmbito do sistema de reservas, com o objetivo de minimizar as perdas de receitas decorrentes dos cancelamentos e *no-show* de passageiros no momento do embarque. O *overbooking* diz respeito à comercialização de assentos acima da capacidade real de um vôo, e não ao fato de se materializar, na hora do embarque, um número de passageiros acima do limite. A maioria dos vôos das empresas aéreas é sempre comercializada com alguma taxa de *overbooking*, mas nem todos os vôos resultam em preterição por excesso de passageiros, a empresa aérea deverá acomodar os passageiros portadores de bilhete de passagem em outro vôo, próprio ou de congênere, que ofereça um serviço equivalente para o mesmo destino, no prazo máximo de quatro horas do horário estabelecido no bilhete de passagem. Caso este prazo não possa ser cumprido, o usuário poderá optar entre viajar em outro vôo (por endosso) ou reembolso do bilhete de passagem. De acordo com o Termo de Compromisso, a compensação dos passageiros rejeitados poderá ser feita por uma das seguintes formas:

- ✦ Crédito compensatório (ressarcimento);

- ✦ *upgrade* de classe;

- ✦ passagem aérea adicional;

- ✦ outras compensações acordadas entre as partes.

A materialização do *overbooking* é extremamente prejudicial à imagem das companhias aéreas, que podem perder vantagens competitivas importantes caso o seu *overbooking* esteja mal configurado e exponha os passageiros desnecessariamente às situações de preterição nos vôos. Empresas aéreas com gestão de qualidade investem pesadamente na habilidade de efetuar previsões acuradas do *no-show*, para evitar problemas com sua imagem no mercado.

**Princípio 2 – Ambiente de Negócios com Baixo Risco Regulatório**

Como vimos ao longo deste trabalho, o transporte aéreo já é naturalmente alvo de todo tipo de instabilidades e "choques exógenos" advindos tanto da demanda quanto da oferta. Para complicar, existe a economia como um tudo, cada vez mais globalizada e sujeita a crises. Existem os atentados terroristas, as crises financeiras globais, as desvalorizações cambiais, etc. Em um ambiente naturalmente instável como esse, o regulador deve buscar minimizar o risco regulatório e conter a sua ânsia reformista. Mudanças para melhor? Sim, reformas regulatórias bem-intencionadas são sempre bem-vindas, mas somente as cuidadosamente planejadas e antecipáveis pelos agentes.

Outra coisa: vamos dizer um "não" à regulação discricionária. Impor mecanismos de regulação esporádicos, contigenciais, apenas com o efeito de "apagar incêndios", não é sábio procedimento. Ainda mais se o intuito for suavizar as pressões competitivas, pois isso pressupõe a existência de pisos de lucratividade na operação, um conceito incompatível com a livre concorrência. Intervir no mercado, quer seja "congelando" a oferta, quer seja impondo restrições à mobilidade de ajuste das variáveis estratégicas de todas as companhias aéreas operadoras, é impor essa regulação discricionária, isto é, aquela que emerge apenas quando da conveniência do regulador. Estes mecanismos prejudicam a estabilidade das regras e ao ambiente de negócios como um todo. O regulador já sabe, pelo conteúdo apresentado ao longo deste livro, que a lucratividade do setor é cíclica e que, de tempos, em tempos, haverá pressões reformistas em prol do "setor como um todo". Deve ter cuidado com esses movimentos e estar atento para o risco de captura advindo do poder do *lobby* das empresas.

Isso não impede que haja controles antitruste e incentivos de política industrial no mercado, que, em geral, são dispositivos que se aplicam a um subconjunto do mercado (ex. uma firma à beira da insolvência ou uma firma que pratica preços predatórios), e não a todo o mercado, prejudicando a dinâmica competitiva, o investimento planejado e a inovação. Aliás, já está comprovado que o antitruste tem papel importantíssimo na maior credibilidade regulatória e na maior indução de investimentos nos diversos setores de um País.

**Princípio 3 – Metas de Acesso ao Transporte Aéreo**

Quanto mais pessoas, mais localidades e mais regiões do Brasil tiverem acesso ao transporte aéreo, melhor. Mais bem-estar econômico estará sendo gerado, o que é um indicador de desenvolvimento, sem contar nos *spillovers* que o transporte aéreo proporciona. Melhorar a acessibilidade ao setor de transporte aéreo significa não apenas focar no estímulo a competição via preço – onde pessoas que não tinham o modal em sua cesta de consumo são incorporadas, o que é extremamente importante. Mas significa também cuidar do marco regulatório das infra-estruturas aeroportuárias e garantir um fluxo de investimentos que mantenha as obras e reformas em pequenos aeroportos pelo Brasil afora. Com a malha aeroportuária bem cuidada e com os incentivos adequados à operação da aviação regional e da aviação geral, teremos alavancada a cobertura ao longo do território nacional.

> *Acessibilidade ao setor aéreo significa a possibilidade de ter o transporte por modal aéreo na "cesta de consumo" de um dado indivíduo ou família. Indivíduos cuja renda não permite participar do mercado – mesmo que com uma freqüência reduzida de viagens anuais – ou cuja localidade não pertence à zona de influência de nenhum aeroporto, estão totalmente à margem do setor aéreo. Incrementar a acessibilidade significa atrair uma parcela desses indivíduos para a carteira de clientes das companhias aéreas. Significa, portanto, incrementar a universalização da indústria e potencializar os seus efeitos de alavancagem do bem-estar econômico.*

Quinze milhões de pessoas vivem atualmente na Região Norte[38], região onde o transporte terrestre é, em muitos casos, extremamente incipiente. Sabendo que o monitoramento e o desenvolvimento da Amazônia são questões-chave e estratégicas para o futuro do País e sua trajetória rumo ao desenvolvimento sustentável, tem-se que pensar o transporte aéreo também à luz dessa problemática. Questões como o acesso a pequenas comunidades, ou a regiões consideradas estratégicas (*public service offers*, PSO), podem ser tratadas no âmbito de uma política industrial, que avalie as possibilidades de geração de recursos orçamentários e financiamentos (mas nunca subsídios-cruzados), desde que resguardados os princípios relativos ao livre mercado.

**Princípio 4 – Desregulação Econômica com Fiscalização Técnica**

O setor deve ser livre para produzir seus resultados econômicos, o que significa que as companhias aéreas devem ter garantida a liberdade de encontrar as melhores estratégias de abordagem dos consumidores de forma maximizar seu desempenho, seja visando o lucro ou a expansão do *market share*. Potenciais interessados em operar no transporte aéreo devem ser rigorosamente inspecionados do ponto de vista técnico, mas deve-se tomar cuidado com requisitos econômicos à entrada para que os mesmos não se transformem em verdadeiras barreiras com prejuízo às alternativas de consumo dos passageiros. Em suma, uma vez cuidada da fiscalização das operações e das condições de segurança de vôo, deve-se garantir o livre acesso, livre mobilidade e a liberdade estratégica.

**Livre Acesso e Livre Mobilidade**

A entrada e a saída de operadoras na indústria do transporte aéreo deve ser livre, resguardados a). os requisitos de regulação técnica e certificação; b). o objetivo de minimização de descontinuidades no provimento do serviço, quando possível; c). os direitos de consumidor dos usuários. As concessões de operadoras seriam por prazo determinado, obedecendo, numa primeira etapa, a um processo seletivo geral, que coincidiria com o preenchimento dos requisitos de certificação da companhia aérea interessada. A saída de operadoras em dificuldade deve ser vista como benéfica para a competição, devendo ser desenvolvidos mecanismos facilitadores e formas de suavização de seus impactos de curto prazo nas relações de consumo, trabalhistas e com os credores. Os limites de participação de capital estrangeiro devem ser revistos e ampliados, medida esta que potencializaria a alavancagem de novos empreendimentos, dinamizando o setor. Empresas pequenas poderiam ter limites de participação de capitais estrangeiros maiores do que empresas grandes, o que fomentaria a contestabilidade aos mercados por meio do maior vigor financeiro daquelas.

Livre mobilidade é indicativa da livre entrada e saída de operadoras já certificadas nos mercados aéreos existentes (ligações), também resguardadas as condições de prestação de serviço regular (prazos mínimos para saída) e infra-estrutura disponível. Deve ser vedado o uso de critérios econômicos para coibir a livre mobilidade das operadoras, como, por exemplo, a necessidade de estudos de viabilidade econômica observada em um passado recente.

> *Livre acesso e livre mobilidade significam que é possível a uma operadora interessada ingressar no setor aéreo e escolher operar as rotas que melhor lhe convier no sentido de maximização de seus lucros. Novas entrantes saudáveis financeiramente podem ser extremamente importantes na indução de eficiência econômica no setor aéreo, pois incrementam a contestabilidade aos mercados. Há que se garantir, entretanto, que a liberdade estratégica não sirva de mecanismo para práticas anticompetitivas das firmas instaladas (incumbentes). Os controles antitruste, mais do que a regulação, devem ser usados para cuidar desses potenciais problemas.*

## Liberdade Estratégica

Este inclui o regime de liberdade tarifária, que foi consagrado na Lei de Criação da ANAC. Mas garantir a liberdade estratégica é muito mais do que garantir apenas a liberdade tarifária. Envolve o livre ajuste de todas as variáveis estratégicas pelas companhias no mercado – freqüências de vôo, tamanho de aeronaves, propaganda, decisões de fusões, alianças, acordos de compartilhamento, etc. Deve-se resguardar, obviamente, as condições de regulação técnica e o mínimo de formalismo regulatório – prazos e procedimentos de análise de HOTRANs, por exemplo –, que garantam a previsibilidade das ações do regulador em face às ações e pleitos das empresas.

**Princípio 5 – Eficiência Econômica e Regras Transparentes nas Alocações de Infra-Estrutura**

Uma vez certificadas as operadoras, o princípio da Livre Mobilidade garante que qualquer uma delas possa ter acesso a todo e qualquer mercado. Entretanto, pode haver conflitos de interesses por conta de sobreposição de malhas, situação que é agravada na existência de aeroportos congestionados e com regras de *slots*. Há, como visto, que se definir uma forma de alocar os recursos escassos de forma a atender as necessidades das firmas reguladas que, ao mesmo tempo, se paute pela eficiência econômica e legalidade, evitando-se os chamados *grandfather rights*, isto é, a concessão sem critérios e que se perpetua *ad eternum*. Mecanismos de mercado para a alocação de *slots*, onde se introduza a concorrência pelo acesso à infra-estrutura escassa, são desejáveis.

**Princípio 6 – Todo Poder à Coleta e Disseminação de Dados**

A primeira tarefa das autoridades regulatórias em um ambiente desregulado é entender o mercado para acompanhá-lo e antecipar os impactos de possíveis ações e mecanismos regulatórios. Sem um adequado entendimento do que se passa no mercado, um regulador fica à mercê das argumentações dos regulados e, portanto, exposto às tentativas de captura dos mesmos. A captura do regulador pelos regulados se materializa na situação em que a regulação acaba por favorecer os regulados em detrimento do consumidor. Pior: o regulador mal informado não sabe o que se passa com o consumidor – justamente o seu "público-alvo", a razão de ser de todo o aparato regulatório e da própria existência da autoridade.

Entretanto, a função de acompanhamento econômico incorre em sérios riscos caso a qualidade da informação não seja a melhor possível. É claro que não existe monitoramento governamental de empresas privadas que isento do problema da assimetria de informação regulador-regulados – isto é, da não-observabilidade perfeita de custos, preços, perfil do consumidor transportado, etc. Por outro lado, **não existe regulação de qualidade em regime de fortes assimetrias de informação**. A qualidade da regulação está fortemente atrelada à qualidade da informação disponível ao regulador. Os dados devem ser, sempre que possível, publicados e disponibilizados em formato de consulta (*query*) no *website* do regulador.

Os responsáveis pela coleta e disseminação dos dados relativos ao setor devem ser, na estrutura hierárquica da ANAC, funcionalmente tão importantes quanto os próprios condutores da regulação. Deve também haver a formação de uma base de dados sistêmica com informações sistematizadas do setor, de forma a incrementar a qualidade do monitoramento econômico e reduzir as assimetrias de informação regulador-regulados.

É fundamental que haja coleta periódica de amostras de bilhetes aéreos, no estilo das pesquisas realizadas atualmente pelo Departamento de Transportes do governo dos Estados Unidos. Deve também haver o aperfeiçoamento das coletas e bases de dados estatísticos atualmente existentes, com maior investimento em sistemas, bem como a disponibilização sistemática e irrestrita dos dados entre os órgãos conveniados e, com algumas restrições (definidas de forma transparente), à sociedade como um todo. O conceito do que é "informação estratégica" deve ser revisto e delimitado temporalmente, de forma a assegurar o direito de acesso da sociedade como um todo aos dados e à história do setor, resguardados os interesses das companhias aéreas no mercado.

*A Base de Dados de número DB1B representa a chamada "Airline Origin and Destination Survey" do Departamento de Transportes dos Estados Unidos. Trata-se, seguramente, de uma das fontes de dados de maior abrangência e qualidade do transporte aéreo mundial. A DB1B ilustra bem a importância que as autoridades norte-americanas conferem não apenas à transparência pública, mas também à necessidade de redução das assimetrias de informação entre regulador e regulados. A DB1B é composta por meio de uma amostra de dez por cento de todos os bilhetes comercializados pelas companhias aéreas dos Estados Unidos. Os dados incluem todos os detalhes do itinerário e do bilhete, sendo fonte de informação inestimável para a elaboração de planos de ação regulatória e de políticas públicas, projeções de demanda e de saturação da malha aeroportuária, planos de investimento em aeroportos e em linhas aéreas, estudos de mercado, investigações antitruste, dentre outros inúmeros benefícios. A base de dados é quase toda pública e pode ser consultada no site www.transtats.bts.gov.*

## Princípio 7 – Binômio Embasamento Técnico-Transparência da Tomada de Decisão

A tomada de decisão das autoridades – seja com relação ao antitruste, à política industrial, etc. – deve estar adequadamente embasada em métodos de investigação e previsão. Deve também haver o máximo de transparência, e o atual mecanismo de consulta pública, com exposição de motivos pelo regulador, deve ser mantido e aperfeiçoado.

Deve-ser buscar a formação de uma rede de cooperação regulador-centros de pesquisa, com convênios específicos, de forma que o meio acadêmico possa participar dos debates quanto ao acompanhamento do mercado desregulado. Estudos acadêmicos, quando bem embasados em dados reais e de qualidade, podem auxiliar em muito na tomada de decisão. Como a maior parte das questões relativas à tomada de decisão em políticas públicas, em geral, envolve agilidade, e, por outro lado, os estudos acadêmicos costumam demandar tempo considerável, torna-se fundamental que os convênios sejam firmados e as demandas por estudos sejam planejadas, antecipando as possíveis necessidades futuras de estudos; também é fundamental o acesso irrestrito às bases de dados pelas entidades conveniadas. Exemplo de estudos que poderiam ser realizados: estimação de elasticidade-preços nos diversos mercados – informação crucial para uma autoridade que deseja testar a hipótese de predação, cartelização, ou fazer estudos de projeção da demanda por um aeroporto, por exemplo. Importante também é serem criados simpósios regulador-centros de pesquisa, onde haja maior troca de informação, apresentação e divulgação dos estudos já realizados.

## Princípio 8 – Coordenação das Políticas Setoriais

Esquemas de coordenação (*coordination scheme*) são tema conhecido em macroeconomia, mas carecem de maior implementação no âmbito das políticas setoriais. No caso do transporte aéreo, existem três tipos de autoridades responsáveis pelo andamento do setor: as responsáveis pela condução da Política Regulatória (ANAC e Ministério da Defesa/SAC), da Política Industrial (BNDES, Ministério do Turismo, dos Transportes e da Fazenda), e da Política de Defesa da Concorrência (SEAE, SDE e CADE). É fundamental que haja coordenação entre essas políticas, quer seja no âmbito das decisões estratégicas do CONAC (Conselho de Aviação Civil), quer seja no âmbito das decisões táticas relativas ao sistema aéreo como um todo. Deve haver a disseminação de dados do setor aéreo entre os diversos entes do sistema de autoridades, no sentido de permitir uma maior coordenação das ações governamentais.

## Princípio 9 – Flexibilização e Constituição do Marco Regulatório das Infra-Estruturas

Os "Apagões Aéreos" de 2006 e 2007 tiveram como conseqüência principal a escassez de infra-estrutura aeroportuária e de controle de espaço aéreo. Há um nítido descolamento entre o ritmo de crescimento do transporte aéreo – este movido pela dinâmica do mercado – e o ritmo dos investimentos nas infra-estruturas – estas quase sempre movidas pelos ânimos governamentais de plantão. É fundamental haver uma "Flexibilização" também das infra-estruturas, de forma que o espírito empresarial e o empreendedorismo possa ser encampado também nesses setores. Há que se constituir um verdadeiro Marco Regulatório das Infra-Estruturas Aeroportuária e de Controle do Espaço Aéreo, que garanta a qualidade da gestão e um fluxo permanente de investimentos nessas importantes áreas do setor aéreo nacional.

## Princípio 10 – Livre Mercado, Sim, Mas com Defesa da Concorrência e do Consumidor

O acompanhamento de defesa da concorrência deve ser contínuo e multi-institucional. Deve ser realizado pela ANAC em conjunto com o Sistema Brasileiro de Defesa da Concorrência – composto por Secretaria de Acompanhamento Econômico (SEAE), Secretaria de Direito Econômico (SDE), e Conselho Administrativo de Defesa Econômica (CADE). O regulador do transporte aéreo deve ser incumbido do acompanhamento continuado das práticas de mercado das companhias aéreas operadoras, estabelecendo, para isso, os convênios com os demais órgãos do Sistema Brasileiro de Defesa da Concorrência. A Lei de Criação da ANAC, prevê, em seu Art. 6º, que "*Com o objetivo de harmonizar suas ações institucionais na área da defesa e promoção da concorrência, a ANAC celebrará convênios com os órgãos e entidades do Governo Federal, competentes sobre a matéria*". Adicionalmente, no Parágrafo único do Art. 6º, prevê que "*Quando, no exercício de suas atribuições, a ANAC tomar conhecimento de fato que configure ou possa configurar infração contra a ordem econômica, ou que comprometa a defesa e a promoção da concorrência, deverá comunicá-lo aos órgãos e entidades referidos no caput deste artigo, para que adotem as providências cabíveis*".

Um arcabouço onde a ANAC tenha também funções de acompanhamento antitruste é mais eficiente do que um arcabouço alternativo onde ela não tenha essa competência. Isso porque com certeza há ganhos em economias de escala e custo de transação, dado que a agência está muito mais envolvida com a rotina econômica do setor, e que seus quadros são formados de pessoal mais experiente no trato de questões setoriais específicas.

Em suma, deve-se buscar eliminar barreiras a entrada e incrementar a contestabilidade aos mercados. Adicionalmente, aplica-se o controle de condutas em paralelo, de forma a sinalizar às operadoras que existe uma autoridade acompanhando a rotina econômica do setor.

---

### *Pouso: O Nascimento e Morte de um Gigante* - *Code share Varig-TAM*

O período de 2003 e 2005 foi marcado pelo anúncio das duas então maiores empresas aéreas do País, Varig e Tam, de que pretendiam se tornar uma única só. A fusão entre as duas companhias seria precedida, por um acordo de compartilhamento de aeronaves, um acordo do tipo code share. Um acordo de "*code share*" (comparilhamento de código de vôo) é uma prática comercial e operacional comumente utilizada por grandes empresas aéreas pelo mundo afora, e se refere ao fato de que um vôo operado por uma companhia aérea será comercializado conjuntamente como se fosse um vôo de mais uma ou mais empresas. Na prática, reserva-se espaço em um avião - um determinado número de assentos - para que a aliada utilize, em troca de disponibilidade de assentos nos aviões da aliada.

No período do *code share* Varig-TAM, observou-se os maiores índices de concentração da história do mercado doméstico brasileiro, segundo a escala HHI - que vai de zero a dez mil. O acordo foi assinado em 06 de fevereiro de 2003, em um em um protocolo de entendimento aventando a possibilidade de constituição de uma nova empresa holding de capital aberto. Esse foi o início de um planejamento de fases para o que intencionavam finalizar com uma fusão de fato entre as duas maiores empresas de transporte aéreo no Brasil a época. Iniciado em mais precisamente em março de 2003, o compartilhamento foi criado num momento de dificuldades financeiras para as duas companhias, advindas da entrada da Gol e da crise que se seguiu ao 11 de setembro de 2001, e da fortíssima apreciação do dólar de 2002. O acordo visava ser a primeira etapa de um processo de fusão entre as duas companhias. Com a parceria, houve importantes ganhos de eficiência, dado que ambas reduziram seus custos operacionais. Do ponto de vista estratégico, a TAM foi extremamente beneficiada, pois na época era a segunda maior empresa aérea do País (atrás da Varig), conseguindo inverter os papéis e liderar até hoje o transporte aéreo no Brasil. A lucratividade também se inverteu, como atesta a Figura 51.

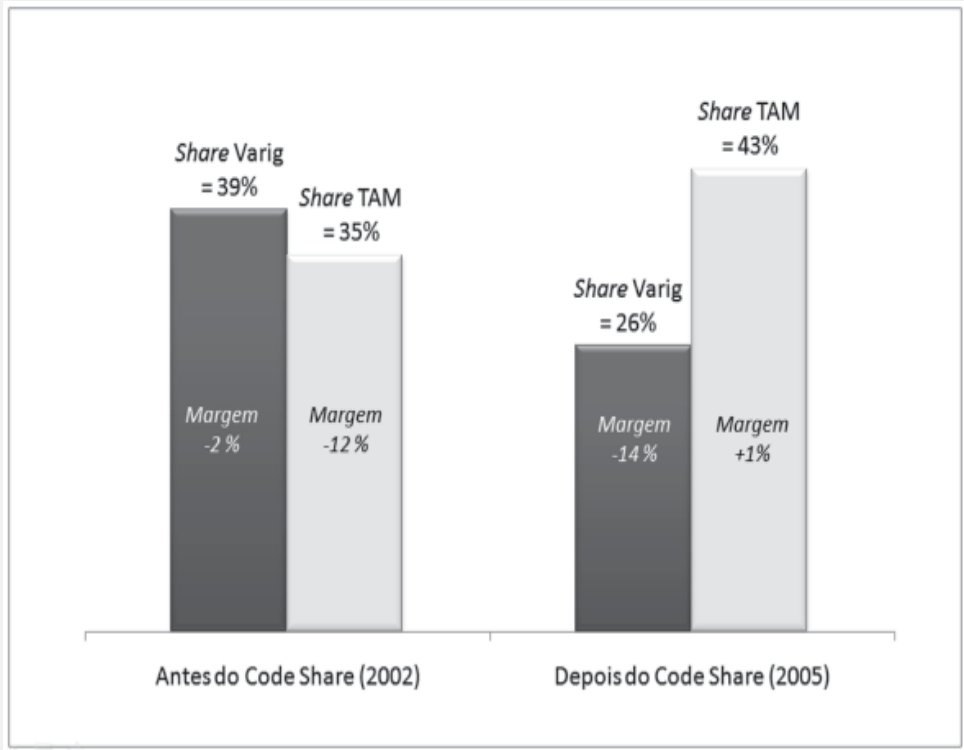

Figura 51 – Desempenho das Empresas Antes e Após o *Code Share*

O acordo durou pouco tempo, entretanto. E tampouco levou à fusão. Em maio de 2005 foi oficialmente extinto, sendo seguido por importantes processos competitivos e guerras de preços, que estimularam fortemente a demanda naquele ano. No estudo do NECTAR "Simulação de Fusões: Aplicação ao Transporte Aéreo", o pesquisador Moisés Vassallo investiga os efeitos do Code share Varig-TAM. O estudo demonstra que no período pós-fusão houve uma elevação real de 16% nos preços da Varig e 11% nos preços da Tam. O estudo elabora um modelo que permitiria à autoridade antitruste antever os aumentos mesmo antes da assinatura do acordo.

# Notas

[12] Fonte: Secretaria dos Transportes do Governo do Estado de São Paulo (2005).

[13] Fonte: Anuários Estatísticos da Agência Nacional de Aviação Civil (vários anos).

[14] Índice com média anual igual a 100. Fonte: Anuários Estatísticos da Agência Nacional de Aviação Civil (2000-2007) e cálculos próprios.

[15] Fonte: www.darinlee.net, todos os direitos reservados. Cálculos de *market share* de número de passageiros para o mercado doméstico.

[16] Fonte: Boguslaski, Ito e Lee (2004).

[17] Fonte: ANAC, Dados Comparativos Avançados (2001-2005). *Market share* em RPK.

[18] Igual ao número de viagens anuais dividido por 365 dias vezes 24 horas vezes 60 minutos

[19] Fonte: ANAC, vide nota de rodapé anterior.

[20] Fonte: The New York Times Company, com todos os direitos reservados.

[21] Considerando-se um cálculo com a etapa média de passageiro reportada no Anuário Estatístico da ANAC (2007), vol I, e considerando-se o número de viagens por ano mediano da Pesquisa de Origem e Destino do Transporte Rodoviário e Aéreo do Estado de São Paulo, realizado pela Secretaria dos Transportes do Governo do Estado de São Paulo (2005).

[22] Curiosamente, o texto original da lei que criou a ANAC continha dispositivos que fortaleciam o *grandfathering*, e foram vetados por orientação do Ministério da Fazenda: "*Art. 48. [Caput Vetado] Os contratos de concessão em vigor relativos às outorgas de serviços aéreos cujos vencimentos se verifiquem antes de 31 de dezembro de 2010 ficam automaticamente prorrogados até aquela data.*" E "*Art. 48. § 2º [vetado] Enquanto forem atendidas as exigências regulamentares de prestação de serviço adequado, ficam mantidos os eslotes atribuídos às empresas concessionárias de serviços aéreos*".

[23] Lei nº 11.182, de 27 de Setembro de 2005.

[24] Fonte: Anuários Estatísticos do Departamento de Aviação Civil (Volumes I e II, 1974-2003) e Informativo PL-3 (2004) e cálculos próprios. As novas companhias entrantes, a partir dos anos 1990, foram as seguintes: Abaeté, Gol, Helisul (depois Tam), Interbrasil (Transbrasil), Itapemirim, Meta, Oceanair, Pantanal, Passaredo, Penta, Presidente, Puma, Rico, Tavaj, Taf, Total, Trip.

[25] Fonte: Anuários Estatísticos do DAC e ANAC (Volumes I e II, 1974-2007 e Dados Comparativos Avançados até Outubro de 2008) e cálculos próprios.

[26] Fonte: Anuários Estatísticos do DAC e ANAC e cálculos próprios.

[27] Ibid.

[28] Ibid.

[29] Ibid.

[30] Ibid.

[31] Ibid.

[32] Ibid.

[33] Ibid.

[34] Ibid.

[35] Salgado, L. H., Pereira, E. A. e Oliveira, A. V. M. (2007) Organização Industrial do Turismo. Série de estudos sobre o turismo, disponível em http://works.bepress.com/lucia_salgado/8

[36] Base de Microdados da Pesquisa de Caracterização e Dimensionamento do Turismo Doméstico de 2006.

[37] Fonte: HOTRAN/ANAC, IBGE, IPEADATA e NECTAR. "Pre-Liberalização" = período entre 1998 e 2000; "Pós-Liberalização" = período entre 2006 e 2008.

[38] Fonte: Ipeadata, 2006. Estimativas das populações residentes, calculadas com data de referência em 1º de julho de cada ano civil.

# Bibliografia